\documentclass[aps,prd,12pt,notitlepage,nofootinbib,tightenlines,options)]{revtex4-1}
\usepackage{amsmath}
\usepackage{bm}
\usepackage{times}
\usepackage{braket}
\usepackage{color}
\usepackage{epsfig}
\usepackage{slashed}
\usepackage{hyperref}
\usepackage{float}
\usepackage{diagbox}
\usepackage{multirow}
\definecolor{Red}{rgb}{1.,0.,0.}

\definecolor{Blue}{rgb}{0.,0.,1.}

\definecolor{nicered}{rgb}{0.7,0.1,0.1}
\definecolor{nicegreen}{rgb}{0.1,0.5,0.1}

\hypersetup{colorlinks,citecolor=nicegreen,linkcolor=nicered}

\begin{document}

\newcommand{\beq}{\begin{eqnarray}}
\newcommand{\eeq}{\end{eqnarray}}
\newcommand{\non}{\nonumber \\ }
\newcommand{\psl}{ p \hspace{-2.0truemm}/ }
\newcommand{\qsl}{ q \hspace{-2.0truemm}/ }
\newcommand{\epsl}{ \epsilon \hspace{-2.0truemm}/ }
\newcommand{\nsl}{ n \hspace{-2.2truemm}/ }
\newcommand{\vsl}{ v \hspace{-2.2truemm}/ }

\newcommand{\jpsi}{ J/\Psi }
\newcommand{\cala}{ {\cal A} }
\newcommand{\calb}{ {\cal B} }


\def \cpc{ { Chin. Phys. C } }
\def \ctp{ { Commun. Theor. Phys. } }
\def \csb{{ Chin. Sci. Bull. } }
\def \sbu{{ Sci. Bull.  } }

\def \epjc{{ Eur. Phys. J. C} }
\def \ijmpa{ { Int. J. Mod. Phys. A } }
\def \jhep{{ JHEP } }
\def \jpg{ { J. Phys. G} }
\def \mpla{ { Mod. Phys. Lett. A } }
\def \npb{ { Nucl. Phys. B} }
\def \plb{ { Phys. Lett. B} }
\def \ppnp{ Prog.Part. $\&$ Nucl. Phys. }
\def \pr{ { Phys. Rep.} }
\def \prd{ { Phys. Rev. D} }
\def \prl{ { Phys. Rev. Lett.}  }
\def \ptp{ { Prog. Theor. Phys. }  }
\def \zpc{ { Z. Phys. C}  }

\def \thl {{\theta_\ell}}
\def \thK {{\theta_{K^\ast}}}
\def \re{\text{Re}}
\def \im{\text{Im}}
\def \eff{{\text{eff}}}
\def\Sin{\text{sin}}
\def\Cos{\text{cos}}

\title{Study of  \texorpdfstring{$\bar{B}_s\to K^{(*)}\ell^+ \ell^-$ } decays  in  the PQCD factorization approach with the lattice QCD input}
\author{Su-Ping  Jin$^{1}$ }  \email{2223919088@qq.com}
\author{Xue-Qing Hu$^{1}$ } \email{hu-xueqing@qq.com}
\author{Zhen-Jun Xiao$^{1,2}$  } \email{xiaozhenjun@njnu.edu.cn}
\affiliation{1.  Department of Physics and Institute of Theoretical Physics,
Nanjing Normal University, Nanjing, Jiangsu 210023, People's Republic of China,}
\affiliation{2. Jiangsu Key Laboratory for Numerical Simulation of Large Scale Complex Systems,
Nanjing Normal University, Nanjing 210023, People's Republic of China}
	\date{\today}
\begin{abstract}
In this paper,  we studied systematically the semileptonic decays $\bar{B}_s \to K^{(*)} \ell^+ \ell^-$ with $l^-=(e^-,\mu^-,\tau^-)$ by using the perturbative QCD (PQCD)
and the  ``PQCD+Lattice" factorization approach, respectively.
We first evaluated all relevant form factors  $F_i(q^2)$ in the low $q^2$ region using the PQCD approach, and we  also take the available Lattice QCD results
at the end point $q^2_{max}$ as additional inputs to improve  the extrapolation  of  the form factors  to the high $q^2$ region.  We calculated
the branching ratios and other twelve kinds of physical observables: ${\cal A}_{FB}(l)$,  $P_L$, $F_L^{K^*}$ and the  angular observables
$P_i $ with $i=(1,2,3)$ and $P^\prime_j$ with $j=(4,5,6,8)$.
From our studies, we find the following points:
(a) for $ \bar{B}_s \to K  l^+ l^-$ decays,   the PQCD and ``PQCD+Lattice"  predictions for branching ratios (BRs) ${\cal B}(\bar{B}_s \to K  l^+ l^-)$, the ratios of the
BRs $R_K^{e\mu}$ and $R_K^{\mu\tau }$, and the longitudinal polarization asymmetry  of the leptons $P_L$ agree well  within errors;
(b)  the  PQCD and ``PQCD+Lattice"   predictions for the CP averaged branching ratio ${\cal B}(\bar{B}_s \to K^* \mu^+ \mu^-)$ are
$(3.17^{+0.95}_{-0.78})\times 10^{-8}$ and $(2.48^{+0.56}_{-0.50})\times 10^{-8}$ respectively, which  agree well with the LHCb measured
value  $(2.9\pm 1.1) \times 10^{-8}$ and  the light-cone sum rule (LCSR) prediction;
(c) for the ratios  $R_{K^\ast}^{e\mu }$ and $R_{K^\ast}^{\mu \tau}$,    the PQCD and ``PQCD+Lattice" predictions agree well with each other
and have a small  error less than $10\%$;
(d) for the direct CP asymmetries $\cala_{CP}$ of all considered decay modes, they  are always very small as expected: less than $5\%$ in magnitude; 
(e) for  the angular observables $P_{1,2,3}$ and $P^\prime_{4,5,6,8}$ ,   our theoretical  predictions  for each kind of lepton
are consistent within errors;
(f) the theoretical predictions of  the angular observables $P_3$ and $P'_{6}$ are less than $10^{-2}$ in size,  but the magnitude of
$P_{1,2}$ and $P'_{4,5}$ are larger than $0.2$;  and
(g) the PQCD and ``PQCD+Lattice"  predictions of  the binned values of all considered observables  in the two $q^2$-bins $[0.1-0.98]$GeV$^2$ and
$[1.1-6]$GeV$^2$ generally agree with each other  and  are also consistent with the LCSR results within errors.
We believe that above  predictions could be tested by future LHCb and Belle-II experiments.
\end{abstract}

\maketitle

\section{Introduction}\label{sec:1} 

The lepton flavor universality (LFU), as one of the distinctive hypotheses of the standard model (SM),  requires the same kinds of  couplings
between the gauge bosons and the three families of leptons except for mass effects.
However, the recently reported $R_{K}$ and $R_{K^{\ast}}$ anomalies bring a primary hint of the LFU violation.
The measured values of the ratios $R_{K}$ and $R_{K^{\ast}}$, defined as the ratios of the branching fractions (BRs)  ${\cal B}(B \to  K^{{(\ast)}} \mu^+ \mu^-)$
and ${\cal B}(B \to  K^{{(\ast)}} e^+ e^-)$ \cite{Wei:2009zv} ,
are clearly smaller than the SM predictions \cite{Ali:1999mm,Beneke:2001at,Chen:2001ri,Ali:2006ew,Bobeth:2008ij,Egede:2008uy,Altmannshofer:2008dz}:
 the deviation is about 2.6$\sigma$ for $R_{K}$ and 2.3$\sigma$ for $R_{K^{\ast}}$
\cite{Aaltonen:2011cn,Lees:2012tva,Aaij:2014ora,Khachatryan:2015isa,Aaij:2016flj,Aaij:2017vbb}.
 In addition, the LHCb experiment first observed the so-called $P^\prime_5$ anomaly, a sizeable discrepancy at 3.7 $\sigma$ between the measurement
 and the SM prediction in one bin for the angular observables $P^\prime_5$ \cite{Aaij:2013qta,Aaij:2015oid}.

If the above mentioned anomalies are indeed the signal of the LFU violation in $b \to s \ell^+ \ell^-$ decays,
it must appear in the similar process  $b \to  d \ell^+ \ell^-$,  because they are the same kinds of  flavor-changing neutral current (FCNC) transitions at the quark level
with the differences of CKM matrix elements ( $V_{td}$ vs $V_{ts}$ ) and the masses ( $m_d$ vs $m_s$).
As a consequence of the GIM mechanism \cite{gim70},  the flavor structure of the SM theory permits  the FCNC to arise at the loop level only,
leaving some space for heavy new degress of freedom to contribute to these rare processes \cite{Ciuchini:2017mik}.
With the same quark level $b\to d \mu^+ \mu^-$ transitions,  the exclusive $B^\pm\to \pi^\pm \mu^+\mu^-$ and $B_s^0\to \bar{K}^{*0} \mu^+\mu^- $ decays
have been measured recently by LHCb experiment\cite{jhep10-034,Aaij:2018jhg} :
\beq
\calb(B^\pm \to \pi^\pm \mu^+\mu^-)  &=&       (1.83\pm 0.24(stat.)\pm 0.05(syst.))\times 10^{-8}, \label{eq:exp1}\\
\calb(B^0_s \to \overline{K}^{*0} \mu^+ \mu^-)&=&(2.9 \pm 1.0(stat.) \pm 0.2(syst.) \pm0.3(norm.))\times 10^{-8},  \label{eq:exp2}
\eeq
they agree well with those currently available SM predictions as given for example in
Refs.~\cite{prd77-014017,prd89-094021,prd90-013002,prd92-074020,prl115-152002,Wang:2012ab,Wang:2013ix,Khodjamirian:2017fxg,Kindra:2018ayz} .

In this paper ${K}^{*0}$ denotes a vector $K^{\ast 0}(892)$ meson, which is reconstructed in the $K^+\pi^-$ final state experimentally
by selecting candidates within 100 \rm{MeV}$/c^2$ of the mass \cite{pdg2018,Aoki:2019cca}.
In LHCb experiment, however, no attempt is made to separate the vector $K^{\ast 0}$ from the S-wave or other broad contributions which may present in the
selected $K^+\pi^-$ pair \cite{Aaij:2017vbb}.
Fortunately, the S-wave fraction contribution to the $B^0 \to \overline{K}^{*0} \mu^+ \mu^-$ mode has been measured
by the LHCb and found to be small \cite{Aaij:2016flj}.
For the $B_s$ case, the S-wave contamination of the $B^0_s \to \overline{K}^{*0} \mu^+ \mu^-$ decay is also unknown now
and  assumed to be small to that of the $B^0 \to \overline{K}^{*0} \mu^+ \mu^-$ decay.
Specifically, the S-wave fraction of $F_S(\overline{B}^0 \to \overline{K}^{\ast 0} \mu^+\mu^-)=(3.4\pm0.8)\%$ in the $K^+\pi^-$ system \cite{Aaij:2018jhg}.
Theoretically, the authors of Ref.~\cite{Doring:2013wka} found the S-wave contribution will modify differential decay widths by about $10\%$
in the process of $\overline{B}^0 \to K^- \pi^+ \ell^+ \ell^-$.

Analogous to the ratios $R_K$ and $R_{K^{\ast}}$ for $B\to K^{(\ast)} l^+l^-$ decays as defined in Refs.~\cite{Wei:2009zv, Ali:1999mm,Beneke:2001at,
Chen:2001ri,Ali:2006ew,Bobeth:2008ij,Egede:2008uy,Altmannshofer:2008dz, Aaltonen:2011cn,Lees:2012tva,Aaij:2014ora,Khachatryan:2015isa,
Aaij:2016flj,Aaij:2017vbb}, we can define the similar ratios
of the BRs $R^{e\mu }_{s,K}$ and     $R^{e\mu }_{s,K^{\ast}}$ for the $\bar{B}_s \to  K^{(\ast)} \ell^+ \ell^-$ decays:
\beq
R_{s, K^{(\ast)} }^{e\mu } =\frac{\calb(\bar{B}_s\to K^{(*)} \mu^+\mu^-) }{\calb(\bar{B}_s\to K^{(*)} e^+e^-)  }. \label{eq:rkbs1}
\eeq
Similarly,   we can also define the ratios $R^{\mu \tau}_{s,K}$ and   $R^{\mu\tau }_{s,K^{\ast}}$ in the following form:
\beq
R_{s, K^{(\ast)} }^{ \mu \tau}=\frac{\calb(\bar{B}_s\to K^{(*)} \tau^+\tau^-) }{\calb(\bar{B}_s\to K^{(\ast)} \mu^+\mu^-)  }. \label{eq:rkbs2}
\eeq
These new ratios  $R_{s, K^{(\ast)} }^{e\mu } $ and $ R_{s, K^{(\ast)} }^{ \mu \tau} $ ,  together with the ratios $R_{K}$ and $R_{K^\ast }$,
can help us to examine the $b \to (s,d) \ell^+ \ell^-$ transitions in great details.

Unlike the well studied $B \to K^{(*)} \ell^{+}\ell^{-}$ decays,  the semileptonic  $\bar{B}_s\to K^{(*)} \ell^{+}\ell^{-}$ decays  have not caught much attention
partially due to their lower branching ratios and the lack of the relevant experimental measurements.
In recent years,  these decays have been studied by several authors  for example
in Refs.~\cite{Wang:2012ab,Wang:2013ix,Khodjamirian:2017fxg,Kindra:2018ayz}, and  the
first measured branching ratio as listed in Eq.~(\ref{eq:exp2})  was  reported last year by LHCb Collaboration \cite{Aaij:2018jhg}.
Besides the measurements for the branching ratios, a precise angular reconstructions of the polarized $K^*$ in $\bar{B}_s\to K^{(*)} \ell^{+}\ell^{-}$ decays
was discussed in Ref.~\cite{Altmannshofer:2008dz}.
Recently, the predictions of several angular observables for  the $\bar{B}_s\to K^{*}\ell^{+}\ell^{-}$ decays were provided using the light cone sum rule (LCSR)
and the Lattice QCD method  in Ref.~\cite{Kindra:2018ayz}.

By using  the perturbative QCD (PQCD) factorization approach \cite{pqcd1,pqcd2,li2003},   the semileptonic $\bar{B}_s \to K  \ell^+\ell^+$ decays have been
studied by us in a previous paper ~\cite{Wang:2012ab}.
We considered the next-to-leading order (NLO) contributions  known at 2012 and presented
our  PQCD predictions of the branching ratios:
\beq
\calb( \overline{B}_s^0\to K^0 \ell^+\ell^- )&=& (1.63^{+0.73}_{-0.58})\times 10^{-8}, \quad l=(e,\mu), \label{eq:pqcdbr1}\\
 \calb(\overline{B}_s^0\to K ^0\tau^+\tau^- )&=& (0.43^{+0.18}_{-0.15})\times 10^{-8}.  \label{eq:pqcdbr2}
\eeq

In this paper, we will make  a systematic  study  for the  semileptonic decays $\bar{B}_s \to (K, K^*)  \ell^+\ell^-$ with $l=(e,\mu,\tau)$,
and present the theoretical predictions of  many new physical observables:
\begin{enumerate}
\item[(1)]
For  $\bar{B}_s \to K  \ell^+\ell^-$  decays,  besides the branching ratios,  we also  calculate its forward-backward asymmetry $\cala_{FB}(q^2)$, the longitudinal lepton
polarization asymmetry $P_L(q^2)$,  the direct CP asymmetry $\cala_{CP}$ and the ratios $R^{e\mu }_{s,K}$ and $R^{\mu \tau }_{s,K}$.

\item[(2)]
For  $\bar{B}_s \to K^*  \ell^+\ell^-$  decays,  we  treat them as  a four body decay  $\bar{B}_s \to K^*(\to K \pi)  \ell^+\ell^-$ described by  four kinematic
variables: the lepton invariant mass squared $q^2$ and three angles $(\thK, \thl,\phi)$.
We define and calculate the full  angular decay distributions, the transverse amplitudes,  the partially integrated decay amplitudes over the angles $(\thK,\thl,\phi)$,
the forward-backward (FB) asymmetry $\cala_{FB}(q^2)$,  the $K^*$ polarization fraction $R_{L,T}(q^2)$ and the longitudinal lepton
polarization asymmetry $P_L(q^2)$, and the ratios $R^{e\mu}_{s,K^*}$ and  $R^{\mu \tau}_{s,K^*}$.
Since we do not know how to calculate the possible S-wave or other broad contributions related with the reconstruction of
$K\pi$ pair \cite{Aaij:2016flj,Aaij:2018jhg},  we add  a $10\%$ uncertainty to the PQCD predictions of the branching ratios as an additional theoretical  error
\cite{Doring:2013wka}, but neglect it in the calculations for other ratios due to the strong cancellation.

\item[(3)]
We  use both the PQCD factorization approach and  ``PQCD+Lattice" approach to determine the values and their $q^2$-dependence of
the $\bar{B}_s\to K^{(*)}$ transition form factors. We use  the Bourrely-Caprini-Lellouch (BCL) parametrization method \cite{bcl09,jhep1905-094}
to make the extrapolation for all form factors from the low $q^2$ region to  $q^2_{max}$.
We will calculate the branching ratios and all other physical observables using  the PQCD approach and ``PQCD+Lattice'' approach respectively,
and compare the theoretical predictions obtained based on different models.

\end{enumerate}

The paper is organized as follows: In Sec.~\ref{sec:2}, we give a short review for the kinematics of the $\bar{B}_s \to K^{(*)} \ell^+ \ell^-$
decays including distribution amplitudes of $B_s$ and $K^{(\ast)}$ mesons. Sec.~\ref{sec:3} is devoted to the the theoretical framework including
Hamiltonian and transition form factors based on the PQCD $k_T$ factorization formalism.
In Sec.~\ref{sec:4}, we list all the observables for both types of decays considered in this paper.
Sec.~\ref{sec:5} contains the numerical results of relevant observables and some phenomenological discussions.
We conclude and summarize in the last section.

\section{Kinematics and the wave functions}\label{sec:2}
\begin{figure}[thb]
\centerline{\epsfxsize=14cm \epsffile{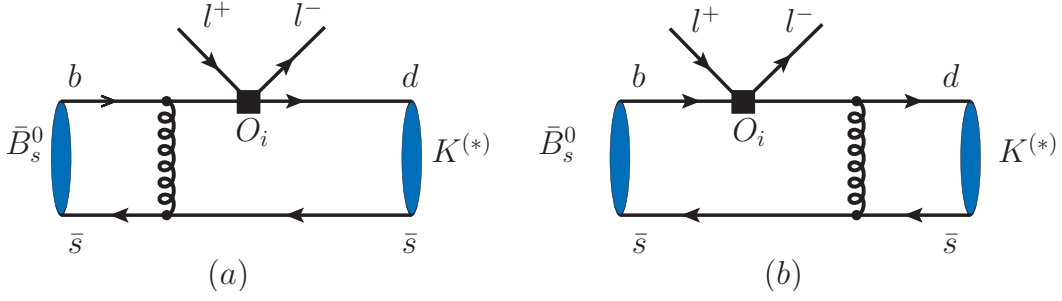}}
 \caption{ The typical Feynman diagrams for the semileptonic decays
$\bar{B}_s^0 \to K^{(*)}\ell^+\ell^-$ in PQCD approach with the flavor-changing neutral current (FCNC) contributions
due to the operators $O_{i}$ denoted as black squares.}
\label{fig:fig1}
\end{figure}
We discuss kinematics of these decays in the large-recoil (low $q^2$) region, where the PQCD factorization approach is applicable
to the considered semileptonic decays involving  $K^{(*)}$ as the final state meson.
In the rest frame of $\bar{B^0_s}$ meson, we define the $\bar{B^0_s}$ meson momentum $p_1$, the $K^{(*)}$ momentum $p_2$ in the light-cone coordinates
as Ref.~\cite{Fan:2013qz}
\beq
\label{eq-mom-p1p2}
p_1&=&\frac{m_{B_s}}{\sqrt{2}}(1,1,0_\bot),\quad p_2=\frac{rm_{B_s}}{\sqrt{2}} (\eta^+,\eta^-,0_\bot),
\eeq
 where the mass ratio $r=m_K/m_{B_s}$ or $m_K^*/m_{B_s}$, and the factor $\eta^\pm$ is defined in the following form:
\beq
\eta^\pm = \eta \pm \sqrt{\eta^2-1}, \quad {\rm with} \quad   \eta =\frac{1}{2r}\left [ 1+r^2-\frac{q^2}{m_{B_s}^2}\right],
\label{eq:eta9}
\eeq
where $q=p_1-p_2$ is the lepton-pair four-momentum.  For the final state $K^*$ meson, its longitudinal  and transverse polarization vector
$\epsilon_{L,T}$ can be written as
\beq
\label{eq-pol-epsilon}
\epsilon_L=\frac{1}{\sqrt{2}}(\eta^+,-\eta^-,0_\bot), \qquad \epsilon_T=(0,0,1).
\eeq
The momenta of the spectator quarks in $B_s$ and $K^{(*)}$ mesons are parameterized as
\beq \label{eq-mom-k1k2}
k_1 =(0,x_1\frac{m_{B_s}}{\sqrt{2}},k_{1\bot}), \quad k_2=\frac{m_{B_s}}{\sqrt{2}}(x_2r\eta^+,x_2r\eta^-,k_{2\bot}).
\eeq
we make the approximation in the small $k_\bot$.

For the $\bar{B^0_s}$ meson wave function, we use the same parameterizations as  in Refs.~\cite{Xiao:2011tx,Wang:2012ab}
\beq
\Phi_{B_s}=\frac{i}{\sqrt{2N_{c}}} (\psl_{B_s} +m_{B_s}) \gamma_5 \phi_{B_s} ({k_1}).
\label{eq:bmeson}
\eeq
Here only the contribution of the Lorentz structure $\phi_{B_s} (k_1)$ is taken into account, since the contribution of the second Lorentz structure $\bar{\phi}_{B_s}$
is numerically small and has been neglected. We adopted the $B_s$-meson distribution amplitude the same as $B$-meson
in the $SU(3)_f$ limit widely used in the PQCD approach
\beq
 \phi_{B_s}(x,b)&=& N_{B_s} x^2(1-x)^2\mathrm{\exp}
\left[ -\frac{m_{B_s}^2\; x^2}{2 \omega_{B_s}^2} -\frac{1}{2} (\omega_{B_s} b)^2\right].
\label{eq:phib}
\eeq
 In order to analyze the uncertainties of theoretical predictions induced by the inputs, one usually take ~$\omega_{B_s}
 =0.50 \pm 0.05$~GeV for $B_s^0$ meson. The normalization factor ~$N_{B_s}$ depends on the values of the shape
 parameter $\omega_{B_s}$ and the decay constant $f_{B_s}$ and defined
through the normalization relation : $\int_0^1dx\; \phi_{B_s}(x,b=0)=f_{B_s}/(2\sqrt{6})$ \cite{Wang:2012ab}.

 For the pseudoscalar $K$ meson , the wave function can be chosen as the same one in Ref.~\cite{Fan:2012kn}:
\beq
\Phi_{K}(p,x)\equiv \frac{i}{\sqrt{6}}\gamma_5 \left [ \psl_K \phi^{A}_{K}(x)
+m^{K}_{0} \phi_{K}^{p}(x)+ \zeta m^{K}_{0} (\nsl \vsl -1)\phi_{K}^{T}(x)\right ], \label{eq:wfk}
\eeq
where $m^{K}_{0}$ and $p$ is the chiral mass and  the momentum  of the meson $K$.
The parameter $\zeta=1$ or $-1$ when the momentum fraction of the quark (anti-quark) of the meson is set to be $x$.
The distribution amplitudes (DA's) of the kaon meson can be found easily in Refs.~\cite{Ball:2006wn,Ali:2007ff,xiao18a,xiao18b}:
\beq
\phi_{K}^A(x) &=&  \frac{3 f_K}{\sqrt{6} } x (1-x)
    \left[1+a_1^{K}C^{3/2}_1(t)+a^{K}_2C^{3/2}_2(t)+ a_4^K C_4^{3/2}(t) \right],\label{eq:piw1}\\
\phi_K^P(x) &=&   \frac{f_K}{2\sqrt{6} }
   \left \{ 1+\left (30\eta_3-\frac{5}{2}\rho^2_{K} \right ) C^{1/2}_2(t)
   -3\left [ \eta_3 \omega_3 + \frac{9}{20}\rho_K^2\left ( 1 + 6a_2^K\right)C_4^{1/2}(t)\right]
   \right \}, \ \  \label{eq:piw2}   \\
\phi_K^T(x) &=&  \frac{f_K(1-2x)}{2\sqrt{6} }
   \left\{ 1+6\left [ 5\eta_3-\frac{1}{2}\eta_3\omega_3-\frac{7}{20}\rho^2_K
   -\frac{3}{5}\rho^2_K a_2^{K} \right ]
   \left (1-10x+10x^2\right )\right \},\quad     \label{eq:piw3}
\eeq
where $t=2x-1$, $f_K$ is the decay constant of kaon meson  and $\rho_K=m_K/m^0_K$ is the mass ratio.
The Gegenbauer moments and other parameters are \cite{Ball:2006wn,Ali:2007ff,xiao18a,xiao18b}:
\beq
a^K_1=0.06 \pm 0.03 ,\quad a^K_2=0.25\pm0.15, \quad a^K_4=-0.015,  \quad \eta_3=0.015, \quad \omega=-3.0.
\label{eq:gb01}
\eeq
The Gegenbauer polynomials appeared in Eqs.~(\ref{eq:piw1},\ref{eq:piw2}) are of the following form  \cite{Ball:2006wn,Ali:2007ff,xiao18a,xiao18b}:
\beq
C^{3/2}_1(t)&=& 3t,  \quad C^{1/2}_2(t)=\frac{1}{2}\left (3t^2-1 \right ), \quad C^{3/2}_2(t)=\frac{3}{2}\left (5t^2-1 \right ), \non
C^{1/2}_4(t)&=& \frac{1}{8} \left (3-30t^2+35 t^4 \right ), \quad C^{3/2}_4(t)=\frac{15}{8} \left (1-14t^2+21t^4 \right ).
\label{eq:cij}
\eeq

 For the light vector meson $K^*$,  the longitudinal and transverse  polarization components can provide the
 contribution. Here we adopt the wave functions of the vector $K^* $  as in Ref.~\cite{Ali:2007ff}:
\beq
\Phi_{K^*}^{||}(p,\epsilon_L)&=& \frac{i}{\sqrt{6}}
\left [\not\! \epsilon_L m_{K^*}\phi_{K^*}(x)+\not\! \epsilon_L\psl\phi^{t}_{K^*}(x)
+m_{K^*}\phi^{s}_{K^*}(x)\right], \label{eq:phiv}\\
\Phi_{K^*}^{\perp}(p,\epsilon_T)&=& \frac{i}{\sqrt{6}}
\left [\not\! \epsilon_T m_{K^*}\phi^{v}_{K^*}(x)+\not\! \epsilon_T\psl\phi^{T}_{K^*}(x)
+m_{K^*}i\epsilon_{\omega\upsilon\rho\sigma}\gamma_5\gamma^{\omega}\epsilon^{v}_{T}n^{\rho}v^{\sigma}\phi^{a}_{K^*}(x)\right],
\label{eq:phivT}
\eeq
where $p$ and $m_{K^*}$ are the momentum and the mass of the $K^*$ meson,  $ \epsilon_L $ and $ \epsilon_T $
correspond to the longitudinal and transverse polarization vectors of the vector meson, respectively.
The  $\phi_{K^*}$ and $\phi_{K^*}^T$  in Eqs.~(\ref{eq:phiv},\ref{eq:phivT}) are the twist-2  DAs \cite{Ali:2007ff}:
\begin{eqnarray}
\phi_{K^*}(x)&=&\frac{3f_{K^*}}{\sqrt{6}} x (1-x)\left[1+a_{1K^*}^{||}C^{3/2}_1(t)+ a_{2K^*}^{||}C_2^{3/2} (t)\right]\;,\\
\phi^T_{K^*}(x)&=&\frac{3f^T_{K^*}}{\sqrt{6}} x (1-x)\left[1+a_{1K^*}^{\perp}C^{3/2}_1(t)+
a_{2K^*}^{\perp}C_2^{3/2} (t)\right], \label{eq:t2-01}
\end{eqnarray}
where $f_{K^*}$ and $f^T_{K^*}$  are the longitudinal and transverse components of the decay constants.
The Gegenbauer moments  in  Eqs.~(\ref{eq:phiv},\ref{eq:phivT})  are  the same ones as those in Ref.~\cite{Ali:2007ff}:
\beq
a_{1K^*}^{||}&=&0.03\pm0.02,\quad a_{2K^*}^{||}=0.11\pm0.09, \non
a_{1K^*}^{\perp}&=&0.04\pm0.03,\quad a_{2K^*}^{\perp}=0.10\pm0.08.
\label{eq:gb02}
\eeq
The twist-3 DAs $\phi^{s,t}_{K^*}$ and $\phi^{v,a}_{K^*}$ in  Eqs.~(\ref{eq:phiv},\ref{eq:phivT}) are defined with the asymptotic form
as in Ref.~\cite{Ali:2007ff}:
\beq
\phi^t_{K^*} = \frac{3f^T_{K^*}}{2\sqrt 6} t^2,   \quad \phi^s_{K^*}=\frac{3f_{K^*}^T}{2\sqrt 6} (-t)~,\quad
\phi^v_{K^*} =  \frac{3f_{K^*}}{8\sqrt 6}(1+t^2),   \quad \phi^a_{K^*}=\frac{3f_{K^*}}{4\sqrt 6} (-t)~,
\label{eq:t301}
\eeq


\section{THEORETICAL FRAMEWORK }\label{sec:3}

\subsection{Effective Hamiltonian for \texorpdfstring{$b \to d \ell^+ \ell^-$}{} decays }\label{sec:3c}

\begin{figure}[thb]
\begin{center}
\centerline{\epsfxsize=9cm\epsffile{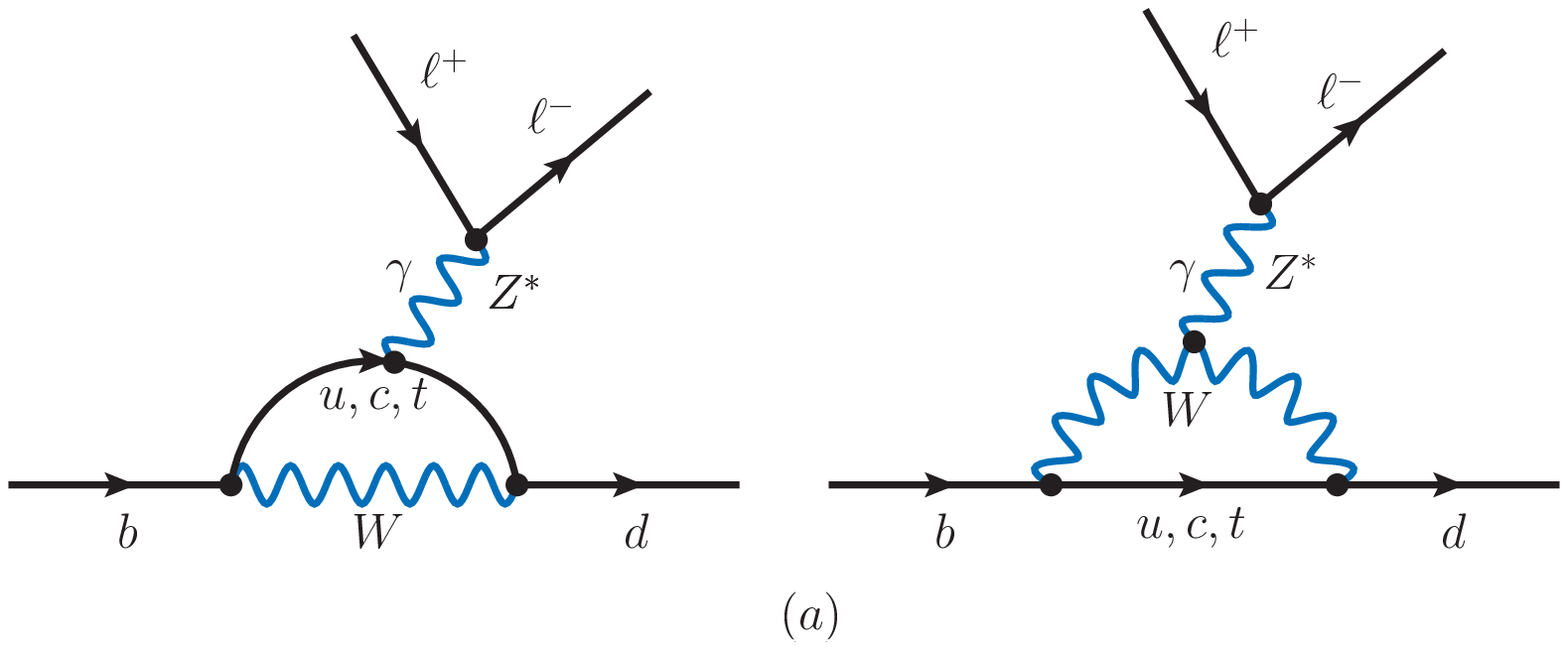}}
\centerline{\epsfxsize=9cm\epsffile{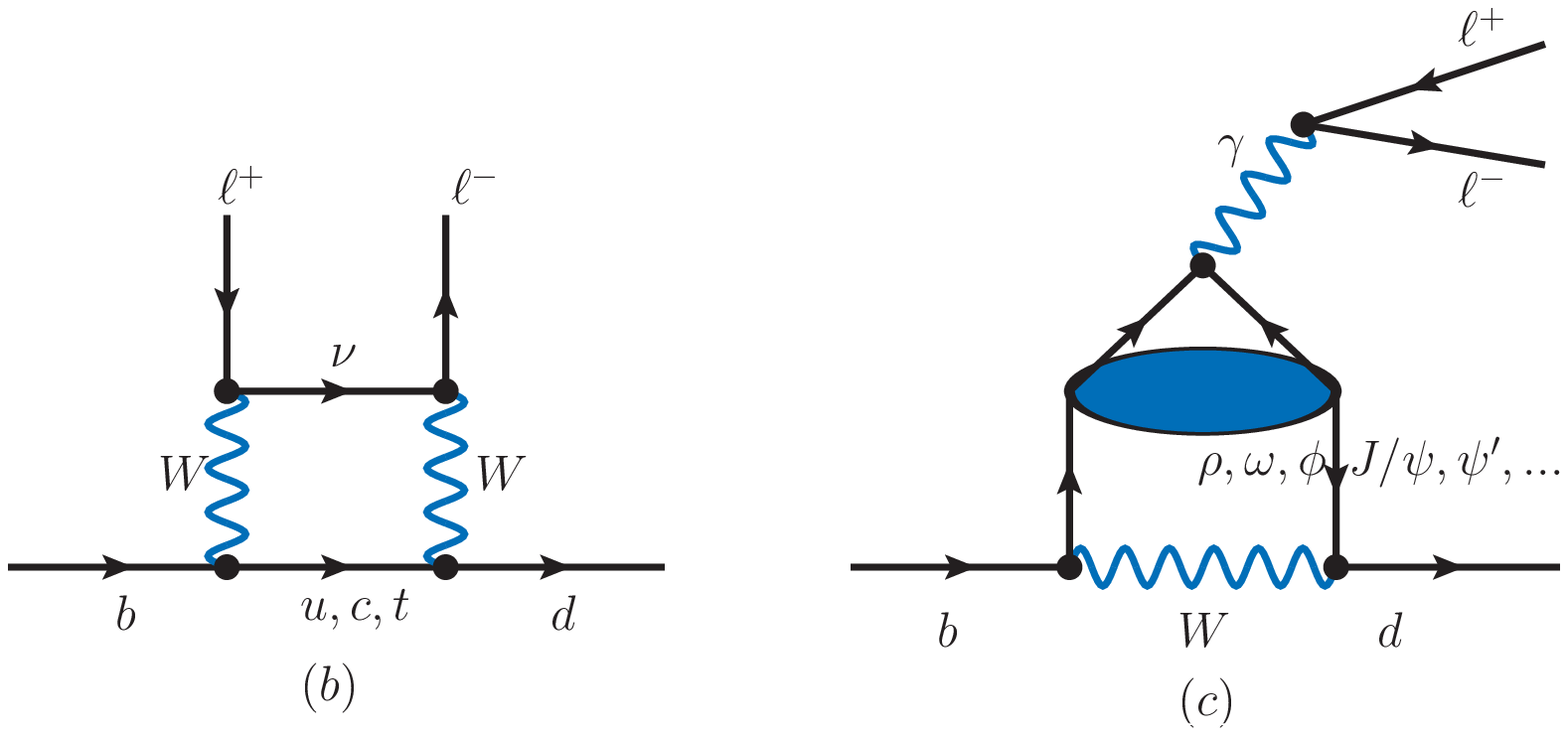}}
\caption{ Typical Feynman loop diagrams:  the $\gamma$-penguin (2a) with  ${\mathcal{O}}_{7}$,
the $z(\gamma)$-penguin (2a) and $W$-box (2b) with ${\mathcal{O}}_{9,10}$,
and the loops (2c) with ${\mathcal{O}}^{u,c}_{1,2}$. }
\label{fig:fig2} 
\end{center}
\end{figure}

For the considered $b \to d \ell^+\ell^-$ transitions, the effective Hamiltonian in the framework of the SM can be written in the following form
~\cite{Li:2008tk,Kindra:2018ayz,Singh:2019hvj,Nayek:2018rcq}:
\beq
    {\cal H}_{\text{eff}} &=& -\frac{G_F}{\sqrt{2}} V_{tb}V_{td}^* \left \{
      C_1(\mu)\mathcal{O}_1^c(\mu)+ C_2(\mu) \mathcal{O}_2^c(\mu)+ \sum_{i=3}^{10}{C}_i(\mu)
    {\mathcal{O}}_i(\mu) \right.  \non
   &&   +\lambda_u \Big [ C_1(\mu)(\mathcal{O}_1^c(\mu)-\mathcal{O}^u_1(\mu))
   + C_2(\mu)\left ( \mathcal{O}_2^c(\mu)-\mathcal{O}^u_2(\mu) \right ) \Big ] \Bigg \} ,
 \label{eq:heff}
\eeq
where $G_F=1.16638\times 10^{-5}{GeV}^{-2}$  is the Fermi constant,
$\lambda_u=V_{ub}V^*_{ud}/(V_{tb}V_{td}^*)$ is a ratio of the CKM elements, $C_i(\mu)$   and
$\mathcal{O}_i(\mu)$ are the Wilson coefficients and the 4-fermion operators at the renormalization scale $\mu$.
In SM, a suitable basis of the operators ${\mathcal{O}}_i(\mu)$ for $b \to d \ell^+\ell^-$
transition is given by the current-current operators ${\mathcal{O}}^{u,c}_{1,2}$,
the QCD penguin operators ${\mathcal{O}}_{3-6}$,
the electromagnetic penguin operator ${\mathcal{O}}_{7}$ and  the chromomagnetic penguin operator  ${\mathcal{O}}_{8}$,
as well as the semileptonic operators ${\mathcal{O}}_{9,10}$ :
\begin{align}
\label{eq:operators}
{\mathcal{O}}^c_{1} &=(\bar d_{\alpha}c_{\beta})_{V-A}(\bar c_{\beta}b_{\alpha})_{V-A},&
{\mathcal{O}}^c_{2} &=(\bar d_{\alpha}c_{\alpha})_{V-A}(\bar c_{\beta}b_{\beta})_{V-A},\non
 {\mathcal{O}}^u_{1} &=(\bar d_{\alpha}u_{\beta})_{V-A}(\bar u_{\beta}b_{\alpha})_{V-A},&
{\mathcal{O}}^u_{2} &=(\bar d_{\alpha}u_{\alpha})_{V-A}(\bar u_{\beta}b_{\beta})_{V-A},\non
{\mathcal{O}}_{3} &=(\bar d_{\alpha}b_{\alpha})_{V-A}\sum_q(\bar q_{\beta}q_{\beta})_{V-A},&
{\mathcal{O}}_{4} &=(\bar d_{\alpha}b_{\beta})_{V-A}\sum_q(\bar q_{\beta}q_{\alpha})_{V-A},\non
{\mathcal{O}}_{5} &=(\bar d_{\alpha}b_{\alpha})_{V-A}\sum_q(\bar q_{\beta}q_{\beta})_{V+A},&
{\mathcal{O}}_{6} &=(\bar d_{\alpha}b_{\beta})_{V-A}\sum_q(\bar q_{\beta}q_{\alpha})_{V+A},\non
{\mathcal{O}}_{7} &=\frac{e m_b}{8\pi^2}\bar d\sigma^{\mu\nu}(1+\gamma_5)bF_{\mu\nu},&
{\mathcal{O}}_{8} &=\frac{g m_b}{8\pi^2}\bar d\sigma^{\mu\nu} T^a(1+\gamma_5)b G^a_{\mu\nu},\non
{\mathcal{O}}_{9} &= \frac{\alpha_{\rm{em}}}{2\pi}(\bar d\gamma^{\mu}(1-\gamma_5)b)(\bar \ell\gamma_{\mu}\ell), &
{\mathcal{O}}_{10}& =\frac{\alpha_{\rm{em}}}{2\pi}(\bar d\gamma^{\mu}(1-\gamma_5)b)(\bar \ell\gamma_{\mu}\gamma_5\ell),
\end{align}
where $T^a$ denotes the generators of the $SU(3)_C$  group and $m_b$ is the running $b$ quark mass in the $\overline{\rm MS}$
scheme;  $F_{\mu\nu}$ and $G^a_{\mu\nu}$ are the  electromagnetic and chromomagnetic tensors, respectively.
The labels ${V\pm A}$  refers to the Lorentz structure $\gamma_{\mu}(1\pm \gamma_5)$.
The dominant contribution to $b \to d \ell^+\ell^-$ transitions are given by ${\mathcal{O}}_{7}$ and ${\mathcal{O}}_{9,10}$,
as well as ${\mathcal{O}}^{u,c}_{1,2}$.
The operator ${\mathcal{O}}_{7}$ corresponds to the $\gamma$-penguin diagram,
as shown in Fig.~\ref{fig:fig2}(a).
The operators ${\mathcal{O}}_{9,10}$  describe the sum of the contributions
from the  Z- and $\gamma$-penguin in Fig.~\ref{fig:fig2}(a) and the W box diagrams
in Fig \ref{fig:fig2}(b).
The current-current operators ${\mathcal{O}}^{u,c}_{1,2}$ involve a long-distance (LD)
contribution, which origins in the real
$u\bar u$,$d\bar d$ and $c\bar c$ intermediate states,
namely the $(\rho, \omega,\phi)$ and $J/\psi$ family in Fig \ref{fig:fig2}(c),
coupled to the lepton pair via the virtual photon. This contribution is proportional
to ${C}_{9}$ and can be absorbed into an
effective Wilson coefficient ${C}^{\text{eff}}_{9}$~\cite{Khosravi:2014hqa}.

Here we neglect the contribution from subleading chromomagnetic penguin, quark-loop
and annihilation diagrams because these effects are highly suppressed \cite{Kindra:2018ayz}.
Hence the decay amplitude for $b\to d l^+l^-$ loop transition can be decomposed as
 \begin{eqnarray}
 {\cal A}(b\to dl^+ l^-)&=&\frac{G_F}{2\sqrt{2}}\frac{\alpha_{\rm{em}}}{\pi}V_{tb}V^*_{td}\bigg\{
 C_9^{\rm{eff}}(q^2)
 [\bar d \gamma_{\mu}(1-\gamma_5)b][\bar l\gamma^{\mu}l] + C_{10}[\bar d\gamma_{\mu}(1-\gamma_5)b]
 [\bar l\gamma^{\mu}\gamma_5l]\non
 &&- 2m_bC_7^{\rm{eff}}\big[\bar d i\sigma_{\mu\nu}\frac{q^{\nu}}{q^2}
 (1+\gamma_5)b\big][\bar l\gamma^{\mu}l] \bigg\},\label{eq:Ampbtos}
 \end{eqnarray}
where $C_7^{eff}(\mu)$ and $C_9^{eff}(\mu)$ are the effective Wilson coefficients,
defined as in Refs.~\cite{Chen:2001zc,Wang:2012ab}
\beq
C_7^{\rm{eff}}(\mu)&=&C_7(\mu)+C^{\prime}_{b\to  d\gamma}(\mu), \label{eq:c7eff}\\
C_9^{\rm{eff}}(\mu,q^2)&=&C_9(\mu)+Y_{\rm{pert}}(\hat{s})+Y_{\rm{res}}(q^2).
\label{eq:c9eff}
\eeq
The analytic expressions for all Wilson coefficients in the NLO approximation can be
found easily in Ref.~\cite{Buchalla:1995vs}.
The numerical values of the NLO Wilson coeffients $C_i(\mu)$ at three different renormalization scales
$\mu=(m_b/2, m_b,3 m_b/2)$ are listed in Table \ref{tab:table1}.  Note that the Wilson coefficient $C_{10}$
is independent of the $\mu$ scale and $C_{9}(\mu)$ is relatively sensitive to the choice of $\mu$.

\begin{table}
 \caption{The values of  the Wilson coefficients $C_i(\mu)$ in NLO level  at three different renormalization scales  $\mu=(m_b/2, m_b,3 m_b/2)$.}
 \label{tab:table1}
\begin{center}
\begin{tabular}{|c|c|c|c|c|c|c|c|c|c|c|}
\hline\hline
\multicolumn{1}{|c|}{$\mu\!\setminus\! C_i(\mu)$} & ${C}_1$ & ${C}_2$ & ${C}_3(\%)$  & ${C}_4(\%)$  & ${C}_5(\%)$  & ${C}_6(\%)$& ${C}_{7}$
 & ${C}_{8}$ & ${C}_9$ & ${C}_{10}$ \\ \hline
$m_b/2$                                      &  \, $-0.276$ \, & \,  $1.131 \,$ & $2.005$ &  $-4.845$ & $1.375$ & $-5.841$ & \, $-0.329$ \, &  \,$-0.165$  \,&  \,$4.450$  \, &  \,$-4.410$ \,   \\ \hline
$m_b$                                         & \, $-0.175$  \,& \, $1.076 \,$ & $1.258$ & $-3.279$ & $1.112$ & $-3.634$ & \, $-0.302$  \,&  \,$-0.148$  \,&  \,$4.232$ \,  & \, $-4.410$ \,   \\ \hline
$3 m_b/2$                                      & \, $-0.129$  \,& \,  $1.053 \,$ & $0.966$ & $-2.608$ & $0.964$ & $-2.786$ & \, $-0.287$  \,&  \,$-0.139$  \,&  \, $4.029$  \, &  \,$-4.410$ \,   \\ \hline\hline
\end{tabular}
 \end{center}
\end{table}

The term $C^{\prime}_{b \to d\gamma}$ in Eq.~(\ref{eq:c7eff}) is the absorptive part of $b\to d\gamma$ and was given in Ref.~\cite{Chen:2001zc}
\beq
C^{\prime}_{b \to  d\gamma}(\mu)=i\alpha_s\left \{\frac{2}{9}\eta^{14/23}\left [G_I(x_t)-0.1687 \right ]-0.03C_2(\mu) \right \},
\eeq
where $\eta=\alpha_s(m_W)/\alpha_s(\mu)$, $x_t=m_t^2/m_W^2$ and
\beq
G_I(x_t)=\frac{x_t \left( x_t^2-5x_t-2\right )}{8 \left (x_t-1 \right )^3}+\frac{3x_t^2 \ln x_t}{4(x_t-1)^4}.
\eeq

Besides the ordinary Wilson coefficient $C_9(\mu)$, the effective Wilson coefficient $C_9^{\rm{eff}}(q^2)$ in Eq.~(\ref{eq:c9eff})
also contains two additional effective terms $Y_{\rm{pert}}(\hat{s})$ and $Y_{\rm{res}}(q^2)$.
The term  $Y_{\rm{pert}}(\hat{s})$ describes  the short distance contribution from the soft-gluon emission and
the one-loop contribution of the four-quark operators ${\mathcal{O}}_{1}-{\mathcal{O}}_{6}$.
The term $Y_{\rm{res}}(q^2)$ includes the contributions of the virtual resonances described by the Breit-Wigner form prescribed
in Refs.\cite{Ali:1991is,Lim:1988yu,Deshpande:1988bd,ODonnell:1991cdx,Nayek:2018rcq}.
 \beq
Y_{\rm{pert}}(\hat{s}) &=& 0.124\, \omega(\hat{s})+g(\hat{m}_c,\hat{s})C_0
+\lambda_u\left [ g(\hat{m}_c,\hat{s})-g(\hat{m}_u,\hat{s}) \right] (3C_1+C_2) \non
&& - \frac{1}{2}g(\hat{m}_d,\hat{s})(C_3+3C_4) - \frac{1}{2}g(\hat{m}_b,\hat{s})(4C_3+4C_4+3C_5+C_6) \non
&& + \frac{2}{9}(3C_3+C_4 +3C_5+C_6) , \label{eq:ypert}\\
Y_{\rm{res}}(q^2)&=&-\frac{3\pi}{\alpha_{\rm{em}}^2}
\Big[  C_0 \cdot \sum_{V=J/\Psi,\Psi'...}\frac{m_V {\cal B}(V\to l^+l^-)\Gamma_{\rm{tot}}^V} {q^2-m_V^2+im_V\Gamma_{\rm{tot}}^V}\non
&&- \lambda_u g(\hat{m}_u,\hat{s})(3C_1+C_2)\cdot
\sum_{V=\rho,\omega,\phi}\frac{m_V {\cal B}(V\to l^+l^-)
\Gamma_{\rm{tot}}^V}{q^2-m_V^2+im_V\Gamma_{\rm{tot}}^V}\Big],
\label{eq:yres}
\eeq
where $C_0=3C_1+C_2+3C_3+C_4+3C_5+C_6$, $\hat{s}\!=\!q^2/m^2_b$, $\hat{m}_q\!=\!{m_q}/{m_b}$.
In the above expressions, $\omega(\hat{s})$ is the soft-gluon correction to the matrix element of operator $\mathcal{O}_{9}$
and  was given in Refs.~\cite{Nayek:2018rcq,Jezabek:1988ja}
\beq
\omega(\hat{s})&=& \!-\frac{2}{9}\pi^2  +\frac{4}{3} \int^{\hat{s}}_0 \frac{\ln(1-u)}{u}du - \frac{2}{3}\ln(\hat{s})\ln(1-\hat{s})
-\frac{5+4\hat{s}}{3(1\!+\!2\hat{s})}\ln(1-\hat{s})\non
&& - \frac{2\hat{s}(1\!+\!\hat{s})(1\!-\!2\hat{s})}{3(1\!-\!\hat{s})^2(1\!+\!2\hat{s})}\ln\hat{s}\!+\!\frac{5\!+\!9\hat{s}\!
-\!6\hat{s}^2}{6(1\!-\!\hat{s})(1\!+\!2\hat{s})}.
\eeq
The loop coefficient functions $g(\hat{m}_q,\hat{s})$ in Eqs.~(\ref{eq:ypert},\ref{eq:yres})  describe the one-loop  $(q\bar{q})$ contributions to the four-quark operators
${\mathcal{O}}_{1}-{\mathcal{O}}_{6}$, and can be written as the well-known expression \cite{Grinstein:1988me,Misiak:1992bc,Buras:1994dj,Khodjamirian:2010vf}:
\begin{eqnarray}
g(\hat{m}_q,\hat{s})= -\frac{8}{9}\ln(\hat{m}_q)+\frac{8}{27}+\frac{4}{9}x
 -\frac{2}{9}(2+x)\sqrt{|1-x|} \times
	\begin{cases}
		2 \arctan \frac{1}{\sqrt{x-1}} \hspace{-0.1in}&\!\!, x>1 \vspace{0.02in}\\
		 \ln|\frac{1+\sqrt{1-x}}{1-\sqrt{1-x}}| - i\pi \hspace{-0.1in}&\!\!, x<1,
	\end{cases}
\end{eqnarray}
where $x\equiv4\hat{m}_q^2/\hat{s}$.

In Ref.~\cite{Khodjamirian:2010vf}, the authors employed the dispersion approach to compute the charm-loop effect in a form of the correction
to the Wilson coefficient $C_9$.  By fitting the whole dispersion relation to the OPE result at $q^2<<4 m^2_c$,  the authors found  that there exists a destructive interference
between the $J/\psi$ and $\psi(2S)$ states.
According to their opinion, a reliable prediction for the charm-loop effect above $\psi(2S)$ based on QCD is hard to make.
Although the actual effect depending on the interference of many charmonium states cannot be reliably constrained by OPE, yet it could be
considered as small in this region.

The term $Y_{\rm{res}}$ in Eq.~(\ref{eq:yres}) denotes the long-distance resonance contributions from
those $B_s\to K^{(*)}V \to K^{(*)} (V \to \ell^+ \ell^-) $ transitions ,
where $V$ stands for the possible intermediate resonance states decaying to lepton pairs:
\begin{enumerate}
\item[(1)]
The charmless light vector mesons $V=(\rho,\omega,\phi)$. The kinematic region where the light resonances
 ($\rho,\omega,\phi$) contribute is typically not excluded from the experimental analyses because their effects on branching fractions
 and other physical observables might be substantial \cite{Du:2015tda}.

\item[(2)]
The $c\bar{c}$ charmonia  $V_{c \bar c}=\psi(1S,2S,3770,4040,4160,4415)$.
The two lowest charmonium states $\psi(1S)$ and $\psi(2S)$ (i.e.  $J/\psi$ and $\psi'$),  whose masses are below the open charm threshold ($D \bar{D}$),
have  tiny width and can induce large breaking of quark-hadron duality.
Hence, the narrow charmonia resonance regions are routinely rejected in the  theoretical and experimental analysis.
For the four higher charmonium resonances, however, they are broad and overlapping throughout
the high-$q^2$ regions. One usually make the integration over the full high-$q^2$ range.
\end{enumerate}

As reported in Ref.~\cite{Aaij:2013pta},  a resonance above $\psi(2S)$ compatible with the $\psi(4160)$ has been observed
by LHCb in $B \to K \mu^+ \mu^-$ decay.  Consequently, nearly all available contribution about the $J^{PC}=1^{--}$ charmonium resonances
above the open charm threshold should be taken into account ~\cite{Lu:1997bu}.
In Table \ref{tab:table2}, we list the properties of all considered intermediate resonance states:  their mass, width, and branching fractions of the
leptonic decay channel $V\to l^+l^-$~\cite{pdg2018}. For the case $\ell=\tau$, only the fraction of $J/\psi(2S) \to \tau^+\tau^-$ does not vanish ,
which equals $3.1\times10^{-3}$ from Ref.\cite{pdg2018}.

\begin{table}[thb]
\caption{The masses,  decay widths and branching fractions of the dilepton decays of  the vector charmonium states~\cite{pdg2018}. }
\label{tab:table2}
\begin{center}
\begin{ruledtabular}
\begin{tabular}{cccc}
$V$ & Mass[\rm{GeV}] &  $\Gamma_{\rm tot}^V$[\rm{MeV}] &${\cal BR}(V\to l^+ l^-)$ with $l=e,\mu$
\\
\hline $\rho(770)$ & $0.775$ & $149.1$ & $4.63\times10^{-5}$
\\
 $\omega(782)$ & $0.782$ & $8.490$ & $7.38\times10^{-5}$
\\
 $\phi(1020)$ & $1.019$ & $4.249$ & $2.92\times10^{-4}$
\\
\hline $J/\psi(1S)$ & $3.096$ & $0.093$ & $5.96\times10^{-2}$
\\
$\psi(2S)$ & $3.686$ & $0.294$ & $7.96\times10^{-3}$
\\
$\psi(3770)$ & $3.773$ & $27.2$ & $9.60\times10^{-6}$
\\
$\psi(4040)$ & $4.039$ & $80$ & $1.07\times10^{-5}$
\\
$\psi(4160)$ & $4.191$ & $70$ & $6.90\times10^{-6}$
\\
$\psi(4415)$ & $4.421$ & $62$ & $9.40\times10^{-6}$
\\
\end{tabular}
\end{ruledtabular}
\end{center}
\end{table}

\subsection{ \texorpdfstring{ $B_s \to K, K^* $}{} transition form factors}\label{sec:FFs}

The $B_s\to K$ transition can be induced by the vector current $V^\mu$ and the tensor currents $T^{\mu\nu}$:
\beq
\langle K(p_2)| V^{\mu}|B_s(p_1)\rangle &=&f_1(q^2) p_1^{\mu}+ f_2(q^2) p_2^{\mu},   \label{eq:f1f2} \\
\langle K(p_2)|T^{\mu\nu}|B_s(p_1)\rangle &=& i\frac{2}{m_{B_s}+m_K}\left[p^{\mu}_2q^\nu-q^\mu p^{\nu}_2\right]F_T(q^2),
\label{eq:FT}
\eeq
where $V^{\mu}=\bar{d}\gamma^{\mu}b$ and $ T^{\mu\nu}=\bar{d}\sigma^{\mu\nu}b$,  and  $q=p_1-p_2$ is the momentum carried
off by the lepton pairs  and  $\sigma^{\mu\nu}=i[\gamma^{\mu},\gamma^{\nu}]/2$.

The $B_s\to K$ transition form factors $F_+(q^2)$ and $F_0(q^2)$ can be written as a combination of  the auxiliary form factors
$f_1(q^2)$ and $f_2(q^2)$ in Eq.~(\ref{eq:f1f2}):
\beq
F_+(q^2)&=&\frac12\left[f_1(q^2)+f_2(q^2)\right], \label{eq:fpq2}\\
F_0(q^2)&=& F_+(q^2) + \frac12 \left[f_1(q^2)-f_2(q^2)\right] \frac{q^2}{m_{B_s}^2-m_K^2}.
\label{eq:f0q2}
\eeq
We also have the relation $F_+(0)=F_0(0)$ in order to smear the pole at $q^2=0$.

Using the well-studied wave functions as given in Sec.~\ref{sec:2}, we  calculated the three $B_s\to K$ form factors $f_1(q^2), f_2(q^2)$ and $F_T(q^2)$ in
the PQCD factorization approach:
\beq
f_1(q^2)&=& 16\pi m_{B_s}^2C_F\int dx_1 dx_2\int b_1 db_1 b_2 db_2\phi_{B_s}(x_1)\non
&& \times  \Bigl\{ \Big[\!-\!x_2r^2\phi^A_{K}(x_2)\!+\!r_0\phi^P_{K}(x_2)\!-\!\frac{\eta\!+\!2x_2r}{\sqrt{\eta^2\!-\!1}}r_0\phi^T_{K}(x_2)\Big]
\!\cdot \!H_1(t_1)\non
&& + \Big[ \left (x_1(\eta\!+\!\sqrt{\eta^2\!-\!1})\!-\!r\!+\!\frac{x_1}{2\sqrt{\eta^2\!-\!1}} \right ) r\phi^A_{K}(x_2)\!\non
&& - x_1 \left ( 1\!+\!\frac{\eta}{\sqrt{\eta^2\!-\!1}} \right )r_0\phi^P_{K}(x_2)\Big]\!\cdot \!H_2(t_2)\Bigr \}, \label{eq:f1q2}
\eeq
\beq
f_2(q^2)&=&16\pi m_{B_s}^2C_F\int dx_1 dx_2\int b_1 db_1 b_2 db_2\phi_{B_s}(x_1)\non
&& \times \Bigl\{\Big[\!(1+x_2r\eta)\phi^A_{K}(x_2)\!-\!2x_2r_0\phi^P_{K}(x_2)\!
-\!\frac{1\!-\!2x_2r\eta}{r\sqrt{\eta^2\!-\!1}}r_0\phi^T_{K}(x_2)\Big]\!\cdot \!H_1(t_1)\non
&& +\Big[\!-\!\frac{x_1}{2}\left (1\!+\!\frac{\eta}{\sqrt{\eta^2\!-\!1}} \right )\phi^A_{K}(x_2)\!
+\! \left (2\!+\!\frac{x_1}{r\sqrt{\eta^2\!-\!1}} \right )r_0\phi^P_{K}(x_2)\Big]
\cdot H_2(t_2)\Bigr \}, \label{eq:f2q2}\\
F_T(q^2)&=& 8\pi m_{B_s}^2C_F(1+r)\int dx_1 dx_2\int b_1 db_1 b_2 db_2\phi_{B_s}(x_1)\non
&& \times \Bigl\{\Big[\phi^A_{K}(x_2)\!-\!x_2r_0\phi^P_{K}(x_2)\!
-\!\frac{1\!+\!x_2r\eta}{r\sqrt{\eta^2\!-\!1}}r_0\phi^T_{K}(x_2)\Big]\!\cdot \!H_1(t_1)\non
&& + \Big[\!-\!\frac{x_1}{2}\left (1\!+\!\frac{\eta}{\sqrt{\eta^2\!-\!1}} \right )\phi^A_{K}(x_2)\!
+\! \left (2\!+\!\frac{x_1}{r\sqrt{\eta^2\!-\!1}} \right )r_0\phi^P_{K}(x_2)\Big]\!\cdot \!H_2(t_2)\Bigr \}.
\label{eq:fTq2}
\eeq
where  $C_F=4/3$ is a color factor, $r_0=m^0_K/m_{B_s}$,  $r=m_k/m_{B_s}$,  $\eta$ is defined in Eq.~(\ref{eq:eta9}),  and  the function $H_i(t_i)$ in the following form
\beq
H_i(t_i)=h_i(x_1,x_2,b_1,b_2)\cdot\alpha_s(t_i)\exp \left[-S_{ab}(t_i)\right],\quad for \quad i=(1,2).
\label{eq:hiti}
\eeq
The explicit expressions of the hard functions $h_{1,2}(x_1,x_2,b_1,b_2)$ , the hard scales $t_{1,2}$ and the Sudakov factors $S_{ab}(t_i)$
will be given in Appendix~\ref{sec:app1}.

For the vector meson $K^*$ with polarization vector $\epsilon^{*}$, the relevant  form factors for $B_s \to K^*$ transitions are
$V(q^2)$ and $A_{0,1,2}(q^2)$ of the vector and axial-vector currents, and  $T_{1,2,3}(q^2)$ of the tensor currents.
In the PQCD factorization approach,  these seven form factors  of $B_s \to K^{\ast} \ell^+\ell^-$ decays
can be calculated and  written in the following form:
\beq
V(q^2)&=& 8\pi m_{B_s}^2C_F(1\!+\!r)\int dx_1 dx_2\int b_1 db_1 b_2 db_2\phi_{B_s}(x_1)\non
&& \times \Bigl\{\Big[\!-\!x_2r\phi^v_{K^*}(x_2)\!+\!\phi^T_{K^*}(x_2)\!+\!\frac{1\!+\!x_2r\eta}{\sqrt{\eta^2\!-\!1}}\phi^a_{K^*}(x_2)\Big] \!\cdot \!H_1(t_1)\non
&& \!+\!\Big[ \left (r\!+\!\frac{x_1}{2\sqrt{\eta^2\!-\!1}} \right ) \phi^v_{K^*}(x_2)
\!-\!\frac{x_1-2r\eta}{2\sqrt{\eta^2\!-\!1}}\phi^a_{K^*}(x_2)\Big]\!\cdot \!H_2(t_2)\Bigr \}, \label{eq:Vq2}
\eeq
\beq
A_0(q^2)&=& 8\pi m_{B_s}^2C_F\int dx_1 dx_2\int b_1 db_1 b_2 db_2\phi_{B_s}(x_1) \times \Bigl\{\Big[ \left (1\!+\!x_2r(2\eta\!-\!r) \right )\phi_{K^*}(x_2)\non
&& +\!(1\!-\!2x_2)r\phi^t_{K^*}(x_2)\!
+\!\frac{(1\!-\!r\eta)-2x_2r (\eta\!-\!r)}{\sqrt{\eta^2\!-\!1}}\phi^s_{K^*}(x_2)\Big]\!\cdot \!H_1(t_1)\non
&& \!+\!\Big[\Big[\frac{x_1}{\sqrt{\eta^2\!-\!1}} \left (\frac{\eta\!+\!r}{2}\!-\!r\eta^2 \right )\!+
\!\left (\frac{x_1}{2}\!-\!x_1r\eta\!+\!r^2 \right ) \Big]\phi_{K^*}(x_2)\non
&& -\!\Big [ \frac{x_1(1\!-\!r\eta)+2r(r\!-\!\eta)}{\sqrt{\eta^2\!-\!1}}\!-\!x_1r\Big]\phi^s_{K^*}(x_2)\Big ]
\!\cdot \!H_2(t_2)\Bigr \},\label{eq:A0q2}
\eeq
\beq
A_1(q^2)&=&16\pi m_{B_s}^2C_F\frac{r}{1\!+\!r}\int dx_1 dx_2\int b_1 db_1 b_2 db_2\phi_{B_s}(x_1)\non
&& \times \Bigl\{\Big[ (1\!+\!x_2r\eta)\phi^v_{K^*}(x_2)
\!+\! (\eta\!-\!2x_2r)\phi^T_{K^*}(x_2)\!+\!x_2r\sqrt{\eta^2\!-\!1}\phi^a_{K^*}(x_2)\Big]\!\cdot \!H_1(t_1)\non
&& \!+\!\Big[\left (r\eta\!-\!\frac{x_1}{2} \right )\phi^v_{K^*}(x_2)\!+\!\left (r\sqrt{\eta^2\!-\!1}\!+\!\frac{x_1}{2}\right )
\phi^a_{K^*}(x_2)\Big]\!\cdot \!H_2(t_2)\Bigr \},\label{eq:A1q2}
\eeq
\beq
A_2(q^2)&=&\frac{(1\!+\!r)^2(\eta\!-\!r)}{2r(\eta^2\!-\!1)}A_1(q^2)-8\pi m_{B_s}^2C_F\frac{1\!+\!r}{\eta^2\!-\!r}\int dx_1 dx_2\int b_1 db_1 b_2 db_2\phi_{B_s}(x_1)\non
&& \times \Bigl\{\Big[\big[\eta \left (1\!-\!x_2r^2\right )\!+\!r \left (x_2(2\eta^2\!-\!1)\!-\!1 \right )\big]\phi_{K^*}(x_2)
\!+\!\Big[1\!+\!2x_2r^2\!-\!(1\!+\!2x_2)r\eta\Big] \phi^t_{K^*}(x_2)\non
&& +\!r(1\!-\!2x_2)\sqrt{\eta^2\!-\!1}\phi^s_{K^*}(x_2)\Big]\!\cdot \!H_1(t_1)\non
&& +\!\Big[\Big[ \left ( r\eta\!-\!\frac{1}{2} \right )x_1\sqrt{\eta^2\!-\!1}\!-\!\Big [ r \left (r\eta\!-\!1\!-\!x_1\eta^2\right )\!+\!\frac{x_1(r\!+\!\eta)}{2}\Big] \Big] \phi_{K^*}(x_2)\non
&&+\!\Big[  x_1(r\eta\!-\!1)\!+\!(x_2\!-\!2)r\sqrt{\eta^2\!-\!1}\Big] \phi^s_{K^*}(x_2)\Big]\!\cdot \!H_2(t_2)\Bigr \}, \label{eq:A2q2}
\eeq
\beq
T_1(q^2)&=&8\pi m_{B_s}^2C_F\int dx_1 dx_2\int b_1 db_1 b_2 db_2\phi_{B_s}(x_1)\times \Bigl\{\Big[(1\!-\!2x_2)r\phi^v_{K^*}(x_2)\! \non
& +& \left (1\!+\!2x_2r\eta\!-\!x_2r^2 \right )
\phi^T_{K^*}(x_2)\!+\!\frac{1\!+\!2x_2r^2\!-\!(1\!+\!2x_2)r\eta}{\sqrt{\eta^2\!-\!1}}\phi^a_{K^*}(x_2)\Big ]\!\cdot \!H_1(t_1)\non
&+& \Big[ \Big[ \left (1\!-\!\frac{x_1}{2} \right )r-\frac{x_1(r\eta\!-\!1)}{2\sqrt{\eta^2\!-\!1}}\Big ]\phi^v_{K^*}(x_2)
+\Big[ \frac{r(\eta\!-\!r)}{\sqrt{\eta^2\!-\!1}}\!+\!\frac{x_1}{2}\left (r+\frac{r\eta\!-\!1}{\sqrt{\eta^2\!-\!1}}\right ) \Big ] \phi^a_{K^*}(x_2)\Big ]\non
&& \!\cdot \!H_2(t_2)\Bigr \},\quad
\label{eq:T1q2}
\eeq
\beq
T_2(q^2)&=&16\pi m_{B_s}^2C_F\frac{r}{1\!-\!r^2}\int dx_1 dx_2\int b_1 db_1 b_2 db_2\phi_{B_s}(x_1)\non
&& \times \Bigl\{\Big[(1\!-\!(1+2x_2)r\eta\!+\!2x_2r^2)\phi^v_{K^*}(x_2)\!\non
&& +\!\Big[x_2r\eta(2\eta\!-\!r)\!-\!x_2r\!+\!\eta\!-\!r\Big]\phi^T_{K^*}(x_2)
+\!(1\!-\!2x_2)r\sqrt{\eta^2\!-\!1}\phi^a_{K^*}(x_2)\Big]\!\cdot \!H_1(t_1)\!\non
&& +\!\Big[\Big[\frac{x_2}{2}(1\!+\!\frac{\eta}{\sqrt{\eta^2\!-\!1}})(r\eta\!-\!1)\!+
\!(r\!+\!\frac{x_1}{2\sqrt{\eta^2\!-\!1}})(\eta\!-\!r)\Big]\phi^v_{K^*}(x_2)\non
&& \!+\!\Big[(1\!-\!\frac{x_1}{2})r\sqrt{\eta^2-1}\!+\!\frac{x_1}{2}(1\!-\!r\eta)\Big]\phi^a_{K^*}(x_2)\Big]
\!\cdot \!H_2(t_2)\Bigr \},\label{eq:T2q2}
\eeq
\beq
T_3(q^2)&=&\frac{(1\!-\!r)^2(\eta\!+\!r)}{2r(\eta^2\!-\!1)}T_2(q^2)-8\pi m_{B_s}^2C_F\frac{1\!-\!r^2}{\eta^2\!-\!1}\int dx_1 dx_2\int b_1 db_1 b_2 db_2\phi_{B_s}(x_1)\non
&&\hspace{-1cm}\times \Bigl\{\Big[\frac{\eta^2\!-\!(1\!+\!2x_2)r\eta\!+\!2x_2r^2}{\eta\!-\!r}\phi_{K^*}(x_2)
\!+\!(1\!+\!x_2r\eta)\phi^t_{K^*}(x_2)\!+\!x_2r\sqrt{\eta^2\!-\!1}\phi^s_{K^*}(x_2)\Big]\!\cdot \!H_1(t_1)\non
&&\hspace{-1cm} +\!\Big[\big[r\!-\!\frac{x_1}{2}(\eta\!+\!\sqrt{\eta^2\!-\!1})\big]\phi_{K^*}(x_2)
\!+\!(x_1\!+\!2r\sqrt{\eta^2\!-\!1})\phi^s_{K^*}(x_2)\Big]\!\cdot \!H_2(t_2)\Bigr \},\label{eq:T3q2}
\eeq
where $r=m_{K^*}/m_{B_s}$,  the twist-2 DAs $(\phi_{K^*},\phi^T_{K^*})$ and other four twist-3 DAs are defined
in Eqs.~(\ref{eq:phiv},\ref{eq:phivT},\ref{eq:t301}), the functions $H_{1,2}(t_{1,2})$ are the same ones as those defined in Eq.~(\ref{eq:hiti}) for
$B_s \to K$ transition, but with a replacement of $r=m_{K}/m_{B_s}$ by $r=m_{K^*}/m_{B_s}$.

\section{Observables for \texorpdfstring{$B_s \to K^{(\ast)} \ell^+ \ell^-$ decays}{}}\label{sec:4}

\subsection{Observables for \texorpdfstring{$B_s \to K \ell^+ \ell^-$ decays}{}}

Within the SM operator basis, the  decay amplitude of $B_s \to K\ell^+\ell^-$ decay  can be written in the following form~\cite{Bobeth:2011nj}:
\begin{align}
\mathcal{A}\left(\bar{B^0_s} \rightarrow K \ell^{+} \ell^{-}\right)=\frac{G_{\mathrm{F}} \alpha_{em}}{\sqrt{2} \pi} V_{t b}
V_{t d}^{*} \left[\delta_{V} p_1^{\mu}\left(\bar{\ell} \gamma_{\mu} \ell\right)+\delta_{A} p_1^{\mu}\left(\bar{\ell}
\gamma_{\mu} \gamma_{5} \ell\right)+\delta_{P}\left(\bar{\ell} \gamma_{5} \ell\right)\right]
\end{align}
with
\beq
 \delta_A & =& C_{10} F_+(q^2), \non
  \delta_V & =& C_9^{\rm{eff}}F_+(q^2)+C_{7}^{\rm{eff}}\frac{2(m_b-m_d)}{m_{B_s}+m_K}F_T(q^2), \non
   \delta_P & =& -m_\ell C_{10}\left\{F_+(q^2)+\frac{m^2_{B_s}-m^2_K}{q^2}\left[F_+(q^2)-F_0(q^2)\right] \right \},
\eeq
where $p_1^\mu$ denotes the four-momentum of the $B_s$-meson and $m_\ell$ is the lepton mass.

Based on the matrix elements of the operators in terms of the form factors, we obtain the double differential decay rate for
$\bar B_s\to K\ell^+\ell^-$ with respect to $q^2$ and $\thl$ with lepton flavor $\ell$~\cite{Bobeth:2007dw},
\beq
\frac{d^2\Gamma}{dq^2d\cos\thl} = a_\ell(q^2) + b_\ell(q^2) \cos\thl + c_\ell(q^2) \cos^2\thl \,.
\eeq
The angle $\theta_\ell$ is defined as the angle between the $\bar{B_s}$-direction and the $\ell^-$-direction in the $\ell^+\ell^-$
rest frame.  The corresponding angular coefficients $a_\ell$, $b_\ell$ and $c_\ell$ can be written as\cite{Bobeth:2007dw,Bobeth:2011nj}
\beq
a_\ell(q^2) &=&  \mathcal{N} \Big[ q^2 |\delta_P|^2 + \frac{\lambda}{4}(|\delta_A|^2 + |\delta_V|^2) + 4m_\ell^2 m_{B_s}^2 |\delta_A|^2\non
&& + 2m_\ell(m_{B_s}^2 - m_K^2 + q^2){\rm Re}(\delta_P\delta_A^*) \Big] \,, \\
b_\ell(q^2) &=& 0 \,, \\
c_\ell(q^2) &=&-\frac{\mathcal{N} \lambda \beta_\ell^2}{4}(|\delta_A|^2 + |\delta_V|^2).
\eeq
with the factor $ \mathcal{N} $,
\beq
\mathcal{N} &=& \frac{G_F^2 \alpha_{em}^2 |V_{tb} V_{td}^*|^2}{2^9\pi^5m_{B_s}^3}\beta_\ell \sqrt{\lambda}.
\eeq
where $\beta_l =\sqrt{1- 4\hat m_l^2}$ with $\hat m_l=m_l/\sqrt{q^2} $, $\alpha_{em}=1/137$ is the fine structure constant, $m_\ell$ means the lepton mass and
$\lambda = \lambda(m_{B_s}^2, m_{K}^2, q^2) $ is the K\"allen function:  $ \lambda (a,b,c) = a^2+ b^2 +c^2- 2(ab+bc+ca)$.

Integration over the polar angle $\thl$ leads to the expression for the differential decay rate,
\beq\
{\frac{d\Gamma}{dq^2}} =  2a_\ell(q^2)+{\frac{2}{3}}c_\ell(q^2). \label{eq:DDR1}
\eeq
We see that the linear dependence on $\cos\thl$  is lost after integration over $\thl$, consequently , the lepton forward-backward asymmetry 
$\mathcal{A}_{\mathrm{FB}}$  will also become zero,
\beq
\mathcal{A}_{\mathrm{FB}}\left(q^{2}\right)=\frac{\int_{0}^{1} \frac{d^{2}
\Gamma}{d q^{2} d \cos \theta_{\ell}} d \cos \theta_{\ell}-\int_{-1}^{0} \frac{d^{2} \Gamma}{d q^{2} d \cos \theta_{\ell}}
d \cos \theta_{\ell}}{d \Gamma / d q^{2}}=\frac{b_\ell(q^2)}{d \Gamma / d q^{2}}=0.
\eeq

Another observable of interest that we calculate is the longitudinal polarization asymmetry $P_L(q^2)$ of the leptons defined as ~\cite{Singh:2019hvj}:
\begin{eqnarray}
P_L(q^2)=\frac{1}{d\Gamma/dq^2}\left[\frac{d\Gamma^{h_{\ell}=-1}}{dq^2}-\frac{d\Gamma^{h_{\ell}=+1}}{dq^2}\right],
\end{eqnarray}
where $h_{\ell}=+1(-1)$ implies a right(left)-handed charged lepton $\ell^-$ in the final state. For the $\bar B_s\to K\ell^+\ell^-$ decay, the lepton polarization is given by~\cite{Singh:2019hvj}:
\begin{eqnarray}
P_L(q^2)={\frac{1}{3}}\cdot \frac{G_F^2 \alpha_{em}^2 |V_{tb} V_{td}^*|^2}{256\pi^5m_{B_s}^3}
\cdot\frac{ \beta^2_\ell \lambda^{3/2} \cdot {\text {Re}}\{\delta_V \delta^\ast_A \}}{d \Gamma / d q^{2}}.
\end{eqnarray}
 For the CP-conjugated mode $B^0_s \to \bar{K} \ell^{+} \ell^{-}$,  the decay amplitude and physical observables are obtained by making the substitution
 ${\cal A}\to {\cal \bar{A}}$, i.e., by making the complex conjugation of the CKM factor involved for  the $\bar{B}_s^0$ decay modes.
Analogous to  Ref.~\cite{Genon:2013vna},  we  also define the direct CP asymmetry of the considered $B_s \to K l^+l^-$ decays in the following form:
\beq
\cala_{CP}(q^2) = \frac{d\Gamma(\bar{B}_s \to K l^+l^-) /dq^2 - d\bar{\Gamma}(B_s \to \bar{K} l^+l^- ) /dq^2}
{ d\Gamma(\bar{B}_s \to K l^+l^-) /dq^2 + d\bar{\Gamma}(B_s \to \bar{K} l^+l^- ) /dq^2 } .
\label{eq:acp01}
\eeq

\subsection{Observables for \texorpdfstring{$B_s \to K^{\ast} \ell^+\ell^-$}{} }

For a four body decay, $B_s\to K^\ast(\to \pi K)\ell^+\ell^-$, the decay distribution can be completely described in terms of four kinematic
variables\cite{Genon:2012zf,Genon:2013vna,Bobeth:2008ij}: the lepton invariant mass squared ($q^2$) and
three angles $\thK$, $\thl$, and $\phi$. The angle $\thK$ is the angle between  the direction of flight of $K$
and the $B_s$ meson in the rest frame of $K^*$, $\thl$ is the angle made by $\ell^-$ with respect to
the $B_s$ meson in the dilepton rest frame and $\phi$ is the azimuthal angle between the two planes formed by dilepton and $\pi K$.
The full angular decay distribution of $\bar{B_s}\to K^\ast(\to \pi^- K^+)\ell^+\ell^-$ is given by \cite{Kindra:2018ayz,Altmannshofer:2008dz,Becirevic:2011bp},
\beq
  \frac{d^4\Gamma}{dq^2\, d\cos\thK\, d\cos\thl\, d\phi} =
   \frac{9}{32\pi} I(q^2, \thK, \thl, \phi)\,, \label{eq:d4Gamma}
\eeq
where the functions $ I(q^2, \thK, \thl, \phi)$ are of the following form~\cite{Altmannshofer:2008dz}:
\beq
 I(q^2, \thl, \thK, \phi)& =& \sum\limits_{i} I_{i}(q^2){f_{i}(\thK,\thl,\phi)} \non
    & =&       I_1^s \sin^2\thK + I_1^c \cos^2\thK       + (I_2^s \sin^2\thK + I_2^c \cos^2\thK) \cos 2\thl \non
    & +&  I_3 \sin^2\thK \sin^2\thl \cos 2\phi       + I_4 \sin 2\thK \sin 2\thl \cos\phi \non
    & + & I_5 \sin 2\thK \sin\thl \cos\phi       + I_6^s \sin^2\thK \cos\thl       + I_7 \sin 2\thK \sin\thl \sin\phi
\non
    & + & I_8 \sin 2\thK \sin 2\thl \sin\phi       + I_9 \sin^2\thK \sin^2\thl \sin 2\phi\,.  \label{eq:angulardist}
\eeq
For the CP-conjugated mode $B_s\to \bar{K}^\ast(\to \pi^+ K^-)\ell^+\ell^-$, the corresponding expression of the angular decay distribution is
\beq
  \frac{d^4\bar{\Gamma}}{dq^2\, d\cos\thK\, d\cos\thl\, d\phi} =
   \frac{9}{32\pi} \bar{I}(q^2, \thK, \thl, \phi)\,, \label{eq:d4Gammabar}
\eeq
The function $ \bar{I}(q^2, \thK, \thl, \phi)$ is obtained from Eq.~(\ref{eq:angulardist}) by the substitution~\cite{Altmannshofer:2008dz}:
\beq
 I_{1,2,3,4,7} \to \bar{I}_{1,2,3,4,7}, \quad I_{5,6,8,9} \to -\bar{I}_{5,6,8,9},\label{eq:replacement}
\eeq
where $\bar{I}_i$ are obtained by making the complex conjugation for all weak phases in $I_i$.
The minus sign in Eq.~(\ref{eq:replacement}) is a result of the convention that, under the previous definitions of three angles, a CP transformation
interchanges  the lepton and antilepton, leading to the modification $\thl \to \thl-\pi$ and $\phi \to -\phi$.

The angular coefficients $I_i$ of the distributions in above equation can be written in terms of the transverse amplitudes~\cite{Matias:2012xw,Aaij:2015oid}.
For the massless case there are six such complex amplitudes: ${\cal A}_0^{R,L}$, ${\cal A}_\|^{R,L}$ and ${\cal A}_\perp^{R,L}$.
For the massive case an additional complex amplitude ${\cal A}_t$ is required.
In Table \ref{tab:table3},  we show the expressions for those angular coefficients $I_i(q^2)$ and the corresponding angular factor ${f_{i}(\thK,\thl,\phi)}$ as those
defined in Refs.~\cite{Matias:2012xw,Aaij:2015oid}.

\begin{table}[thb]
\caption{The explicit expressions of the angular coefficients $I_i(q^2)$ and $f_i$ appeared in Eq.~(\ref{eq:angulardist}).  }
\label{tab:table3}
\begin{center}
\begin{tabular}{|l|l|l|}
\hline
$i$ & $I_{i}(q^2)$ & $f_{i}$ \\
\hline
$1s$ & $(\frac{3}{4}\!-\!\hat{m}^2_\ell)\left[ |{\cal A}_{\parallel}^{\rm L}|^{2}\!+\! |{\cal A}_{\perp}^{\rm L}|^{2} \!+\!|{\cal A}_{\parallel}^{\rm R}|^{2} \!+\! |{\cal A}_{\perp}^{\rm R}|^{2}\right]\!+\!4 \hat{m}^2_\ell {\rm Re}\left[ {\cal A}_{\perp}^{\rm L}{\cal A}_{\perp}^{\rm R\ast}\! +\!{\cal A}_{\parallel}^{\rm L}{\cal A}_{\parallel}^{\rm R\ast}\right] $ & $\sin^{2}\thK$ \\
$1c$ &  $|{\cal A}_{0}^{\rm L}|^{2} + |{\cal A}_{0}^{\rm R}|^{2} +4 \hat{m}^2_\ell \left[|{\cal A}_{t}|^{2}+2{\rm Re}[{\cal A}_{0}^{\rm L}{\cal A}_{0}^{\rm R\ast}]\right]$ & $\cos^{2}\thK$ \\
$2s$ & $\frac{1}{4} \beta^2_\ell \left[ |{\cal A}_{\parallel}^{\rm L}|^{2} + |{\cal A}_{\perp}^{\rm L}|^{2}  + |{\cal A}_{\parallel}^{\rm R}|^{2} + |{\cal A}_{\perp}^{\rm R}|^{2}\right]$ & $\sin^{2}\thK\cos 2\thl$ \\
$2c$ &  $-\beta^2_\ell \left[|{\cal A}_{0}^{\rm L}|^{2} +  |{\cal A}_{0}^{\rm R}|^{2}\right]$ & $\cos^{2}\thK\cos 2\thl$ \\
3 & $\frac{1}{2} \beta^2_\ell \left[ |{\cal A}_{\perp}^{\rm L}|^{2} - |{\cal A}_{\parallel}^{\rm L}|^{2}  + |{\cal A}_{\perp}^{\rm R}|^{2} - |{\cal A}_{\parallel}^{\rm R}|^{2} \right]$  & $\sin^{2}\thK \sin^{2}\theta_{\ell} \cos 2\phi$ \\
4 & $ \sqrt{\frac{1}{2}} \beta^2_\ell {\rm Re}({\cal A}_{0}^{\rm L} {\cal A}_{\parallel}^{{\rm L}\ast} + {\cal A}_{0}^{\rm R} {\cal A}_{\parallel}^{{\rm R}\ast}) $  & $\sin 2\thK \sin 2\thl \cos \phi$ \\
5 &  $\sqrt{2} \beta_\ell {\rm Re}({\cal A}_{0}^{\rm L} {\cal A}_{\perp}^{{\rm L}\ast} - {\cal A}_{0}^{\rm R} {\cal A}_{\perp}^{{\rm R}\ast})$ &  $\sin 2\thK \sin \thl \cos \phi$ \\
$6s$ &  $2 \beta_\ell {\rm Re}({\cal A}_{\parallel}^{\rm L} {\cal A}_{\perp}^{{\rm L}\ast} - {\cal A}_{\parallel}^{\rm R} {\cal A}_{\perp}^{{\rm R}\ast})$ &  $\sin^{2}\thK \cos \thl$ \\
7 &  $\sqrt{2} \beta_\ell {\rm Im}({\cal A}_{0}^{\rm L} {\cal A}_{\parallel}^{{\rm L}\ast} - {\cal A}_{0}^{\rm R} {\cal A}_{\parallel}^{{\rm R}\ast})$ &  $\sin 2\thK \sin \thl \sin \phi$ \\
8 &  $\sqrt{\frac{1}{2}} \beta^2_\ell {\rm Im}({\cal A}_{0}^{\rm L} {\cal A}_{\perp}^{{\rm L}\ast} + {\cal A}_{0}^{\rm R} {\cal A}_{\perp}^{{\rm R}\ast})$&  $\sin 2\thK \sin 2\thl \sin \phi$ \\
9 &  $\beta^2_\ell {\rm Im}({\cal A}_{\parallel}^{{\rm L}\ast}{\cal A}_{\perp}^{\rm L}  +  {\cal A}_{\parallel}^{{\rm R}\ast}{\cal A}_{\perp}^{\rm R}) $ &  $\sin^{2}\thK \sin^{2}\thl \sin 2\phi$ \\
\hline
\end{tabular}
\end{center}
\end{table}

The seven transversity amplitudes  ${\cal A}_0^{R,L}$, ${\cal A}_\|^{R,L}$, ${\cal A}_\perp^{R,L}$ and ${\cal A}_t$, in turn,  can be expressed in terms of the relevant
$B_s \to K^{\ast} \ell^+\ell^-$ form factors~\cite{Egede:2008uy,Altmannshofer:2008dz}:
\begin{align} \label{eq:transversity}
{\cal A}_{\perp}^{\rm L,R}=&- N_\ell\sqrt{2 N_{K^*}} \sqrt{\lambda} \left[(C_9^{\rm{eff}}\mp C_{10})\frac{V(q^2)}{m_{B_s}+m_{K^*}}+2 \hat{m}_b C_7^{\rm{eff}} T_1(q^2)\right],\\
{\cal A}_{\parallel}^{\rm L,R}=&N_\ell\sqrt{2 N_{K^*}}
\Big [ (C_9^{\rm{eff}}\mp C_{10})(m_{B_s}+m_{K^*})A_1(q^2)+2 \hat{m}_b C_7^{\rm{eff}} (m^2_{B_s}-m^2_{K^*})T_2(q^2) \Big ],\\
{\cal A}_{0}^{\rm L,R}=&\frac{N_\ell\sqrt{N_{K^*}}}{2m_{K^*} \sqrt{q^2}}\Big\{(C_9^{\rm{eff}}\!\mp\! C_{10})
\left [ (m^2_{B_s}\!-\!m^2_{K^*}\!-\!q^2)(m_{B_s}\!+\!m_{K^*})A_1(q^2)
\!-\!\frac{\lambda}{m_{B_s}\!+\!m_{K^*}}A_2(q^2) \right ]\non
&+2 m_b C_7^{\rm{eff}} \Big [ (m^2_{B_s}\!+\!3m^2_{K^*}\!-\!q^2)T_2(q^2)\!-\!\frac{\lambda}{m^2_{B_s}\!-\!m^2_{K^*}}T_3(q^2) \Big ]\Big\},\\
{\cal A}_{t}=2&N_\ell\sqrt{N_{K^*}}\frac{\sqrt{\lambda}}{\sqrt{q^2}}C_{10}A_0(q^2).
\end{align}
 where the factors $N_\ell$ and $N_{K^*}$ are of the following form:
 \beq
 N_\ell=\frac{i \alpha_{em} G_F}{4\sqrt{2}\pi}V_{tb}V^\ast_{td},\quad
 N_{K^*}=\frac{8\beta_\ell\sqrt{\lambda}q^2}{3\cdot256\pi^3m^3_{B_s}}.
 \eeq
 with $\lambda\equiv(m^2_{B_s}\!-\!m^2_{K^*}\!-\!q^2)^2\!-\!4m^2_{K^*} q^2$,
$\beta_\ell=\sqrt{1\!-\!4m_\ell^2/q^2}$ and $\hat{m}_b=m_b/q^2$.

Analogous to Ref.~\cite{Becirevic:2019tpx},  one can  write down three partially integrated decay distributions,
integrating all but one angle at a time. For the CP-conjugated process, we can do the similar operation using the corresponding decay distributions.
\begin{itemize}
\item[(1)]
The $\thK$ distribution :
\beq
{\frac{d^2\Gamma}{dq^2 d\cos\thK}} &=& a_\thK(q^2) + c_\thK(q^2)\cos^2\thK \,, \non
a_\thK(q^2) &=& {\frac{3}{8}} \left( 3I^s_{1} - I^s_{2} \right) \,, \non
c_\thK(q^2) &=& {\frac{3}{8}} \left(3I^c_{1} - 3I^s_{1} - I^c_{2} + I^s_{2} \right) \,.
\eeq
\item[(2)]
The $\thl$ distribution :
\beq
{\frac{d^2\Gamma}{dq^2 d\cos\thl}} &=& a_\thl(q^2) + b_\thl(q^2)\cos\thl + c_\thl(q^2)\cos^2\thl \,, \non
a_\thl(q^2) &=& {\frac{3}{8}} \left( I^c_{1} + 2I^s_{1} - I^c_{2} - 2I^s_{2} \right) \,, \non
b_\thl(q^2)  &=& {\frac{3}{4}} \left( I^s_{6} \right) \,, \non
c_\thl(q^2)  &=&  {\frac{3}{4}} \left( I^c_{2} + 2I^s_{2} \right) \, .
\eeq

\item[(3)]
The $\phi$ distribution :
\beq
{\frac{d^2\Gamma}{dq^2 d\phi}} &=& a_\phi(q^2) + c_\phi^c(q^2)\cos2\phi + c_\phi^s(q^2)\sin2\phi \,, \non
a_\phi(q^2) &=& {\frac{1}{8\pi}} \left( 3I^c_{1} + 6I^s_{1} - I^c_{2} - 2I^s_{2} \right),  \non
c_\phi^c(q^2) &=& {\frac{1}{2\pi}} I_3 \,, \non
c_\phi^s(q^2) &=& {\frac{1}{2\pi}} I_9 \,.
\eeq
\end{itemize}

From the full angular distributions as defined in Eq.~(\ref{eq:angulardist}), we set various coefficients apart and combine
them into diverse quantities normalized to the differential decay rate and other observables\cite{Becirevic:2019tpx}.
Analogous observables are constructed for the CP-conjugated mode, after making the replacements as shown in Eq.~(\ref{eq:d4Gammabar}) and (\ref{eq:replacement}).
\begin{itemize}
\item[(1)]
The differential decay rate:
\begin{equation}
{\frac{d\Gamma}{dq^2}} = {\frac{1}{4}} \left( 3I^c_{1} + 6I^s_{1} - I^c_{2} - 2I^s_{2} \right) \,,
\label{eq:dGdq2}
\end{equation}

\item[(2)]
The lepton forward-backward asymmetry:
\begin{equation}
\mathcal{A}_{\rm FB}(q^2) = { \frac{b_\thl(q^2)}{d\Gamma/dq^2} } = {\frac{3I^s_{6}}{3I^c_{1} + 6I^s_{1} - I^c_{2} - 2I^s_{2}}} \,, \\
\label{eq:AFB}
\end{equation}

\item[(3)]
The $K^*$ polarization fraction:
\begin{equation}
R_{L,T}(q^2) = {\frac{d\Gamma_L/ dq^2}{d\Gamma_T/dq^2}} \,,
\label{eq:RLT}
\end{equation}
where $\Gamma_L$ and $\Gamma_T$ represent the longitudinal and transverse $K^\ast$ polarization decay rates,
\beq
{\frac{d\Gamma_L}{dq^2}} &=& {\frac{2}{3}} \left[ a_\thK (q^2) + c_\thK (q^2) \right] = {\frac{1}{4}} \left( 3I^c_{1} - I^c_{2} \right) \,, \\
{\frac{d\Gamma_T}{dq^2}} &=& {\frac{4}{3}} a_\thK (q^2) = {\frac{1}{2}} \left( 3I^s_{1} - I^s_{2} \right) \,.
\eeq
Alternatively, one can define the quantity $F^{K^\ast}_L$ which is a measure of the longitudinally polarized $K^\ast$'s in the whole
ensemble of ${B_s\to K^\ast \ell\ell}$ decays, which is linked to $R_{L,T}(q^2)$ as:
\beq
\label{eq:FLDstar}
F_L^{K^\ast}(q^2)= \frac{R_{L,T}(q^2)}{1+ R_{L,T}(q^2)} \,=\frac{1}{4}\cdot\frac{3 I^c_{1}-I^c_{2}}{d\Gamma/dq^2}=\,
\frac{3 I^c_{1}-I^c_{2}}{3I^c_{1} + 6I^s_{1} - I^c_{2} - 2I^s_{2}}\, ,
\eeq
where $F_L^{K^\ast}$ is a number obtained by integrating $F_L^{K^\ast}(q^2)$ over the proper  phase space.

\item[(4)]
The direct CP asymmetry can also be defined in the same way as for $B_s \to K l^+l^-$ decays:
\beq
\cala_{CP}(q^2) = \frac{d\Gamma(\bar{B}_s \to K^* l^+l^-) /dq^2 - d\bar{\Gamma}(B_s \to \bar{K}^* l^+l^- )/dq^2}
{ d\Gamma(\bar{B}_s \to K^* l^+l^-) /dq^2 + d\bar{\Gamma}(B_s \to \bar{K}^* l^+l^- )/dq^2 } .
\label{eq:acp02}
\eeq

\end{itemize}

The above observables are constructed from Eqs.~(\ref{eq:d4Gamma},\ref{eq:angulardist}) by integrating over the angles in various ranges.
These observables  which have a form factor dependence in the leading order are called \textit{ form factor dependent }(FFD)
observables  and generally plagued by the large uncertainties of the form factors.
To avoid this problem, a lot of works have been done to construct observables which are theoretically clean in low-$q^2$  region.
Such observables are free from this dependence at the leading order and are called \textit{form factor independent}
(FFI) observables. In this paper, we study both kinds of  observables.

As a necessary and sufficient condition, such FFI observable must be invariant under the symmetry transformations of the transverse amplitudes $A$'s;
we then say that the observable respects the symmetries of the angular distribution. Fortunately, there exists a systematic procedure to construct all
such possible observables as discussed in Ref.~\cite{Matias:2012xw}.

We start defining the following complex vectors \cite{Egede:2010zc},
\beq
n_\|=\binom{A_\|^L}{A_\|^{R*}}\ ,\quad
n_\bot=\binom{A_\bot^L}{-A_\bot^{R*}}\ ,\quad
n_0=\binom{A_0^L}{A_0^{R*}}\ .
\eeq
With these vectors we can construct the products $|n_i|^2= n_i^\dagger n_i$ and $n_i^\dagger\, n_j$,
\begin{align}
|n_\||^2=&|A_\|^L|^2+|A_\|^{R}|^2= \dfrac{2 I^s_{2} -I_3}{\beta_\ell^2}\ , \quad\! &
n_\bot^\dagger\, n_\| =& A_\bot^{L*} A_\|^L-A_\bot^{R} A_\|^{R*}= \dfrac{\beta_\ell I^s_{6} -2 i I_9}{2{\beta_\ell^2}}\ ,\\[5mm]
|n_\bot|^2=&|A_\bot^L|^2+|A_\bot^{R}|^2=\dfrac{2 I^s_{2} +I_3}{\beta_\ell^2}\ , \quad\! &
n_0^\dagger\, n_\|=& A_0^{L*} A_\|^{L}+A_0^{R} A_\|^{R*}= \dfrac{2 I_{4} - i \beta_\ell I_7}{\sqrt{2}\beta_\ell^2}\ ,\\[5mm]
|n_0|^2=&|A_0^L|^2+|A_0^{R}|^2=-\dfrac{I^c_{2}}{\beta_\ell^2}\ , \quad\! &
n_0^\dagger\, n_\bot=& A_0^{L*} A_\bot^{L}-A_0^{R} A_\bot^{R*}=  \dfrac{\beta_\ell I_5 - 2i I_8}{\sqrt{2}\beta_\ell^2}\ .
\end{align}

We examine the following (clean) FFI observables\cite{Matias:2012xw}:
\beq
P_1&=&\frac{|n_\bot|^2-|n_\||^2}{|n_\bot|^2+|n_\||^2}=\frac{ I_{3}}{2 I^s_{2}}\ , \quad
P_2=\frac{{\rm Re}(n_\bot^\dagger\, n_\|)}{|n_\||^2+|n_\bot|^2}=\beta_\ell\frac{I^s_{6}}{8 I^s_{2}}\ ,\\
P_3&=&\frac{{\rm Im}(n_\bot^\dagger\, n_\|)}{|n_\||^2+|n_\bot|^2}=-\frac{I_{9}}{4 I^s_{2}}\ , \quad
P_4=\frac{{\rm Re}(n_0^\dagger\, n_\|)}{\sqrt{|n_\||^2 |n_0|^2}} =\frac{\sqrt{2}I_4}{\sqrt{-I^c_{2}(2 I^s_{2}-I_{3})}}\ ,
\eeq
\beq
P_5&=&\frac{{\rm Re}(n_0^\dagger\, n_\bot)}{\sqrt{|n_\bot|^2 |n_0|^2}} =\dfrac{\beta_\ell I_5}{\sqrt{-2 I^c_{2}(2 I^s_{2}+I_3)}}\ ,\\
P_6&=&\frac{{\rm Im}(n_0^\dagger\, n_\|)}{\sqrt{|n_\||^2 |n_0|^2}} =-\frac{\beta_\ell I_7}{\sqrt{-2 I^c_{2}(2 I^s_{2}-I_3)}}\ ,\\
P_8&=&\frac{{\rm Im}(n_0^\dagger\, n_\bot)}{\sqrt{|n_\bot|^2 |n_0|^2}} =-\frac{\sqrt{2}I_8}{\sqrt{- I^c_{2}(2 I^s_{2}+I_3)}}\ ,
\eeq
The primed observables are also defined in the following form \cite{Kindra:2018ayz}:
\beq
P'_4 & \equiv & P_4 \sqrt{1-P_1}=\frac{I_4}{\sqrt{-I^c_2 I^s_2}}\ ,
P'_5  \equiv  P_5 \sqrt{1+P_1}=\frac{\beta_\ell I_5}{2\sqrt{-I^c_2 I^s_2}}\ ,\\
P'_6 & \equiv & P_6 \sqrt{1-P_1}=-\frac{\beta_\ell I_7}{2\sqrt{-I^c_2 I^s_2}}\ ,
P'_8  \equiv   P_8 \sqrt{1+P_1}=-\frac{I_8}{\sqrt{-I^c_2 I^s_2}}\ .
\eeq
These primed observables $P'_{4,5,6,8}$ are clean and good approximations to $P_{4,5,6,8}$ due to the fact that $P_1\simeq 0$ in the SM.
From the experimental perspective, fitting the primed observables will be simpler and more efficient despite the whole analysis can be performed
directly in terms of the observables $P_{4,5,6,8}$.

 Since the most observables are written in terms of the ratios, $O_i^\ell(q^2)=\mathcal{N}_i^\ell(q^2)/\mathcal{D}_i^\ell(q^2)$ with  $\mathcal{N}$ and $\mathcal{D}$
 being generically a numerator and a denominator, the integrated quantities are then defined as in Ref.~\cite{Becirevic:2019tpx}:
\begin{equation}
\label{eq:integration}
\langle O_i^\ell \rangle =\frac{\displaystyle{\int_{4m_\ell^2}^{q_{\rm max}^2} \mathcal{N}_i^\ell(q^2) \
dq^2} }{\displaystyle{\int_{4m_\ell^2}^{q_{\rm max}^2} \mathcal{D}_i^\ell(q^2) \ dq^2} }\,.
\end{equation}
We also check the physical observables $R^{e\mu }_{K,K^{\ast}}$ and $R^{\mu \tau}_{K,K^{\ast}}$, as defined in Eqs.~(\ref{eq:rkbs1},\ref{eq:rkbs2}),
since the  theoretical uncertainties are largely canceled in the ratio of the branching ratios of $B_s\to K^{(\ast)}\ell^+\ell^-$ decays.

In the region $q^2< 4m^2_{\mu}$,  where only the $e^+ e^-$ modes are allowed, there is a large enhancement of $B_s \to K^\ast e^+ e^-$ due to
the $1/q^2$ scaling of the photon penguin contribution \cite{Aubert:2008ps}.
In order to  remove the phase space effects in the ratio   $R^{e\mu }_{K^*}$ and keep consistent with other analysis
\cite{Hiller:2003js}, we here also use the lower cut of $4 m^2_{\mu}$ for both  the electron and muon modes in the definition of the ratio
 $R^{e\mu }_{K^*}$ as in Ref.~\cite{Hiller:2003js}:
 \beq
R^{e\mu }_{K^*}=\frac{\int^{q^2_{max}}_{4m^2_{\mu}}dq^2\frac{d\calb(B_s\to K^{*} \mu^+\mu^-)}{dq^2} }{\int^{q^2_{max}}_{4m^2_{\mu}}dq^2
\frac{d\calb(B_s\to K^{*} e^+e^-)}{dq^2}  }.
\label{eq:rknew}
\eeq

\section{Numerical results and discussions} \label{sec:5}

In the numerical calculations we use the following input parameters (here masses and decay constants are in units of GeV) \cite{pdg2018,Aoki:2019cca}:
\beq
 \Lambda^{f=4}_{\overline{\rm MS}}&=&0.250, \; \quad \tau_{B_s^0}=1.509 ps,\; \quad m_b=4.8,\; \quad m_W=80.38, \; \non
 m_{K}&=&0.498, \quad m_{K^{*}}=0.892 , \; \quad m_{B_s}=5.367, \; \quad m_{\tau}=1.777,  \;\non
  f_{B_s}&=&0.23 , \; \quad f_{K}=0.16,\quad f_{K^*}=0.217, \; \quad f^T_{K^*}=0.185,\;   \label{eq:inputs}
\eeq
For the CKM matrix elements and angles, we use the values as given in PDG 2018 ~\cite{pdg2018}:
\beq
V_{tb}&=&1.019 \pm0.025, \;\quad V_{ud}=0.97420 \pm 0.00021,  \non
V_{td}&=&|V_{td}|\cdot e^{-i\beta}, \; \quad|V_{td}|=(8.1\pm0.5)\times 10^{-3}, \; \quad \Sin(2\beta)=0.691\pm0.017,\non
V_{ub}&=&|V_{ub}|\cdot e^{-i\gamma},  \; \quad|V_{ub}|=(3.94\pm 0.36)\times 10^{-3},  \; \quad\gamma=(73.5^{+4.2}_{-5.1})^{\circ}.
\label{eq:vij1}
\eeq

\subsection{The form factors}

For the considered semileptonic decays, the differential decay rates and other physical observables strongly rely on the value and the shape of
the relevant form factors $F_{0,+}(q^2)$ and $F_T(q^2)$ for $B_s \to K \ell^+ \ell^-$ decays,  and the form factors $V(q^2),$ $A_{0,1,2}(q^2)$ and $T_{1,2,3}(q^2)$ for
$B_s \to K^* \ell^+ \ell^-$ decays.
These form factors have been calculated in rather different theories or models, such as the relativistic quark model
(RQM) \cite{Faustov:2013ima},  the light cone sum rule (LCSR) \cite{Straub:2015ica,Khodjamirian:2017fxg} and the  covariant confined quark model
(CCQM)\cite{Issadykov:2019vpm}.
For the heavy $B/B_s$ to light meson (such as $K, \pi, \eta^{\prime}, \rho, K^*, etc$) transitions, on the other hand,
the relevant form factors at the low $q^2$ region  have been evaluated successfully by employing the PQCD factorization
approach for example in Refs.~\cite{pqcd1,pqcd2,Kurimoto:2001zj,li2003,Wang:2012ab,Wang:2013ix,xiao18a,xiao18b}.

Since the PQCD predictions for the considered form factors are reliable only at the low $q^{\rm 2}$ region, we usually calculate explicitly
the values of the relevant form factors  at the low  $q^2$ region,  say  $ 0 \leq q^{\rm 2} \leq m_{\rm \tau}^{\rm 2}$,
and then make an extrapolation for all relevant form factors from the low $q^{\rm 2}$ region to the large $q^{\rm 2}$ region by using the
pole model parametrization\cite{cheng2004,wang2009} or other different methods.

In Refs.~\cite{Fan:2015,Hu:2019,Hu:2020}, we developed a new method:  the so-called  ``PQCD+Lattice"  approach.
Here we still use the PQCD approach to evaluate the form factors at the low $q^2$ region, but take those currently available lattice QCD results for the relevant
form factors at the high $q^2$ region as the lattice QCD input to improve  the extrapolation  of  the form factors  up to  $q^2_{max}$.
In Refs.~\cite{Hu:2019,Hu:2020}, we used the Bourrely-Caprini-Lellouch (BCL) parametrization method \cite{bcl09,jhep1905-094} instead of the traditional
pole model parametrization since the BCL method has better convergence.

In Table \ref{tab:table4} and \ref{tab:table5}, we list the values of the lattice QCD results for the relevant $B_s \to K^{*}$ transition form factors at three reference points
of $q^2$ ~\cite{Flynn:2015mha,Horgan:2013hoa} used in this paper.  The systematic uncertainties are included.

\begin{table}[htbp]
\caption{The values for the lattice QCD results of  the relevant $B_s \to K$ transition form factors at three reference  points of $q^2$:
$q^2=17.9,21.2$ GeV$^2$ and $q^2_{max}=(m_{B_s}-m_{K})^2\approx 23.8$GeV$^2$ ~\cite{Flynn:2015mha}. }
\label{tab:table4}
\centering
\setlength{\tabcolsep}{6pt} 
\renewcommand{\arraystretch}{1.5} 
\begin{tabular}{|c|c|c|c|} \hline\hline
\multirow{2}{*}{$FF$} & \multicolumn{3}{c|}{$q^2$ point} \\ \cline{2-4}
 & 17.9 & 21.2 & 23.8 \\\hline
$F_0(q^2)$  &$0.48(5)$  & $0.63(5)$ &  0.80(5)\\\hline
$F_+(q^2)$ &$0.98(7)$  & $1.64(10)$ & 2.76(16) \\\cline{2-4} \hline \hline
\end{tabular}
\end{table}

\begin{table}[htbp]
\caption{The values for the lattice QCD results of  the relevant $B_s \to K^{*}$ transition form factors at three reference points of $q^2$:
$q^2=12,16$ GeV$^2$ and $q^2_{max}=(m_{B_s}-m_{K^*})^2\approx 20$ GeV$^2$ ~\cite{Horgan:2013hoa}.  }
\label{tab:table5}
\centering
\setlength{\tabcolsep}{6pt} 
\renewcommand{\arraystretch}{1.5} 
\begin{tabular}{cccccccc} \hline\hline
 $q^2$  & $V(q^2)$ & $A_0(q^2)$ & $A_1(q^2)$& $A_{2}(q^2)$ & $T_1(q^2)$& $T_2(q^2)$&$ T_{3}(q^2)$ \\  \hline
12 &$0.56(9)$  & $0.84(9)$  &$0.37(3)$  &$0.46(3)$ &$0.61(4)$  &$0.39(3)$ & $0.43(4)$\\
16 &$1.02(8)$  & $1.33(8)$  &$0.45(3)$  &$0.60(5)$ &$0.90(6)$  &$0.47(3)$ & $0.67(5)$\\
20 &$1.99(13)$  & $2.38(16)$  &$0.58(3)$  &$0.85(12)$ &$1.48(10)$  &$0.60(3)$ & $1.10(7)$\\ \hline \hline
\end{tabular}
\end{table}

In this work, we will use both the PQCD factorization approach and the ``PQCD+Lattice" approach to evaluate all relevant form factors
over the whole  range of $q^2$.
\begin{enumerate}
\item[(1)]
In the PQCD approach,  we use the definitions and formulae as given in Eqs.(\ref{eq:fpq2}-\ref{eq:fTq2},\ref{eq:Vq2}-\ref{eq:T3q2} )
to calculate the values of  all relevant form factors $F_{0,+,T}(q^2)$, $V(q^2)$ , $A_{0,1,2}(q^2)$ and $T_{1,2,3}(q^2)$ in the low $q^2$ region:
$0 \leq q^2 \leq m_\tau^2$. We then make the extrapolation for these form factors  to the  large $q^2$ region up to $q_{max}^2 $ by using proper
parametrization method.

\item[(2)]
In the ``PQCD+Lattice " approach,  we  take  the lattice QCD results for the form factors at some large $q^2$ points as inputs
and then  make a combined fit to the PQCD and the lattice QCD results at the low and high $q^2$ region.

\item[(3)]
For both approaches,  we always use the same $z$-series parametrization  as in Refs.~\cite{Kindra:2018ayz,Straub:2015ica}   to make the extrapolation :
\beq
z(q^2)=\frac{\sqrt{t_{+}-q^2}-\sqrt{t_{+}-t_0}}{\sqrt{t_{+}-q^2}+\sqrt{t_{+}-t_0}}
\eeq
where, $t_{\pm}=(m_{B_s}\pm m_{K^{(*)}})^2$ and $t_0=t_{+}(1-\sqrt{1-t_{-}/t_{+}})$. Form factors are parameterized as:
\begin{equation} \label{eq:FFsF}
F_{i}(q^2)=P_i(q^2)\Sigma_k \alpha_k^{i}\left[z(q^2)-z(0)\right]^k.
\end{equation}
where $P_i(q^2)=(1-q^2/m_{R,i}^2)^{-1}$ is a simple pole corresponding to the $\bar{d}b$-resonance with appropriate $J^P$ in the spectrum
and $m_{R,i}$ is the resonance mass:  $m_{R} \to \infty$ for $F_0(q^2)$ (no pole),  $5.279$ GeV for $A_0(q^2)$ couple to $B(0^-)$, 5.325
GeV for $F_{+,T}(q^2)$, $V(q^2)$ and $T_1(q^2)$ couple to $B^*(1^-)$, and 5.724 GeV for rest of the form factors couple to $B_1(1^+)$.

\end{enumerate}

In Table \ref{tab:table6}, as a comparison, we show the centre  values of all relevant form factors in this work and
other theoretical predictions as given in Refs.~\cite{Wang:2012ab,Li:2009tx,Ali:2007ff,Khodjamirian:2017fxg,Faustov:2013ima,Ball:2004rg,Melikhov:2000yu,Lu:2007sg,Wu:2006rd,Su:2011eq}
at the scale $q^2=0$.
The PQCD factorization approach is applied in Refs.~\cite{Li:2009tx,Ali:2007ff} and in Ref.~\cite{Wang:2012ab} with the inclusion
of the NLO corrections. Covariant confined quark model (CCQM) is used in Ref.~\cite{Issadykov:2019vpm}.
Calculations based on Light-cone sum rules (LCSR) in Ref.~\cite{Khodjamirian:2017fxg,Straub:2015ica} with hadronic input parameters
and in Ref.~\cite{Ball:2004rg} with the inclusion of the one-loop radiative corrections.
In Ref.~\cite{Faustov:2013ima}, the authors used the relativistic quark model based on the quasi-potential approach.
In Ref.~\cite{Melikhov:2000yu}, the authors used the quark model and relativistic dispersion approach.
In Ref.~\cite{Lu:2007sg}, the light-cone quark model (LCQM) is utilized based on the basis of the soft collinear effective theory.
The authors of Ref.~\cite{Wu:2006rd} employed  the LCSR in the framework of the heavy quark effective theory.
In Ref.~\cite{Su:2011eq}, the authors evaluated the transition form factors in the six-quark effective Hamiltonian approach.
One can see that there is no significant difference between the theoretical predictions for the
$B_s \to K^{(*)}$ transition form factors evaluated at $q^2=0$ in various models or approaches.

\begin{table}[htbp]
\caption{The theoretical predictions for the centre values of the form factors of the $B_s\to  K^{(*)}$ transitions at $q^2=0$ obtained
by using rather different theories or models.  }
\label{tab:table6}
\centering
\setlength{\tabcolsep}{6pt} 
\renewcommand{\arraystretch}{1} 
\begin{ruledtabular}
\begin{tabular}{c|cc|cccccc}
     & $F_{0,+}(0)$& $F_T(0)$ & $V(0)$ & $A_0(0)$ &$A_1(0)$&$A_2(0)$ &$T_{1,2}(0)$&$T_3(0)$ \\
\hline
This paper &$0.22$  & $0.22$  &$0.24$  &$0.21$ &$0.19$  &$0.19$ & $0.21$& $0.16$\\
PQCD\cite{Wang:2012ab} &0.26&0.28& $-$&$-$&$-$&$-$&$-$&$-$\\
PQCD\cite{Li:2009tx} &$-$&$-$&0.20&0.24& $0.15$& $0.11$& $0.18$& $0.16$ \\
PQCD\cite{Ali:2007ff} &$0.24$&$-$ &$0.21$ &$0.25$ &$0.16$&$-$ &$-$&$-$  \\ \hline
CCQM\cite{Issadykov:2019vpm}   &$-$&$-$&$0.24$  &   $0.18$ & $0.21$ &$0.21$& $0.21$& $0.14$\\
LCSR\cite{Khodjamirian:2017fxg} &0.336&0.320& $-$&$-$&$-$&$-$&$-$&$-$\\
LCSR\cite{Straub:2015ica}   &$-$&$-$&$0.296$  &   $0.314$ & $0.230$ &$-$& $0.239$&$-$\\
LCSR\cite{Ball:2004rg}   &$0.30$  &$-$&   $0.311$ & $0.360$ & $0.233$& $0.181$& $0.260$& $0.136$\\
RQM\cite{Faustov:2013ima}   &$0.284$  & $0.236$  &$0.291$  &$0.289$ &$0.287$  &$0.286$ & $0.238$& $0.122$\\
RDA\cite{Melikhov:2000yu} &0.31 &0.31 & 0.38 &0.37&0.29&0.26&0.32&0.23\\
SCET\cite{Lu:2007sg} &0.290 &0.317 &0.323 &0.279&0.232 &0.210&0.271&0.165\\
HQEFT\cite{Wu:2006rd} &0.296 &0.288 &0.285&0.222&0.227&0.183&0.251&0.169 \\
SQEH\cite{Su:2011eq} &0.260&$-$&0.227&0.280&0.178&$-$&$-$&$-$  \\
\end{tabular}
\end{ruledtabular}
\end{table}

In Table \ref{tab:table7}, we list the PQCD predictions for the form factors $F_{+,0,T}(q^2)$,$V(q^2)$,$A_{0,1,2}(q^2)$ and $T_{1,2,3}(q^2)$
with the corresponding pole and resonance masses, the fitting parametrization constants ($\alpha_0 , \alpha_1, \alpha_2)$  in
Eq.~(\ref{eq:FFsF}) for $B_{s} \to K^{(*)}$ transitions. It is simple to figure out the relation $F_i(0)=\alpha^i_0$ by substituting $q^2$
with zero on the both sides of Eq.~(\ref{eq:FFsF}).
The theoretical errors of the form factors as shown in Table \ref{tab:table7} are the two major errors from the uncertainties of the parameter
$\omega_{B_s}=0.50\pm 0.05$ GeV and the Gegenbauer moments in the distribution amplitudes $a^{K}_i (a^{K^*}_i)$ of the light pseudoscalar (vector)
mesons. The  additional theoretical uncertainties from other input parameters, such as  the decay constants $f_{B_s}, f_K, f_{K^*}, f_{K^*}^T$,
are very small and have been neglected.

\begin{table}[htbp]
\caption{The PQCD predictions for the form factors of $B_s \to K^{(*)} $ transitions.  Form factors $F_{+,0,T}(q^2)$,$V(q^2)$,$A_{0,1,2}(q^2)$ and
$T_{1,2,3}(q^2)$ are fitted by using Eq.~(\ref{eq:FFsF}).}
\label{tab:table7}
\centering
\setlength{\tabcolsep}{6pt} 
\renewcommand{\arraystretch}{1} 
\begin{tabular}{c|c|c|c|c} \hline\hline
PQCD\!   &$\!  B(J^P)$  &$\!  \,\,\alpha_0$ &$\!  \alpha_1$&$\!  \alpha_2$  \\ \hline
$F_+^{B_s\to K}$ \!  &$\!  B^*(1^-)$            &\! ${0.22}^{+0.04}_{-0.06}(\omega_{B_s})\pm {0.005}(a^K_i) $&\! ${-1.21}^{+0.36}_{-0.23}\pm 0.02$&  \! ${0.06}^{+1.74}_{-1.09}\pm 0.14$ \\
$F_0^{B_s\to K}$ \!  &$\!  {\rm no\, pole}$&\!  ${0.22}^{+0.04}_{-0.06}(\omega_{B_s})\pm {0.005}(a^K_i)$&\! ${-1.31}^{+0.40}_{-0.24}\pm 0.03$& \!  ${-0.22}^{+1.73}_{-1.11} \pm  0.13$  \\
$F_T^{B_s\to K}$ \!  &$\  B^*(1^-)$&\!      ${0.22}^{+0.04}_{-0.06}(\omega_{B_s})\pm {0.005}(a^K_i)$&\ ${-1.37}^{+0.40}_{-0.29}\pm 0.03           $& \!  $ {7.05}^{+1.85}_{-1.08} \pm 0.15$  \\
\hline
$V^{B_s\to K^*}$ \!  &$\  B^*(1^-)$&\!      ${0.24}^{+0.05}_{-0.04}(\omega_{B_s})\pm {0.005}(a^{K^*}_i) $&\!  ${-1.87}^{+0.31}_{-0.42}\pm 0.03$& \! $ {7.56}^{+1.22}_{-1.41}\pm 0.13$  \\
$A_0^{B_s\to K^*}$ \ &$\!  B^0(0^-)$&\!  ${0.21}^{+0.04}_{-0.03}(\omega_{B_s})\pm {0.003}(a^{K^*}_i)$ &\!  ${-1.43}^{+0.25}_{-0.29}\pm 0.02$&\!  $ {8.28}^{+1.38}_{-1.03}\pm 0.11$  \\
$A_1^{B_s\to K^*}$ \ &$\!  B_1(1^+)$&\!  ${0.19}^{+0.04}_{-0.03}(\omega_{B_s})\pm {0.003}(a^{K^*}_i)$ &\!  ${-0.64}^{+0.13}_{-0.14}\pm 0.01$&\!  $ {0.35}^{+1.04}_{-1.02}\pm 0.10$  \\
$A_2^{B_s\to K^*}$ \ &$\!  B_1(1^+)$&\!  ${0.19}^{+0.04}_{-0.03}(\omega_{B_s})\pm {0.003}(a^{K^*}_i)$ &\! ${-1.42}^{+0.28}_{-0.31} \pm 0.03$&\!  $ {2.77}^{+1.68}_{-1.07}\pm 0.12$  \\
$T_1^{B_s\to K^*}$ \ &$\!  B^*(1^-)$&\!   ${0.21}^{+0.04}_{-0.03}(\omega_{B_s})\pm {0.003}(a^{K^*}_i)$ &\!  ${-1.52}^{+0.24}_{-0.34}\pm 0.03$&\! $ {9.10}^{+1.35}_{-1.46}\pm 0.13$  \\
$T_2^{B_s\to K^*}$ \ &$\!  B_1(1^+)$&\!  ${0.21}^{+0.04}_{-0.03}(\omega_{B_s})\pm {0.003}(a^{K^*}_i)$ &\!  ${-0.44}^{+0.13}_{-0.08}\pm 0.01$&\! ${4.38}^{+1.33}_{-1.93} \pm 0.16$  \\
$T_3^{B_s\to K^*}$ \ &$\!  B_1(1^+)$&\!  ${0.16}^{+0.03}_{-0.02}(\omega_{B_s})\pm {0.002}(a^{K^*}_i)$ &\!  ${-0.99}^{+0.19}_{-0.27}\pm 0.02$&\!  ${8.12}^{+1.04}_{-1.81} \pm 0.11$  \\
\hline \hline
\end{tabular}
\end{table}

\begin{table}[htbp]
\caption{The ``PQCD+Lattice" predictions for the form factors of $B_s \to K^{(*)}$ transitions. All form factors  are fitted by using Eq.~(\ref{eq:FFsF}).}
\label{tab:table8}
\centering
\setlength{\tabcolsep}{6pt} 
\renewcommand{\arraystretch}{1} 
\begin{tabular}{c|c|c|c|c} \hline\hline
PQCD+Lattice   &$\!  B(J^P)$  &$\!  \,\,\alpha_0$ &$\!  \alpha_1$&$\!  \alpha_2$  \\ \hline
$F_+^{B_s\to K}$ \!  &$\!  B^*(1^-)$            &\! ${0.22}^{+0.04}_{-0.06}(\omega_{B_s})\pm {0.005}(a^K_i) $&\! ${-0.89}^{+0.23}_{-0.36}\pm 0.02$&  \! ${-0.44}^{+0.37}_{-0.44}\pm 0.03$ \\
$F_0^{B_s\to K}$ \!  &$\!  {\rm no\, pole}$&\!  ${0.22}^{+0.04}_{-0.06}(\omega_{B_s})\pm {0.005}(a^K_i)$&\! ${-0.50}^{+0.08}_{-0.15}\pm 0.01$& \!  ${5.07}^{+0.27}_{-0.31} \pm  0.02$  \\
\hline
$V^{B_s\to K^*}$ \!  &$\  B^*(1^-)$&\!  ${0.24}^{+0.05}_{-0.03}(\omega_{B_s})\pm {0.005}(a^{K^*}_i) $&\!  ${-0.13}^{+0.11}_{-0.07}\pm 0.01$&  \! $ {10.14}^{+0.47}_{-0.49}\pm 0.04$  \\
$A_0^{B_s\to K^*}$ \ &$\!  B^0(0^-)$&\!  ${0.21}^{+0.03}_{-0.03}(\omega_{B_s})\pm {0.003}(a^{K^*}_i)$ &\!  ${-3.38}^{+0.17}_{-0.15}\pm 0.02$&\!  $ {-4.13}^{+0.43}_{-0.49}\pm 0.04$  \\
$A_1^{B_s\to K^*}$ \ &$\!  B_1(1^+)$&\!  ${0.19}^{+0.04}_{-0.03}(\omega_{B_s})\pm {0.004}(a^{K^*}_i)$ &\!  ${-0.65}^{+0.27}_{-0.15}\pm 0.02$&\!  $ {-2.66}^{+0.43}_{-0.32} \pm 0.03$  \\
$A_2^{B_s\to K^*}$ \ &$\!  B_1(1^+)$&\!  ${0.19}^{+0.04}_{-0.03}(\omega_{B_s})\pm {0.004}(a^{K^*}_i)$ &\! ${-1.35}^{+0.58}_{-0.55} \pm 0.04$&\!  $ {-3.12}^{+1.96}_{-1.43} \pm 0.15$  \\
$T_1^{B_s\to K^*}$ \ &$\!  B^*(1^-)$&\!  $ {0.21}^{+0.04}_{-0.03}(\omega_{B_s})\pm {0.003}(a^{K^*}_i)$ &\!  ${-1.85}^{+0.30}_{-0.19}\pm 0.02$&\!  $ {-3.23}^{+1.26}_{-1.32}\pm 0.12$  \\
$T_2^{B_s\to K^*}$ \ &$\!  B_1(1^+)$&\!  ${0.21}^{+0.04}_{-0.03}(\omega_{B_s})\pm {0.003}(a^{K^*}_i)$ &\!  ${-0.65}^{+0.29}_{-0.16}\pm 0.02$&\! ${-2.85}^{+0.73}_{-0.55} \pm 0.05$  \\
$T_3^{B_s\to K^*}$ \ &$\!  B_1(1^+)$&\!  $ {0.16}^{+0.03}_{-0.02}(\omega_{B_s})\pm {0.002}(a^{K^*}_i)$ &\!  ${-1.00}^{+0.19}_{-0.15}\pm 0.02$&\!  ${2.82}^{+1.26}_{-1.08} \pm 0.09$  \\
\hline
\hline
\end{tabular}
\end{table}

\begin{figure}[htbp]
\begin{center}
\centerline{\epsfxsize=5cm\epsffile{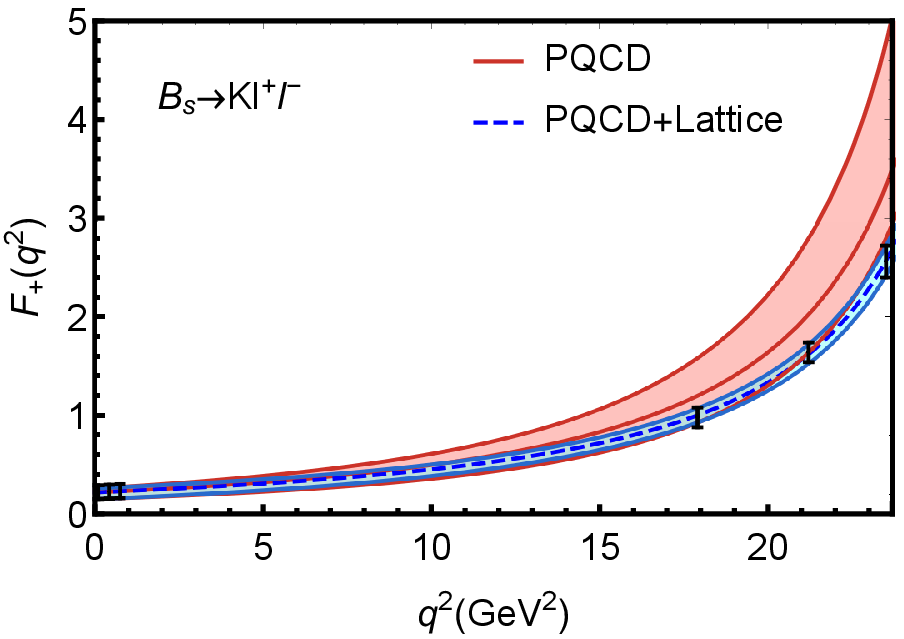}
\epsfxsize=5.2cm\epsffile{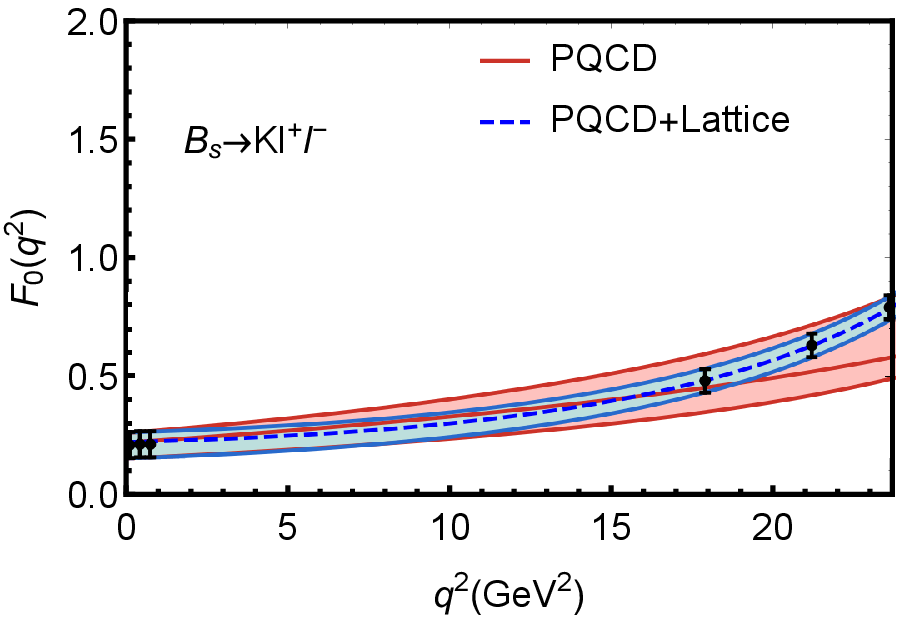}
\epsfxsize=5cm\epsffile{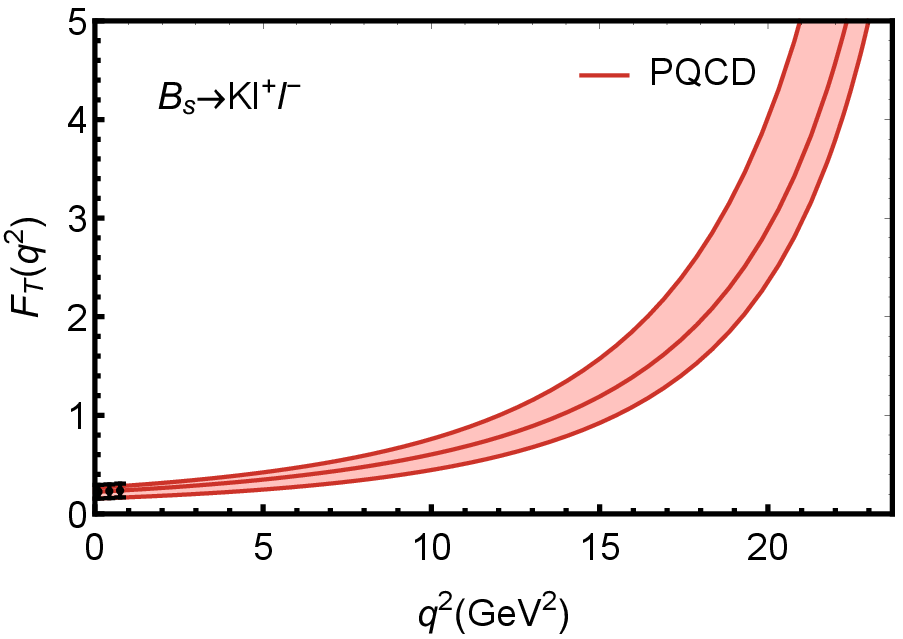}}
\end{center}
\vspace{0cm}
\caption{ The $q^2$-dependence of the form factors $F_{+,0,T}(q^2)$ in the PQCD (red) and ``PQCD+Lattice" (blue) approaches for $B_s \to K$ transition,  while the red (blue)
shaded band shows the major theoretical uncertainty.}
\label{fig:fig3}
\end{figure}

\begin{figure}[htbp]
\begin{center}
\centerline{\epsfxsize=6cm\epsffile{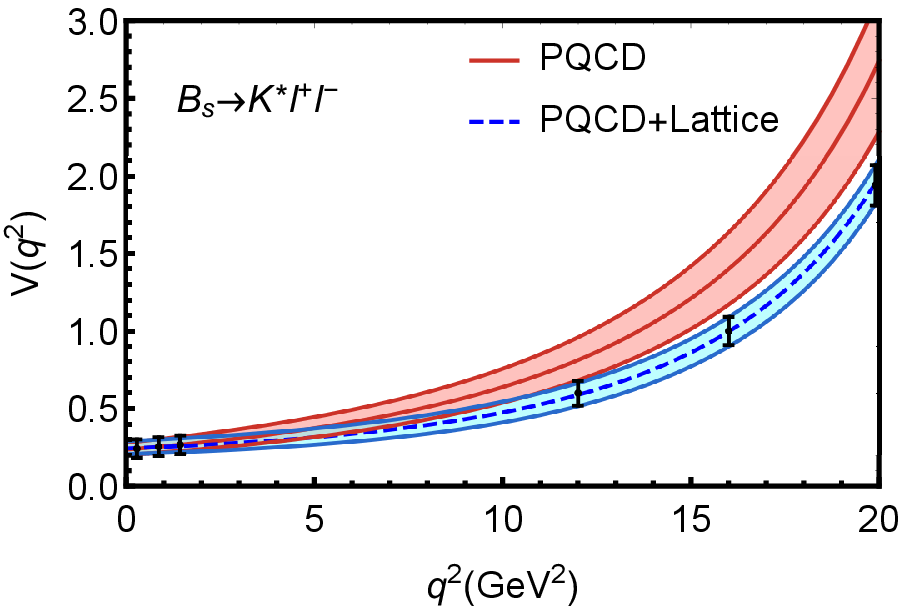}\epsfxsize=6cm\epsffile{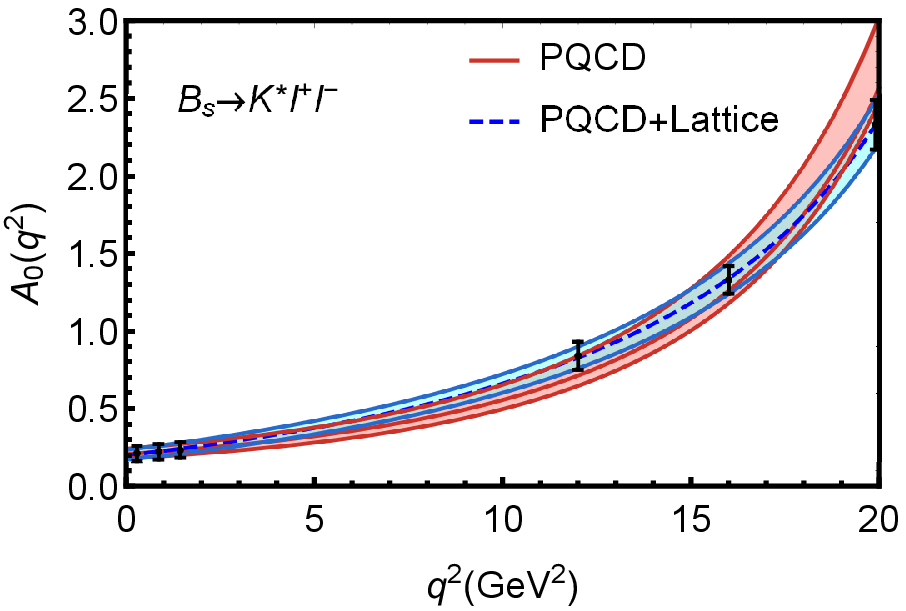} }
\vspace{0.3cm}
\centerline{\epsfxsize=6cm\epsffile{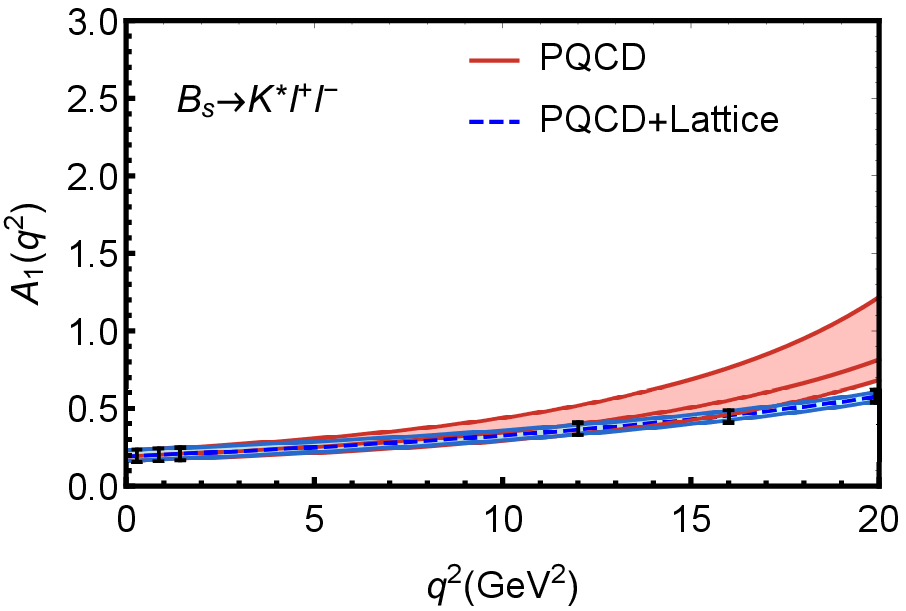}\epsfxsize=6cm\epsffile{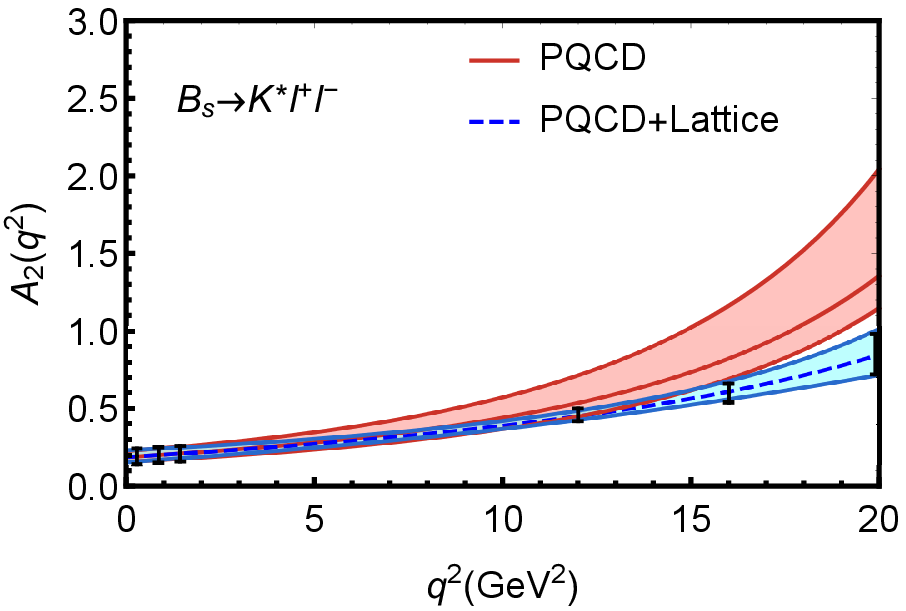}}
\vspace{0.3cm}
\centerline{\epsfxsize=6cm\epsffile{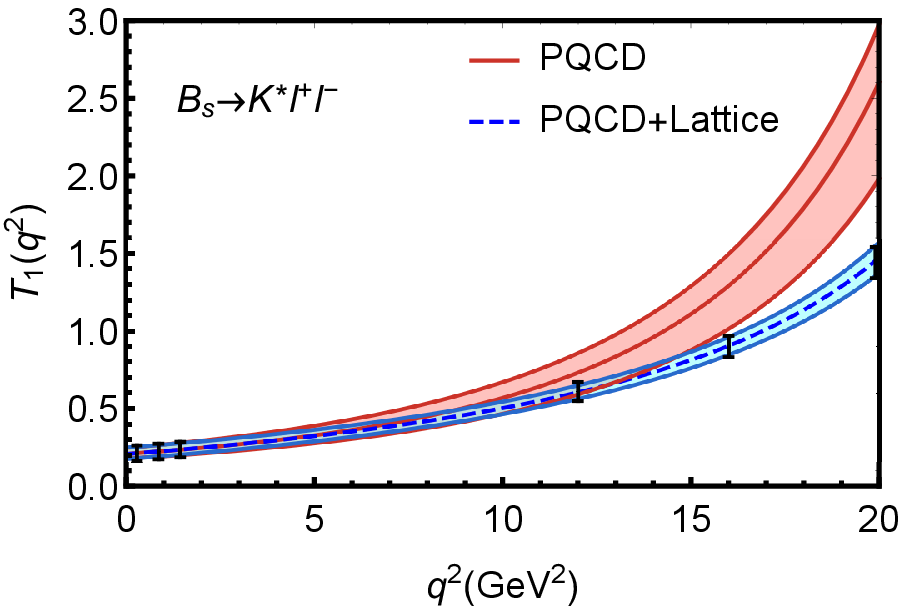}\epsfxsize=6cm\epsffile{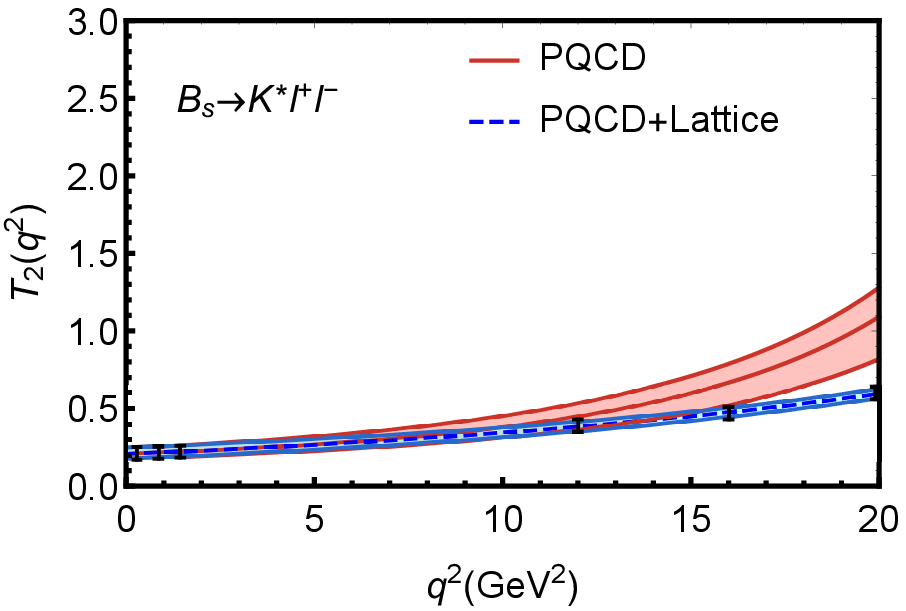}}
\vspace{0.3cm}
\centerline{ \epsfxsize=6cm\epsffile{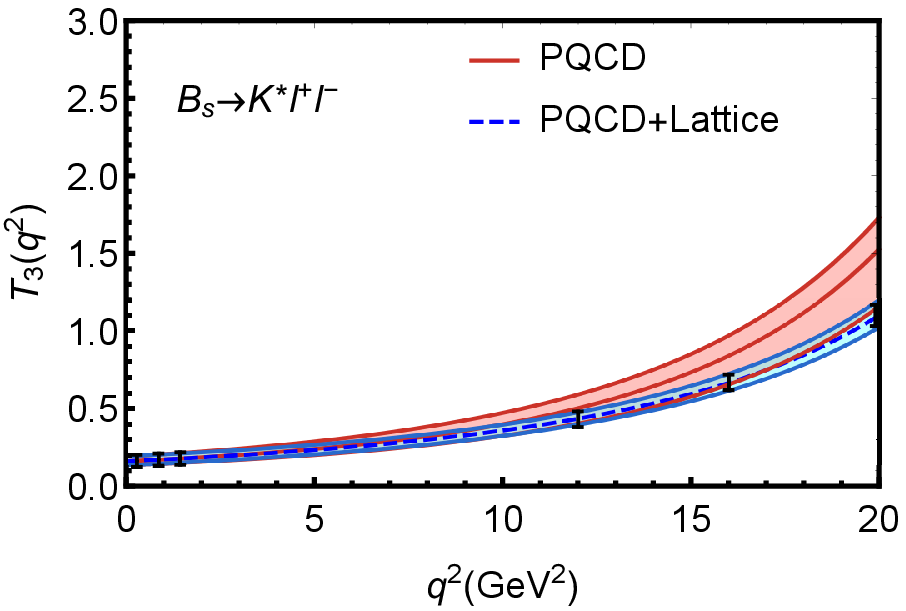} \hspace{6cm} }
\end{center}
\vspace{-0.6cm}
\caption{ The $q^2$-dependence of the form factors $V(q^2)$,$A_{0,1,2}(q^2)$ and
$T_{1,2,3}(q^2)$ in the PQCD (red) and ``PQCD+Lattice" (blue) approaches for $B_s \to K^\ast$ transition, while the red (blue)
shaded band shows the major theoretical uncertainty.  }
\label{fig:fig4}
\end{figure}

In Table \ref{tab:table8}, we list the ``PQCD+Lattice" predictions for the form factors $F_{+,0}(q^2)$, $V(q^2)$, $A_{0,1,2}(q^2)$
and $T_{1,2,3}(q^2)$ by taking into account the lattice QCD results for the form factors at some points of $q^2$ as listed in
Table \ref{tab:table4} and \ref{tab:table5}  from Refs.~\cite{Flynn:2015mha,Horgan:2013hoa}, in a similar way as what we did
in Refs.\cite{Fan:2015,Hu:2019,Hu:2020}. The errors are obtained in the same way as those in Table   \ref{tab:table7}.
The additional form factors $A_{12}(q^2)$ and $T_{23}(q^2)$ can be defined  as the linear combinations of
$A_{1}(q^2)$ and $A_{2}(q^2)$,  $T_{2}(q^2)$ and $T_{3}(q^2)$, together with kinematic variable $\lambda$ as given in Eqs.~(10,11) from
Ref.~\cite{Horgan:2013hoa}.
In Figs.~\ref{fig:fig3} and \ref{fig:fig4}, we show the $q^2$-dependence of the form factors $F_{+,0,T}(q^2)$, $V(q^2)$, $A_{0,1,2}(q^2)$
and $T_{1,2,3}(q^2)$ in the PQCD (the red curves) and  ``PQCD+Lattice¡±  (the blue curves) approaches for $B_s \to K^{(\ast)}$ transitions.
The error bars of the initial PQCD and relevant lattice QCD results  as listed in Table \ref{tab:table4} and \ref{tab:table5} are blackened,
in order to show them clearly.

\subsection{Observables for \texorpdfstring{$B_s \to K \ell^+\ell^-$}{} }

From the differential decay rates as given in Eq.~(\ref{eq:DDR1}),  it is conventional to make the integration over the range of $4m_\ell^2
\!\leq\! q^2\!\leq\!(m_{B_s}-m_{K})^2$.  In order to be consistent with the choices made by experiment collaborations in their data analysis,
however, we have to cut off  the regions of dilepton-mass squared around the charmonium resonances $J/\psi(1S)$ and $\psi(2S)$:
i.e., $8.0\! < \!q^2\! <\! 11.0\, GeV^2$ and $12.5\! <\! q^2\! <\! 15.0 \,GeV^2$  for $\ell=(e,\mu,\tau)$ cases.
The PQCD and  `` PQCD+Lattice" predictions for the branching ratios (Br)  and the longitudinal polarization asymmetry $P_L$ of
the semileptonic decays $\bar{B}_s \to K  \ell^+ \ell^-$ and $B_s \to  \bar{K}  \ell^+ \ell^-$ at three different renormalization scales
$\mu=(0.5m_b,m_b,1.5m_b)$ are listed in Table~\ref{tab:table9} and  \ref{tab:table10} respectively,
where the total theoretical errors  are the combinations of the uncertainties of all relevant input parameters:  $\omega_{B_s}$,  $a_i^K$ and $V_{ij}$.
The direct CP-asymmetries $\cala_{CP}$ are obtained by making integration over  $q^2$ for $\cala_{CP}(q^2)$ as defined in Eqs.~(\ref{eq:acp01},\ref{eq:acp02})
and are also listed in these two tables.
To reduce the large theoretical uncertainties, we also check the physical
observables $R^{e\mu }_{K}$ and $R^{\mu \tau}_{K}$,  as defined in Eqs.~(\ref{eq:rkbs1},\ref{eq:rkbs2}), i.e.,
the ratio of the branching ratios of $B_s\to K\ell^+\ell^-$ decays.
As a comparison,  the previous PQCD  predictions as given in Ref.~\cite{Wang:2012ab} for the decay rates and the ratios
$R^{e\mu }_{K}$ and $R^{\mu \tau}_{K}$  are listed in last  column of Table  \ref{tab:table9} and  \ref{tab:table10}.

In order to show the major theoretical uncertainties from different sources explicitly,
 for instance, we show the PQCD predictions for $\calb(\bar{B}_s \to K \mu^+\mu^-)$  with  the four kinds of errors:
\beq
\calb(\bar{B}_s \to K \mu^+\mu^-)=( 1.24^{+0.69}_{-0.58}(\omega_{B_s}) \pm 0.15(a_i^K) \pm 0.06 (V_{ij})^{+0.01}_{-0.07}(\mu) ) \times 10^{-8},
\label{eq:brexa-01}
\eeq
where  the dominant theoretical error comes from $\omega_{B_s}=0.50\pm 0.05$,  the second one from the Gegenbauer moments
$a_1^K=0.06\pm 0.03$ and $a_2^K=0.25 \pm 0.15$ as given in Eq.~(\ref{eq:gb01}), the third one from the CKM elements $V_{ij}$ in Eq.~(\ref{eq:vij1})
and the last error from the renormalization scale $\mu= (1\pm 0.5) m_b$.  The possible  errors from other input parameters are very small and have been
neglected.

\begin{table}[htbp]
\caption{ The PQCD predictions  for the branching ratios (in unit of $10^{-8}$),  the longitudinal polarization asymmetry $P_L$ and
the ratios $(R_K^{e\mu },R_K^{\mu\tau })$  of the decays $\bar{B_s} \to K\ell^+ \ell^-$ (the first row)
and $B_s \to \bar{K}\ell^+ \ell^-$  (the second row)  at three different renormalization scales.
The direct CP asymmetries $\cala_{CP}$  (in unit of  $10^{-2}$ ) are also listed.  }
\label{tab:table9}
\centering
\setlength{\tabcolsep}{4pt} 
\renewcommand{\arraystretch}{1} 
\begin{tabular}{|c|c|c|c|c|c|}  \hline \hline
Mode&Obs.  & $\mu=0.5m_b$& $\mu=m_b$ & $\mu=1.5m_b$& PQCD\cite{Wang:2012ab} \\ \hline
$\ell=e$ & ${\calb }(\bar{B_s} \to K\ell^+ \ell^-) $& ${1.17}^{+0.68}_{-0.56}$ &  ${1.24}^{+0.70}_{-0.60}$ & ${1.26}^{+0.71}_{-0.61}$ & $ 1.63^{+0.73}_{-0.58}$  \\
\cline{2-6} & ${ \calb}( B_s \to \bar{K}\ell^+ \ell^-) $  & ${1.21}^{+0.68}_{-0.59}$  & ${1.32}^{+0.73}_{-0.64}$ & ${1.63}^{+0.74}_{-0.66}$ & $-$ \\
\cline{2-6} &$P_L (\bar{B_s} \to K\ell^+ \ell^-) $  & $-0.986\pm 0.002$   & $-0.979\pm 0.002$ & $-0.972\pm 0.002$ & $-$ \\
\cline{2-6} & $P_L (B_s \to \bar{K}\ell^+ \ell^-) $& $-0.980\pm 0.002$  & $-0.954 \pm 0.008$ & $-0.937 \pm 0.004$ & $-$ \\
\cline{2-6} & $\cala_{CP}$&  $-1.7\pm 0.9 $  & $-3.1\pm 0.4$ & $-3.8\pm 0.6$ & $-$ \\ \hline
$\ell=\mu$ &${ \calb}$ & ${1.17}^{+0.68}_{-0.56}$ & ${1.24}^{+0.71}_{-0.60}$ &${1.25}^{+0.71}_{-0.61}$  & $ 1.63^{+0.73}_{-0.58}$ \\
\cline{3-6} & & ${1.21}^{+0.68}_{-0.69}$  & ${1.32}^{+0.73}_{-0.64}$ & ${1.35}^{+0.74}_{-0.66}$&$-$ \\
\cline{2-6} &$P_L$& $-0.974\pm 0.002$  & $-0.966 \pm 0.002$ & $-0.959 \pm 0.002 $ & $-$\\
\cline{3-6} &   & $-0.964 \pm 0.003$  & $-0.938 \pm 0.005$ & $-0.920 \pm 0.004$ & $-$ \\
\cline{2-6} & $\cala_{CP}$&  $-1.7\pm 0.9 $  & $-3.1\pm 0.4$ & $-3.8\pm 0.6$ & $-$ \\ \hline
$\ell=\tau$ &${\calb}$ & ${0.380}^{+0.202}_{-0.157}$ &  ${0.382}^{+0.204}_{-0.158}$ & ${0.377}^{+0.203}_{-0.156}$&$0.43^{+0.18}_{-0.15}$  \\
\cline{3-6} & & ${0.382}^{+0.202}_{-0.158}$  & ${0.386}^{+0.206}_{-0.159}$ & ${0.382}^{+0.204}_{-0.158}$ & $-$ \\
\cline{2-6} &$P_L$ & $-0.347 \pm 0.002$  & $-0.348 \pm 0.002$ & $-0.346 \pm 0.002 $ &$-$ \\
\cline{3-6} &   & $-0.348 \pm 0.002 $  & $-0.349 \pm 0.002 $ & $-0.348 \pm 0.001$ & $-$\\
\cline{2-6} & $\cala_{CP}$&  $-0.3\pm 0.2 $  & $-0.5\pm 0.2$ & $-0.7\pm 0.2$ & $-$ \\ \hline
 &$R_K^{e\mu }$ & ${0.996 \pm 0.002}$&${0.996 \pm 0.002}$  &${0.996 \pm 0.002}$ & $1$ \\
\cline{3-6} && ${0.996 \pm 0.002 }$&${0.996 \pm 0.002 }$  &${0.996 \pm 0.002 }$ & $-$ \\
\cline{2-6}&$R_K^{ \mu \tau}$ & ${0.332\pm 0.045}$&${0.325\pm 0.045}$  &${0.322\pm 0.045}$ & $0.26$ \\
\cline{3-6}& &                 ${0.320 \pm 0.046}$&${0.303\pm 0.046}$  &${0.296 \pm 0.046}$ & $-$\\  \hline \hline
\end{tabular}
\end{table}

\begin{table}[htbp]
\caption{  The ``PQCD+Lattice"   predictions  for the branching ratios (in unit of $10^{-8}$),  the longitudinal polarization asymmetry $P_L$ 
and  the direct CP asymmetry $\cala_{CP}$ (in unit of  $10^{-2}$ ), as well as the ratios $(R_K^{e\mu }, R_K^{\mu\tau })$  of the 
decays $\bar{B_s} \to K\ell^+ \ell^-$ (the first row) and $B_s \to \bar{K}\ell^+ \ell^-$ (the second row)  at three different renormalization scales. }
\label{tab:table10}
\centering
\setlength{\tabcolsep}{4pt} 
\renewcommand{\arraystretch}{1} 
\begin{tabular}{|c|c|c|c|c|c|} \hline  \hline
Mode&Obs.  & $\mu=0.5m_b$& $\mu=m_b$ & $\mu=1.5m_b$ &   PQCD\cite{Wang:2012ab} \\ \hline
$\ell=e$ &  ${\calb }(\bar{B_s} \to K\ell^+ \ell^-) $ & ${0.95}^{+0.26}_{-0.31}$ &  ${1.01}^{+0.29}_{-0.33}$ & ${1.03}^{+0.28}_{-0.35}$  & $ 1.63^{+0.73}_{-0.58}$    \\
\cline{3-6} &   ${ \calb }(B_s \to \bar{K}\ell^+ \ell^-) $& ${0.99}^{+0.27}_{-0.33}$  & ${1.10}^{+0.31}_{-0.38}$ & ${1.14}^{+0.32}_{-0.40}$ & $-$ \\
\cline{2-6} &$P_L (\bar{B_s} \to K\ell^+ \ell^-) $  & $-0.983\pm 0.007$  & $-0.975 \pm 0..005$ & $-0.967 \pm 0.006$  & $-$  \\
\cline{3-6} &  $P_L (B_s \to \bar{K}\ell^+ \ell^-) $&       $-0.977 \pm 0.008$  & $-0.949 \pm 0.013$ & $-0.930 \pm 0.010$ & $-$  \\
\cline{2-6} & $\cala_{CP}$&  $-2.0\pm 0.5 $  & $-4.3\pm 1.4$ & $-5.0\pm 0.8$ & $-$ \\ \hline
$\ell=\mu$ &${ \calb }$ & ${0.95}^{+0.25}_{-0.31}$ & ${1.01}^{+0.27}_{-0.33}$ &${1.03}^{+0.28}_{-0.34}$   & $ 1.63^{+0.73}_{-0.58}$  \\
\cline{3-6} & & ${0.99}^{+0.27}_{-0.33}$  & ${1.10}^{+0.31}_{-0.38}$ & ${1.13}^{+0.32}_{-0.40}$   & $-$ \\
\cline{2-6} &$P_L$& $-0.968 \pm 0.008$  & $-0.960 \pm 0.007$ & $-0.953 \pm 0.008$   & $-$ \\
\cline{3-6} &   & $-0.958\pm 0.009$  & $-0.929 \pm 0.012$ & $-0.911 \pm 0.014$  & $-$ \\
\cline{2-6} & $\cala_{CP}$&  $-2.0\pm 0.5 $  & $-4.3\pm 1.4$ & $-5.1\pm 0.8$ & $-$ \\ \hline
$\ell=\tau$ &{\calb } & ${0.365}^{+0.072}_{-0.075}$ &  ${0.368}^{+0.072}_{-0.076}$ & ${0.365}^{+0.072}_{-0.075}$  & $ 0.43^{+0.18}_{-0.15}$  \\
\cline{3-6} & & ${0.366}^{+0.072}_{-0.075}$  & ${0.370}^{+0.073}_{-0.076}$ & ${0.368}^{+0.073}_{-0.076}$  & $-$\\
\cline{2-6} &$P_L$ & $-0.234 \pm 0.020$  & $-0.235 \pm 0.018$ & $-0.233 \pm 0.019$ & $-$   \\
\cline{3-6} &   & $-0.234 \pm 0.020$  & $-0.237 \pm 0.019$ & $-0.236 \pm 0.019$ & $-$  \\
 \cline{2-6} & $\cala_{CP}$&  $-0.1\pm 0.1 $  & $-0.3\pm 0.1$ & $-0.4\pm 0.1$ & $-$ \\ \hline
 &$R_K^{e\mu }$ & ${0.996 \pm 0.002}$&${0.996 \pm 0.002}$  &${0.996 \pm 0.002}$  & $1$  \\
\cline{3-6} && ${0.996 \pm 0.002 }$&${0.996 \pm 0.002 }$  &${0.996 \pm 0.002 }$  & $-$ \\
\cline{2-6}&$R_K^{ \mu \tau}$ & ${0.395\pm 0.081}$&${0.384\pm 0.080}$  &${0.381\pm 0.080}$& $0.26$   \\
\cline{3-6}& &                 ${0.375\pm 0.084}$&${0.350\pm 0.085}$  &${0.341\pm 0.085}$   & $-$ \\  \hline \hline
\end{tabular}
\end{table}

From   Table~\ref{tab:table9} and  \ref{tab:table10} ,  it is easy to find the CP-averaged  decay rates and the direct CP-asymmetries  $\cala_{CP}$
for the considered semileptonic  decays:
\beq
\calb(\bar{B}_s \to K l^+ l^- ) |_{\rm CP-av.}&=&\left \{  \begin{array}{ll}
(1.28^{+0.52}_{-0.48})\times 10^{-8}, &  {\rm PQCD},  \\
(1.06^{+0.22}_{-0.29})\times 10^{-8}, &  {\rm PQCD+Lattice}, \\  \end{array} \right. \label{eq:brav1}, \\
\cala_{CP}  (\bar{B}_s \to K l^+ l^- ) &=&\left \{  \begin{array}{ll}-(3.1^{+0.8}_{-1.5})\times 10^{-2}, &  {\rm PQCD},  \\
-(4.3^{+1.6}_{-2.7})\times 10^{-2}, &  {\rm PQCD+Lattice}, \\  \end{array} \right. \label{eq:acp-n1}
\eeq
for the case of $l=(e,\mu)$, and
\beq
\calb(\bar{B}_s \to K \tau^+ \tau^- ) |_{\rm CP-av.}&=&\left \{  \begin{array}{ll}
(0.38^{+0.14}_{-0.12})\times 10^{-8}, &  {\rm PQCD},  \\
(0.37^{+0.05}_{-0.06})\times 10^{-8}, &  {\rm PQCD+Lattice}, \\  \end{array} \right. \label{eq:brav2}\\
\cala_{CP} (\bar{B}_s \to K \tau^+ \tau^- ) &=&\left \{  \begin{array}{ll} -(0.5\pm 0.3)\times 10^{-2}, &  {\rm PQCD},  \\
-(0.3\pm 0.2)\times 10^{-2}, &  {\rm PQCD+Lattice}, \\  \end{array} \right. \label{eq:acp-n2}
\eeq
for the case of $\tau$ lepton.

\begin{figure}[tbp]
\begin{center}
\centerline{\epsfxsize=7cm\epsffile{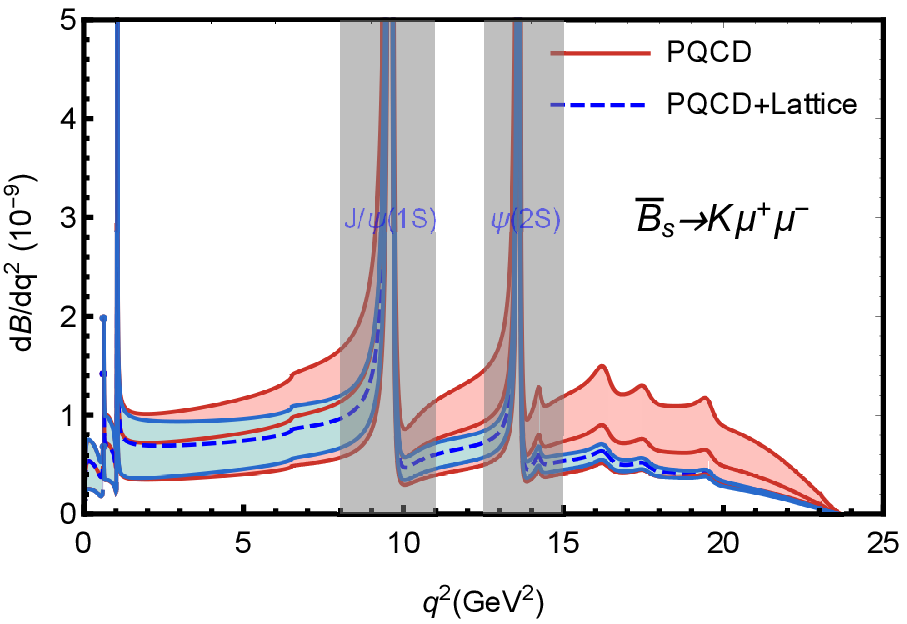}\epsfxsize=7cm\epsffile{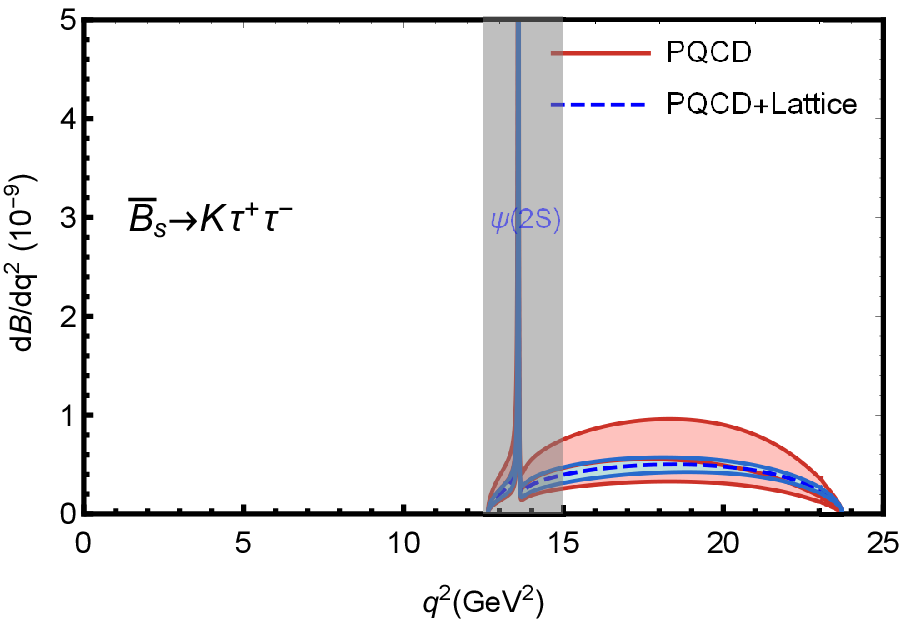} }
\vspace{0.3cm}
\centerline{\epsfxsize=7.2cm\epsffile{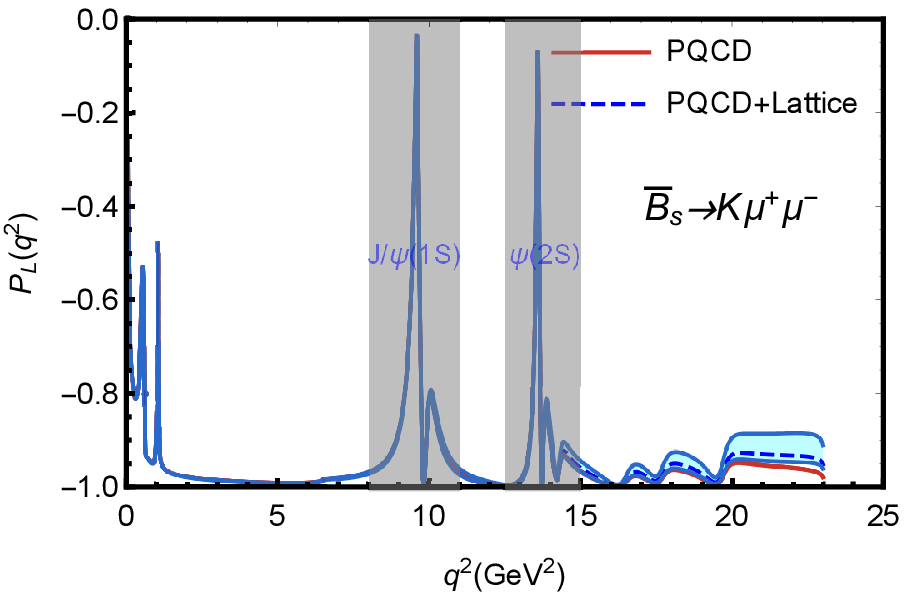}\epsfxsize=7.2cm\epsffile{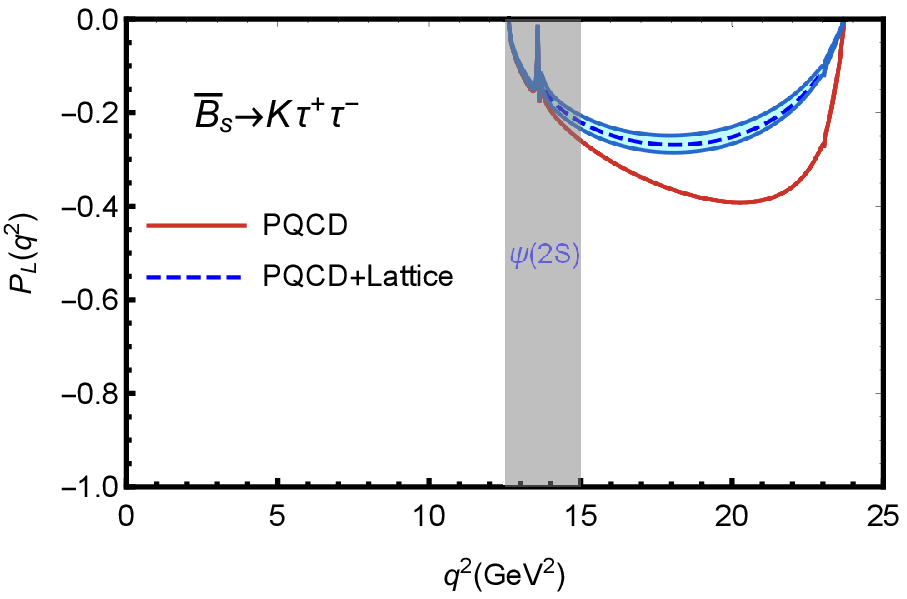} }
\end{center}
\vspace{-0.5cm}
 \caption{The PQCD or ``PQCD+Lattice"  predictions for the $q^2$-dependence of $d\calb/dq^2$ and ${P_L}(q^2)$
for the decays $\bar{B_s} \to K \ell^+ \ell^- $ with $l=\mu,\tau$.  For details see the text. \label{fig:fig5}}
\end{figure}

In Fig.~\ref{fig:fig5}, we show the $q^2$-dependence of the theoretical predictions of  the differential
branching fraction $d\calb/dq^2$  and the longitudinal lepton polarization $P_L(q^2)$ for the decays $\bar{B}_s \to K \ell^+ \ell^- $ with $l=(\mu,\tau)$,
evaluated by using the PQCD  (the red solid curves ) and  the  `` PQCD+Lattice"  (the blue dashed curves) approach, with the choice of the
scale $\mu=m_b$ and $q^2_{max}$ = 23.71 GeV$^2$.
The shaded bands indicate the theoretical error of our predictions due to the uncertainties of the input parameters.
The two vertical  grey blocks are the experimental veto regions \cite{Wei:2009zv}  in order to remove contributions from
$\bar{B}_s \to J/\psi(1S)(\to \ell^+ \ell^-)K $  ( the left-hand band) and
$\bar{B}_s \to \psi^\prime(2S)(\to \ell^+ \ell^-)K $ (the right-hand band) for the $q^2$-dependence of $d\calb/dq^2$ and ${P_L}$.
The figure for the electron mode is almost identical with the one for muon, and therefore not be shown here.

From the numerical results as listed in Table~\ref{tab:table9} and  \ref{tab:table10} ,  one can see the following points:
\begin{enumerate}
\item[(1)]
The theoretical predictions from both PQCD and `` PQCD+Lattice" approaches have a relatively weak dependence on the choice  of  the renormalization scale $\mu$.
The variations of the central values due to the $\mu$-dependence are about $10\%$ in magnitude and smaller than the combined errors
from the uncertainties of other input parameters.

\item[(2)]
Due to the term proportional to $\lambda_u$ in the effective Hamiltonian ${\cal H}_{eff}$ in Eq.~(\ref{eq:heff}), the PQCD and ``PQCD+Lattice" predictions for
the considered observables  for  $\bar{B}_s\to K l^+l^-$ and its CP conjugated mode have a relatively small differences, roughly $(5-15)\%$ in magnitude.
For the CP-averaged branching ratios,   the PQCD  and   ``PQCD+Lattice" predictions agree well within  one-standard deviation.
As generally expected, the direct CP asymmetries $\cala_{CP}$ are very small:  less than $5\%$.

\item[(3)]
For the ratio $R_K^{e\mu }$,  we find $R_K^{e\mu }=0.996(2)$ in  both the PQCD and ``PQCD+Lattice" approaches.
For the ratio $R_K^{\mu\tau }$,  the PQCD prediction is about $0.33$ and a little smaller than the ``PQCD+Lattice"  prediction $R_K^{\mu\tau }\approx 0.39$,
and  they also show a weak $\mu$-dependence:  less than $9\%$ in magnitude for $0.5m_b \leq \mu \leq 1.5 m_b$.
.
\item[(4)]
For the case of $l^-=(e^-,\mu^-)$,  the PQCD and ``PQCD+Lattice " predictions for the values of  $P_L$  are very similar and close to $-1$ in value.
For the $\tau^-$ lepton,  however,  the PQCD  and  ``PQCD+Lattice "  predictions   for  $P_L$  show a moderate difference:   $- 0.35$  against $ - 0.24$.

\end{enumerate}

\subsection{Observables for \texorpdfstring{ $B_s\to K^*\ell^+\ell^-$ }{}}

\begin{table}[htb]
\caption{The PQCD and ``PQCD+Lattice" predictions for the branching ratios (in unit of $10^{-8}$) of the semileptonic decays
$B_s \to K^* \ell^+ \ell^-$  , the lepton forward-backward asymmetry $\cala_{FB}$,  the $K^\ast$ polarization fraction $F_L^{K^\ast}$ ,
the direct CP asymmetry $\cala_{CP}$ (in unit of $10^{-2}$),  and the ratios $R^{e\mu }_{K^\ast}$ and  $ R^{\mu \tau }_{K^\ast}$ .
The theoretical errors from  the input parameters are combined in  quadrature.}
\label{tab:table11}
\centering
\setlength{\tabcolsep}{8pt} 
\renewcommand{\arraystretch}{1} 
\begin{tabular}{l|l|ll | ll}
\hline
\hline
Mode &Obs.&\multicolumn{2}{c|}{PQCD}& \multicolumn{2}{c}{PQCD+Lattice} \\  \cline{3-6}
& &$\bar{B_s} \to {K^\ast}\ell^+ \ell^-$ & $B_s \to \bar{K^\ast}\ell^+ \ell^-$ & $\bar{B_s} \to {K^\ast}\ell^+ \ell^-$ & $B_s \to \bar{K^\ast}\ell^+ \ell^-$ \\
\hline
$\ell=e$ &{$\calb$} & ${3.74}^{+1.38}_{-1.16} $ &${3.68}^{+1.33}_{-1.14}$ & ${3.00}^{+0.91}_{-0.76}$ & ${2.99}^{+0.94}_{-0.78}$  \\
&$\cala_{FB}$         & $-0.244(9)$ &$-0.235(17)$&  $-0.183(10)$ &$-0.176(10)$ \\
&$F_L^{K^\ast}$         & $0.373(2)$  &$0.393(5)$& $0.408(7)$  &$0.432(9)$ \\ \cline{2-6}
&$\cala_{CP}$&\multicolumn{2}{c|}{$0.8\pm 0.3$}  &\multicolumn{2}{c}{$0.2\pm 0.4$} \\  \hline
$\ell=\mu$ &{$\calb$} & ${3.20}^{+1.26}_{-1.00}$ & ${3.14}^{+1.19}_{-0.98}$ &${2.48}^{+0.67}_{-0.59}$& ${2.47}^{+0.70}_{-0.61}$  \\
&$\cala_{FB}$         & $-0.285(17)$ &$-0.275(14)$ & $-0.222(6)$  &$-0.214(6)$ \\
&$F_{L}^{K^\ast}$         & $0.434(12)$  &$0.457(16)$& $0.494(16)$  &$0.522(18)$ \\\cline{2-6}
&$\cala_{CP}$&\multicolumn{2}{c|}{$1.0\pm 0.5$}&\multicolumn{2}{c}{$0.2\pm 0.5$} \\ \hline
$\ell=\tau$ &{$\calb$} & ${0.71}^{+0.35}_{-0.21}$   &${0.72}^{+0.36}_{-0.21}$ &${0.49}\pm0.09$& ${0.50}\pm0.09$  \\
&$\cala_{FB}$         &  $-0.235(33)$  &$-0.232(33)$& $-0.196(5)$ &$-0.194(5)$ \\
&$F_{L}^{K^\ast}$         & $0.420(13)$ &$0.418(13)$ & $0.480(5)$ & $0.478(5)$ \\\cline{2-6}
&$\cala_{CP}$&\multicolumn{2}{c|}{$-0.7\pm 0.3$}&\multicolumn{2}{c}{$-1.0\pm 0.2$}\\  \hline
 &$R_{K^\ast}^{e\mu }$ & $0.993(2)$ & $0.993(2)$ &$0.992(2)$ & $0.992(2)$ \\
&$R_{K^\ast}^{ \mu \tau}$ & ${0.227(30)}$  & $0.238(35)$ &${0.205(20)}$ & $0.211(22)$ \\ \hline \hline
\end{tabular}
\end{table}

\begin{table}[htb]
\caption{The PQCD and ``PQCD+Lattice" predictions for the angular observables $P_i ~(i=1,2,3)$ and $P'_j ~(j=4,5,6,8)$ of the decays
$\bar{B_s} \to {K^\ast}\ell^+ \ell^-$ (the first row) and $B_s \to \bar{K^\ast}\ell^+ \ell^-$ (the second row).
The total uncertainties of the input parameters are combined in  quadrature.}
\label{tab:table12}
\centering
\setlength{\tabcolsep}{6pt} 
\renewcommand{\arraystretch}{1} 
\begin{tabular}{l|ll|ll|ll}
\hline\hline
Obs. &\multicolumn{2}{c|}{$e$ mode}&\multicolumn{2}{c|}{$\mu$ mode}&\multicolumn{2}{c}{$\tau$ mode}\\
\cline{2-7}
 & PQCD & PQCD+Lat. & PQCD & PQCD+Lat. & PQCD & PQCD+Lat.  \\
\hline
$ -P_1 $ & $0.34\pm 0.12$ &$0.37\pm 0.05$ & $0.44\pm 0.12$ &$0.52\pm 0.04$ & $0.62\pm 0.12$ &$0.68\pm 0.02$  \\
$ $ &      $0.33\pm 0.11$ &$0.36\pm 0.05$ & $0.44\pm 0.12$ &$0.52\pm 0.04$ & $0.62\pm 0.12$ &$0.68\pm 0.02$  \\ \hline
$ -P_2 $ & $0.27\pm 0.01$ &$0.21\pm 0.01$ & $0.34\pm 0.03$ &$0.30\pm 0.01$ & $0.36\pm 0.05$ &$0.35\pm 0.01$  \\
$ $ &      $0.26\pm 0.01$ &$0.21\pm 0.01$ & $0.35\pm 0.03$ &$0.31\pm 0.01$ & $0.36\pm 0.05$ &$0.35\pm 0.01$  \\ \hline
$ -P_3 \times 10^{3}$ & $0.17\pm  1.17$ &$0.34\pm0.21$ & $0.19\pm 1.43$ &$0.45\pm 0.29$ & $0.35\pm 0.99$ &$0.91\pm 0.13$  \\
$ $ &                       $1.27\pm 0.99$ &$2.84\pm 0.51$ & $1.68\pm 1.15$ &$4.11\pm 0.52$ & $0.38\pm 1.08$ &$0.98\pm 0.14$  \\ \hline
$ P'_4 $ & $1.06\pm 0.07$ &$1.06\pm 0.04$& $1.10\pm 0.07$ &$1.11\pm 0.03$ & $1.27\pm 0.05$ &$1.29\pm 0.01$  \\
$  $ &     $1.00\pm 0.01$ &$0.99\pm 0.04$ & $1.07\pm 0.07$ &$1.07\pm 0.03$ & $1.27\pm 0.05$ &$1.29\pm 0.01$  \\  \hline
$ -P'_5 $ & $0.57\pm 0.06$ &$0.48\pm 0.02$ & $0.61\pm 0.07$ &$0.52\pm 0.02$ & $0.58\pm 0.10$ &$0.54\pm 0.01$  \\
$  $ &      $0.57\pm 0.05$ &$0.48\pm 0.02$ & $0.61\pm 0.06$ &$0.53\pm 0.02$ & $0.57\pm 0.10$ &$0.54\pm 0.01$  \\ \hline
$ -P'_6\! \times \!10^{3}$ & $0.76\pm 0.15$ &$0.96\pm 0.09$ & $0.68\pm 0.13$ &$0.87\pm 0.08$ & $0.11\pm 0.03$ &$0.12\pm 0.01$  \\
$ $ &                        $0.77\pm 0.14$ &$0.96\pm 0.08$ & $0.70\pm 0.13$ &$0.87\pm 0.07$ & $0.11\pm 0.02$ &$0.12\pm 0.01$  \\ \hline
$ -P'_8\! \times \!10^2$ & $0.61\pm 0.06$ &$0.67\pm 0.07$ & $-0.36\pm 0.27$ &$-0.57\pm 0.26$ & $0.27\pm 0.07$ &$0.37\pm 0.03$  \\
$ $ & $1.33\pm 0.09$ &$1.70\pm 0.06$ & $2.20\pm 0.27$ &$2.83\pm 0.26$ & $0.30\pm 0.07$ &$0.39\pm 0.03$  \\ \hline \hline
\end{tabular}
\end{table}

Analogous to  the cases of $B_s\to K \ell^+\ell^-$ decays,  we  follow the same procedure to calculate the physical observables of
$B_s\to K^*\ell^+\ell^-$ by using the  PQCD and ``PQCD+Lattice" approaches, respectively.  For  $B_s\to K^*\ell^+\ell^-$ decays, however,
much more  physical observables are defined  and studied.

In  Table \ref{tab:table11} we listed the PQCD and `` PQCD+Lattice" predictions for the branching ratios
$\calb(\bar{B}_s\to K^*\ell^+\ell^-)$  and $\calb(B_s\to \bar{K}^*\ell^+\ell^-)$  with
$l=(e,\mu,\tau)$ , the lepton forward-backward asymmetries $\cala_{FB}$,  the longitudinal polarization asymmetries $F_{L}^{K^\ast}$
of the leptons,  the direct CP asymmetries $\cala_{CP}$,   and  the ratios  of the branching ratios $R^{e\mu }_{K^*}$  and  $R^{\mu\tau }_{K^*}$
with the choice of  the scale $\mu=m_b$.
In numerical calculations,  we here use  the mean value of decay rate $\Gamma_{B^0_s}=1.509\times 10^{12} s^{-1}$ \cite{pdg2018}.
For the branching ratios, the extra error from the S-wave pollution up to $10\%$ should be added additionally \cite{Doring:2013wka}.
In Table \ref{tab:table12} we listed the PQCD and `` PQCD+Lattice" predictions for the values of those angular observables
$P_i ~(i=1,2,3)$ and $P'_j ~(j=4,5,6,8)$ in $ l=(e,\mu,\tau)$ mode.
The total errors  of all theoretical predictions  in Table  \ref{tab:table11} and \ref{tab:table12}
are estimated in the same way as that  for the case of $\bar{B}_s \to K l^+l^-$ decays.

From  Table \ref{tab:table11},   it is easy to find the CP-averaged  branching ratio   $\calb(\bar{B}_s \to K^* l^+ l^-)$ for $l=(e,\mu,\tau)$:
\beq
\calb(\bar{B}_s \to K^* e^+ e^- ) |_{\rm CP-av.}&=&\left \{  \begin{array}{ll}
(3.71^{+0.98}_{-0.82}\pm 0.37)\times 10^{-8}, &  {\rm PQCD},  \\
(3.00^{+0.66}_{-0.55}\pm 0.30)\times 10^{-8}, &  {\rm PQCD+Lattice}, \\   \end{array} \right. \label{eq:bravs1}
\eeq
\beq
\calb(\bar{B}_s \to K^* \mu^+ \mu^- ) |_{\rm CP-av.}&=&\left \{  \begin{array}{ll}
(3.17^{+0.89}_{-0.71} \pm 0.32)\times 10^{-8}, &  {\rm PQCD},  \\
(2.48^{+0.50}_{-0.43} \pm 0.25)\times 10^{-8}, &  {\rm PQCD+Lattice}, \\  \end{array} \right. \label{eq:bravs2}
\eeq
\beq
\calb(\bar{B}_s \to K^* \tau^+ \tau^- ) |_{\rm CP-av.}&=&\left \{  \begin{array}{ll}
(0.72^{+0.25}_{-0.17} \pm 0.07)\times 10^{-8}, &  {\rm PQCD},  \\
(0.50 \pm 0.06 \pm 0.05 )\times 10^{-8}, &  {\rm PQCD+Lattice}. \\ \end{array} \right. \label{eq:bravs3}
\eeq
where the second errors come from the  $10\%$ S-wave pollution  as estimated in Ref.~\cite{Doring:2013wka}.

\begin{figure}[htb]
\begin{center}
\centerline{\epsfxsize=7cm\epsffile{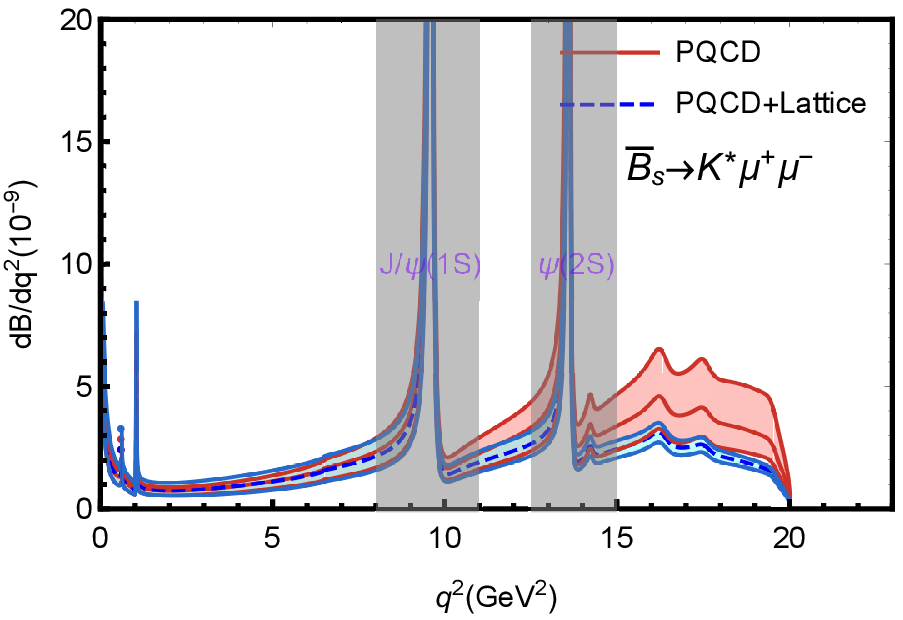} \hspace{0.5cm}\epsfxsize=7cm\epsffile{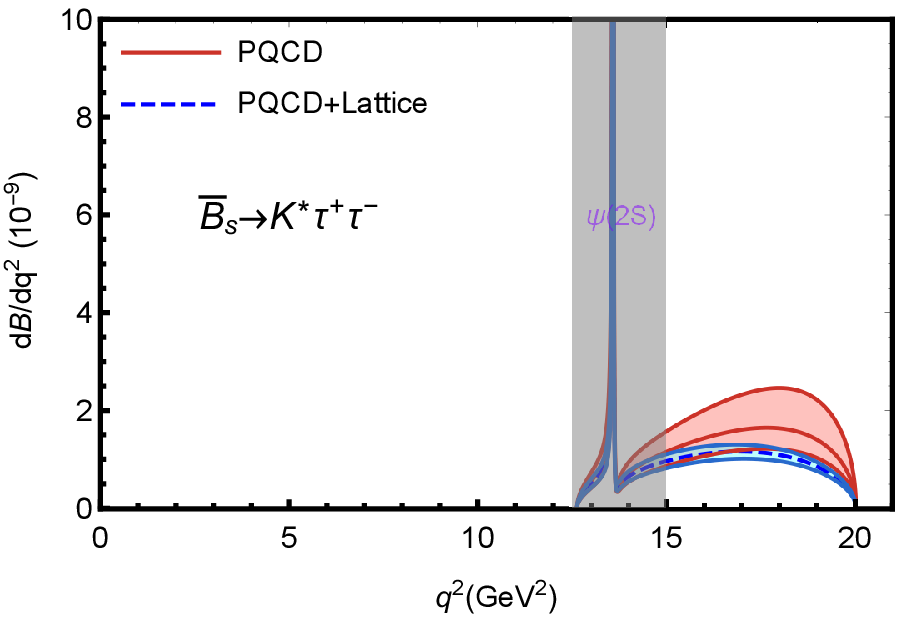} }
\vspace{0.3cm}
\centerline{\epsfxsize=7cm\epsffile{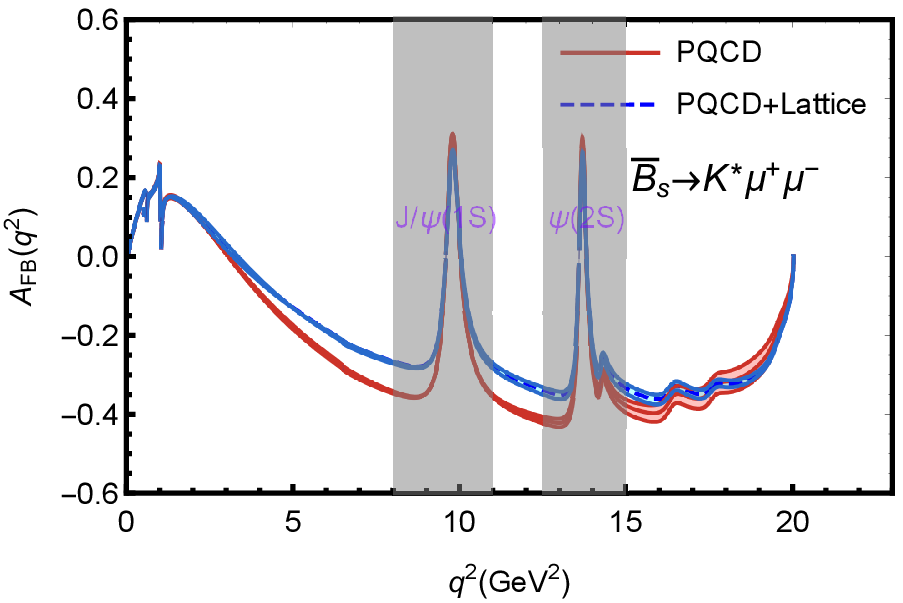}  \hspace{0.5cm} \epsfxsize=7cm\epsffile{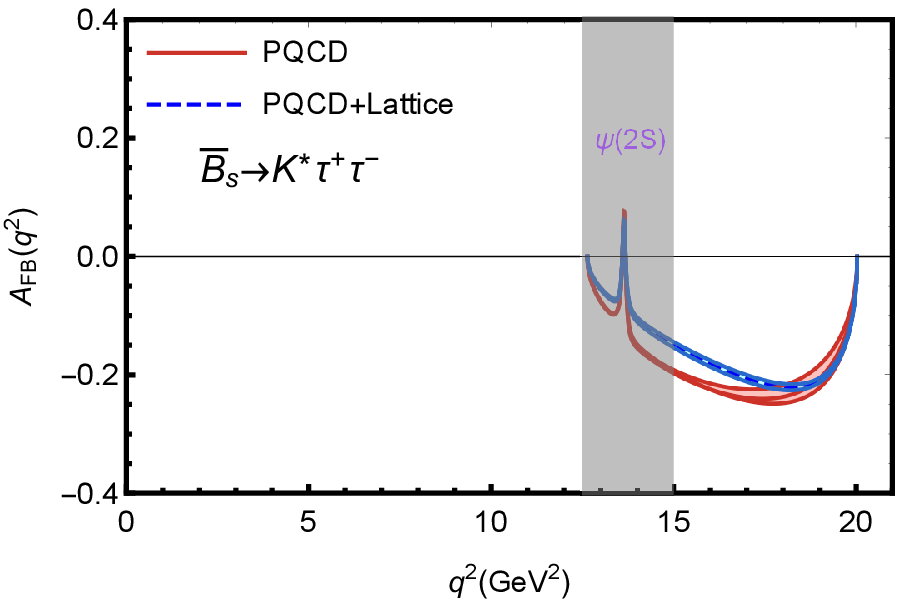} }
\vspace{0.3cm}
\centerline{\epsfxsize=7cm\epsffile{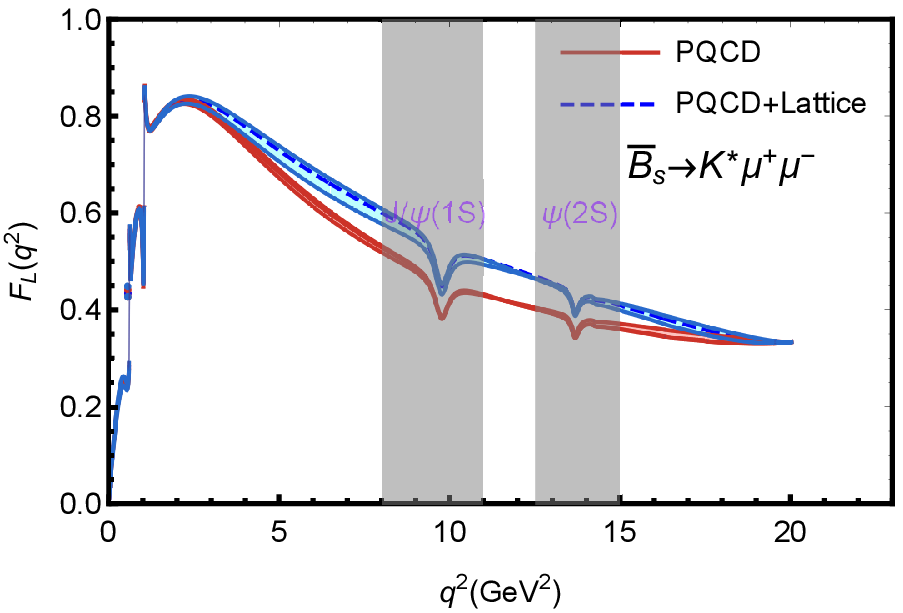}  \hspace{0.5cm} \epsfxsize=7cm\epsffile{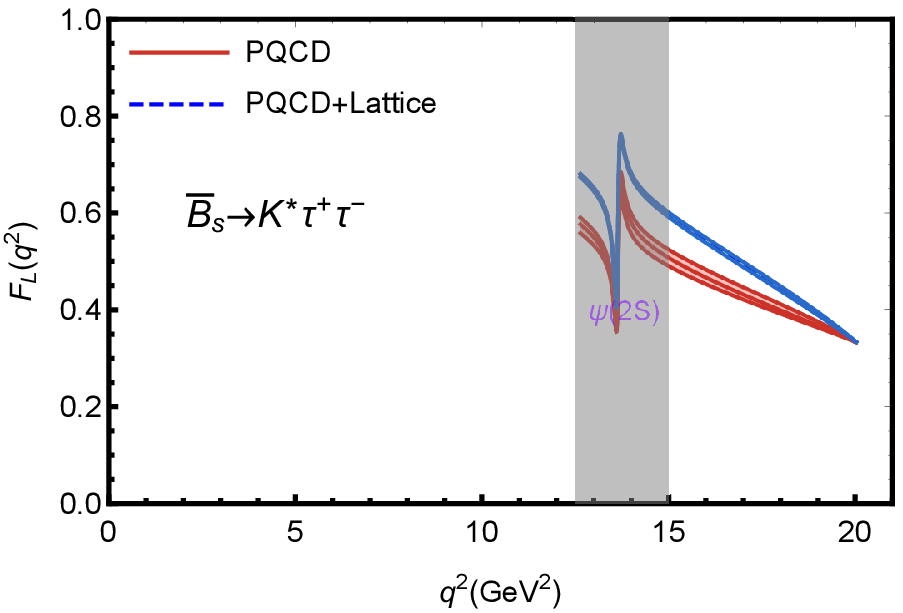} }
\end{center}
\caption{The theoretical predictions for the $q^2$-dependence of $d\calb/dq^2$, $ \cala_{FB}(q^2)$ and $ F^{K^\ast}_{L}(q^2)$
for the decays $\bar{B}_s \to K^\ast \ell^+ \ell^- $ with $\ell=(\mu,\tau)$ in the PQCD and ``PQCD+Lattice" approaches.
For more details see the text. }
\label{fig:fig6}
\end{figure}

In Fig.~\ref{fig:fig6}, we  show the PQCD and the ``PQCD+Lattice" predictions of $q^2$-dependence of the differential decay rate
$d\calb/dq^2$,  the forward-backward asymmetry $\cala_{FB}(q^2)$,  the longitudinal polarization $ F^{K^\ast}_L(q^2)$ for  $B_s \to K^\ast \ell^+ \ell^- $
decays with $\ell=(\mu,\tau)$,   $q^2_{max}$ = 20.02 GeV$^2$ and the renormalization scale $\mu=m_b$.
The red (blue) lines (dashed lines) correspond to the  predictions obtained using the PQCD  (``PQCD+Lattice") approach, while
the shaded  narrow bands (red and blue)  indicate the uncertainty of our predictions due to the variations of the input parameters.
For the cases of the decays $B_s\to K^*\mu^+\mu^-$,  the two vertical grey blocks  show the experimental veto regions \cite{Wei:2009zv}
in order to remove the contributions from  the resonance $ J/\psi(1S)$ (left-hand band) and $\psi^\prime(2S)$  (right-hand band)
to  the  \textit{ form factor dependent}(FFD) observables.
For the case of the decay $B_s\to K^*\tau^+\tau^-$,  on the other hand,   there is one  vertical grey block which shows the experimental veto region
\cite{Wei:2009zv} for the resonance $\psi^\prime(2S)$ only.

\begin{figure}[thb]
\begin{center}
\centerline{\epsfxsize=7cm\epsffile{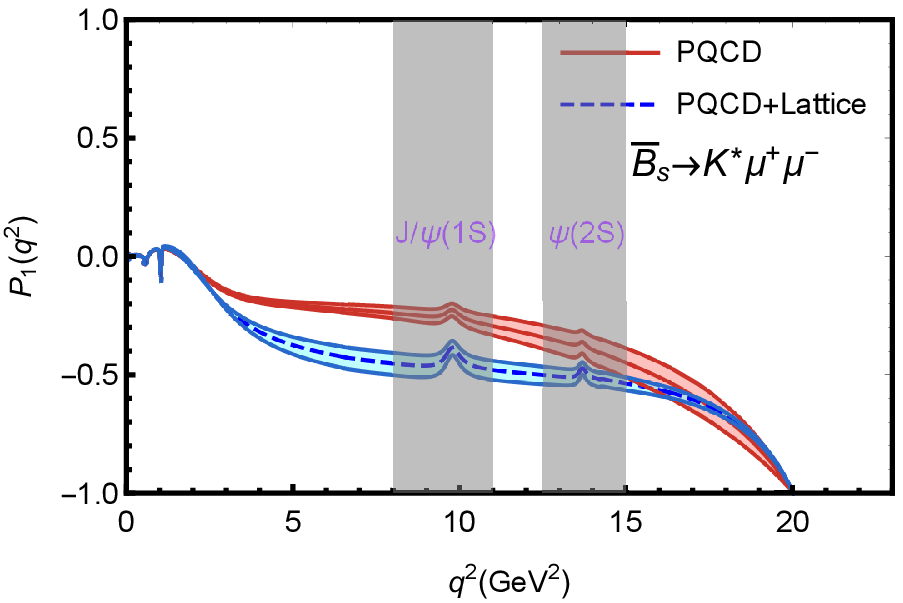}      \hspace{0.5cm} \epsfxsize=7cm\epsffile{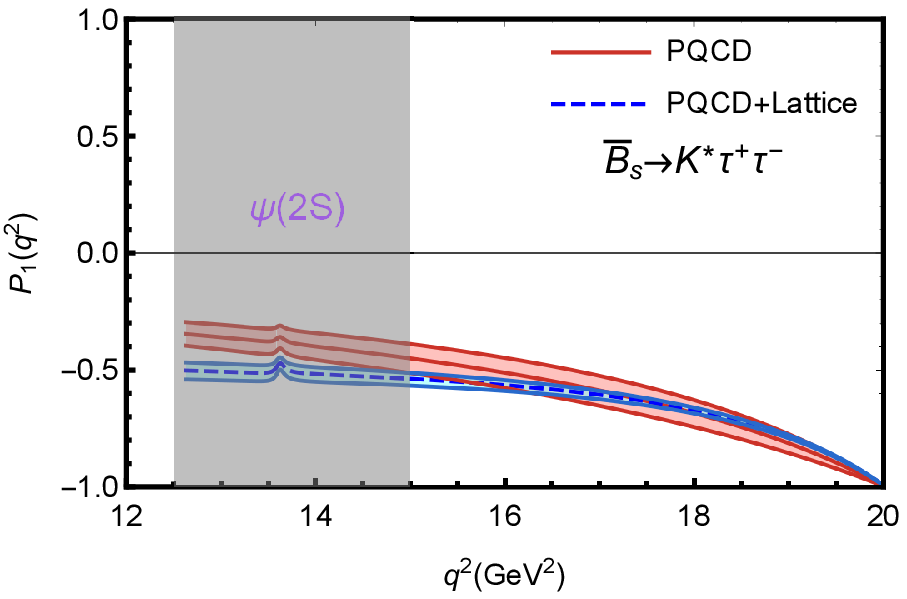} }
\vspace{0.3cm}
\centerline{\epsfxsize=7cm\epsffile{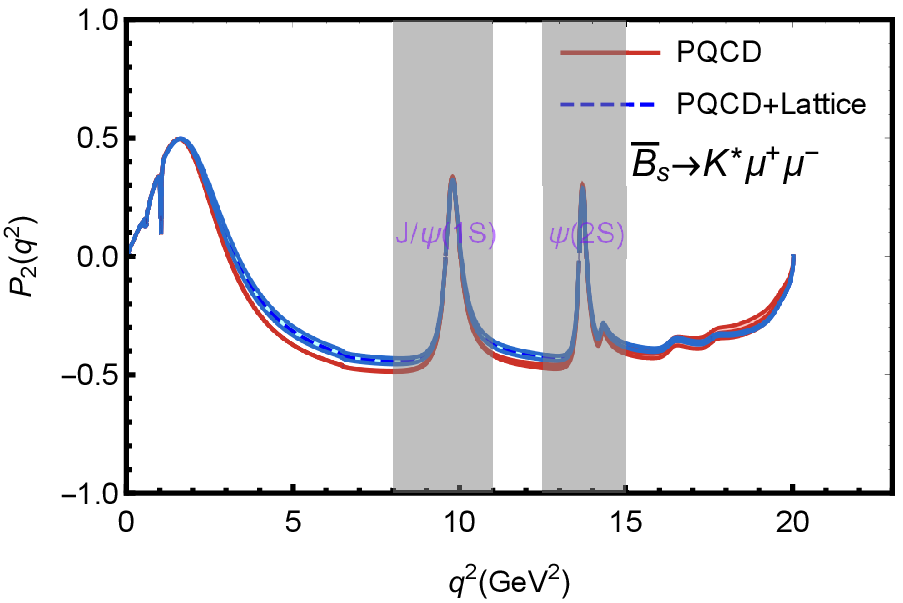}      \hspace{0.5cm}  \epsfxsize=7cm\epsffile{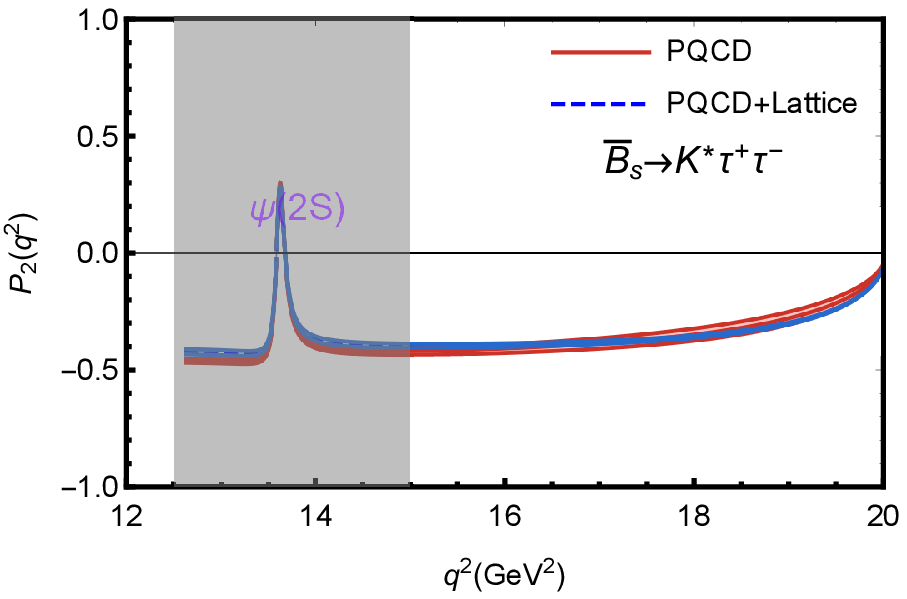} }
\vspace{0.3cm}
\centerline{\epsfxsize=7cm\epsffile{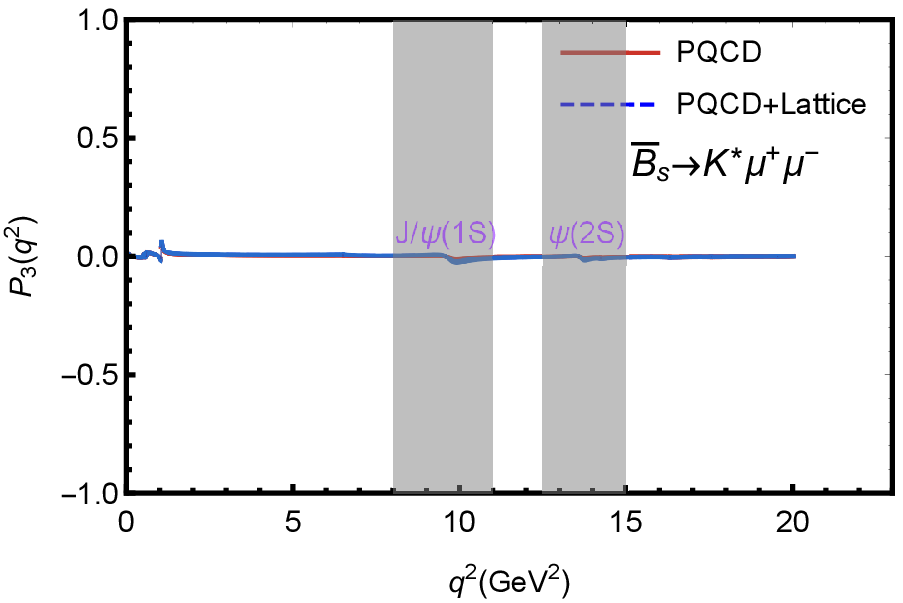}      \hspace{0.5cm} \epsfxsize=7cm\epsffile{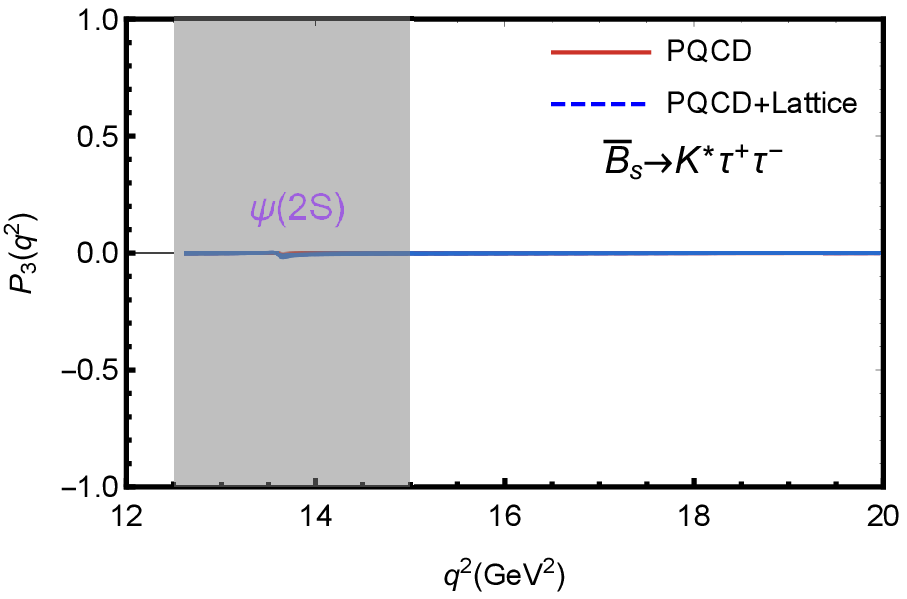} }
\end{center}
\caption{The theoretical predictions for the $q^2$-dependence of the angular observables $P_i ~(i=1,2,3)$  for the decays
$\bar{B}_s \to K^\ast l^+ l^- $ ($\ell=\mu,\tau$) in the PQCD and ``PQCD+Lattice" approaches. For details see the text. }
\label{fig:fig7}
\end{figure}

\begin{figure}[tbp]
\begin{center}
\centerline{\epsfxsize=7cm\epsffile{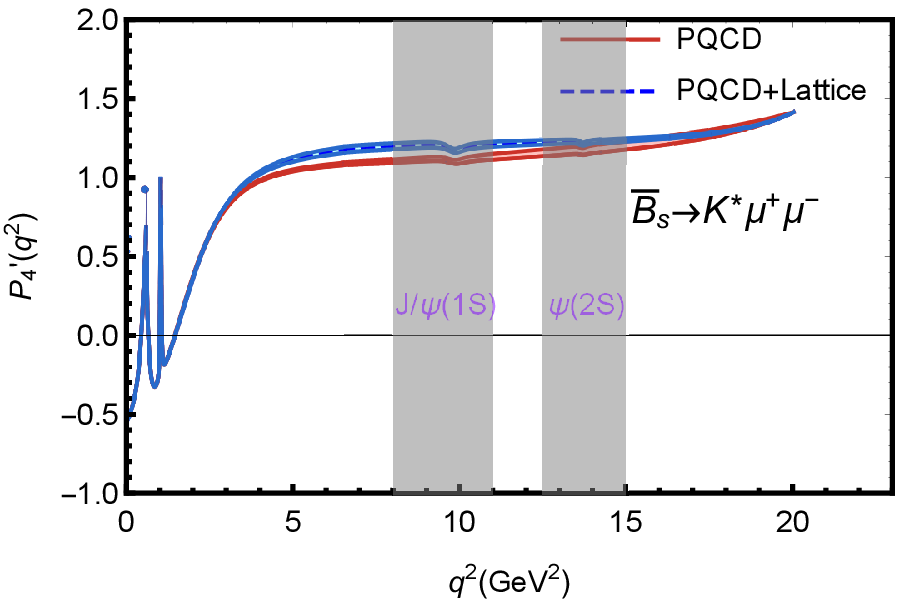}   \hspace{0.5cm}    \epsfxsize=7cm\epsffile{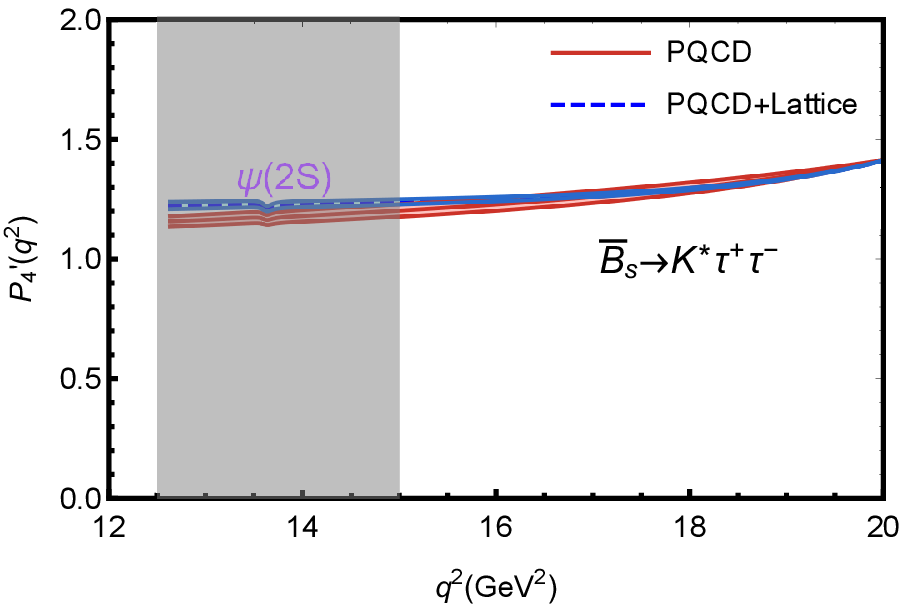} }
\vspace{0.3cm}
\centerline{\epsfxsize=7cm\epsffile{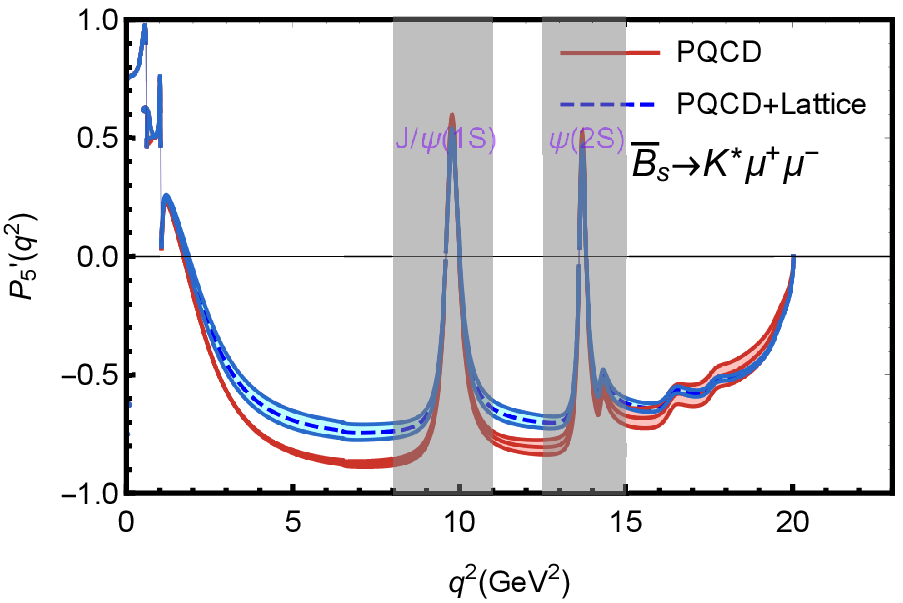}   \hspace{0.5cm}    \epsfxsize=7cm\epsffile{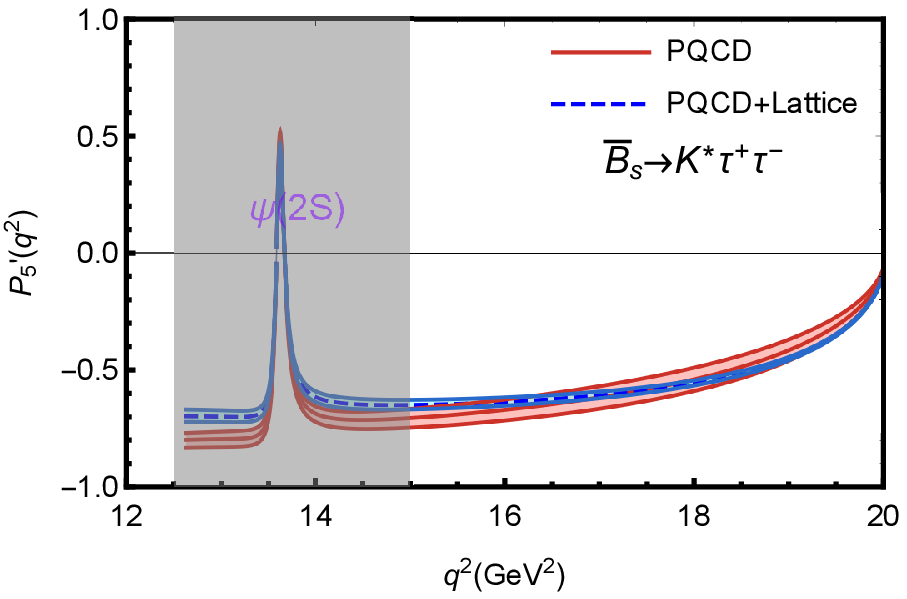} }
\vspace{0.3cm}
\centerline{\epsfxsize=7cm\epsffile{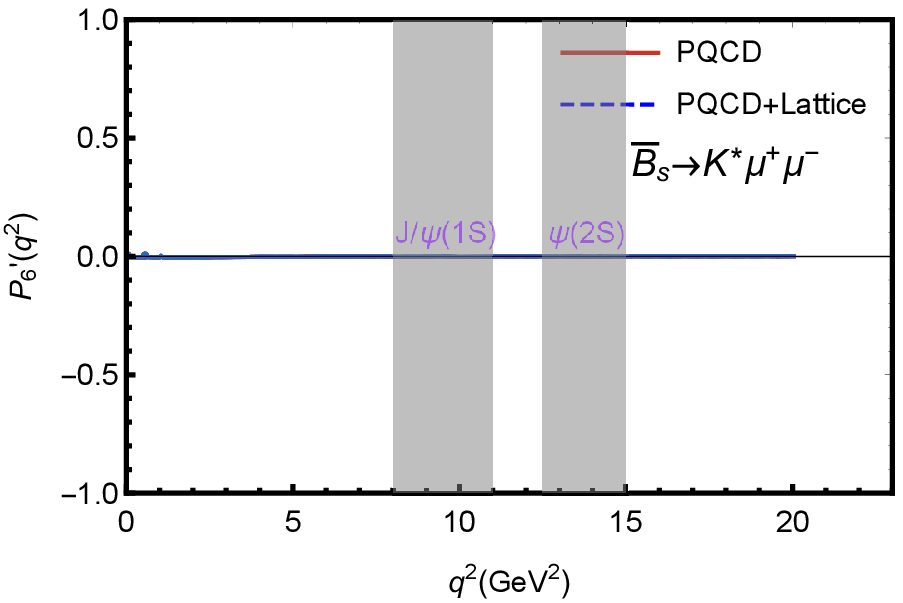}   \hspace{0.5cm}  \epsfxsize=7cm\epsffile{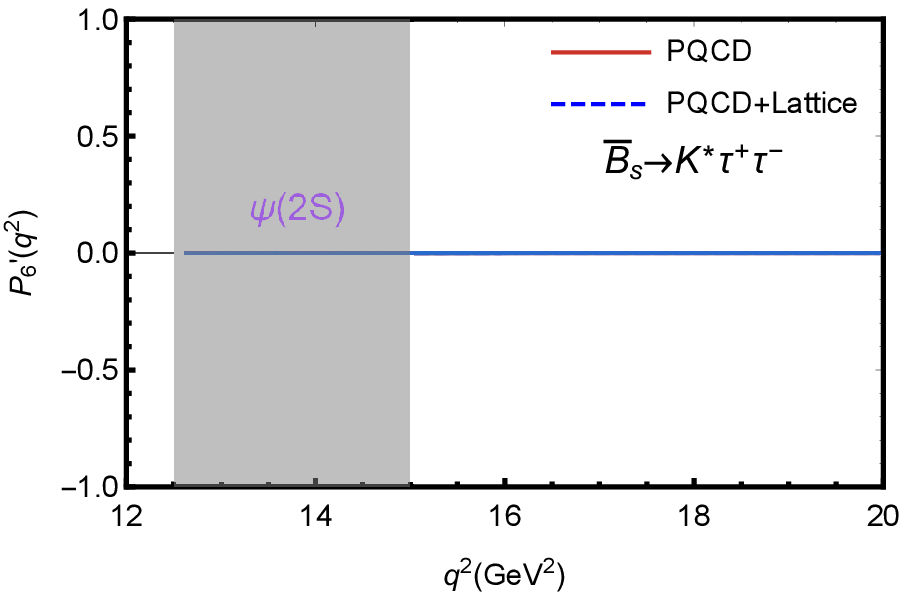} }
\vspace{0.3cm}
\centerline{\epsfxsize=7cm\epsffile{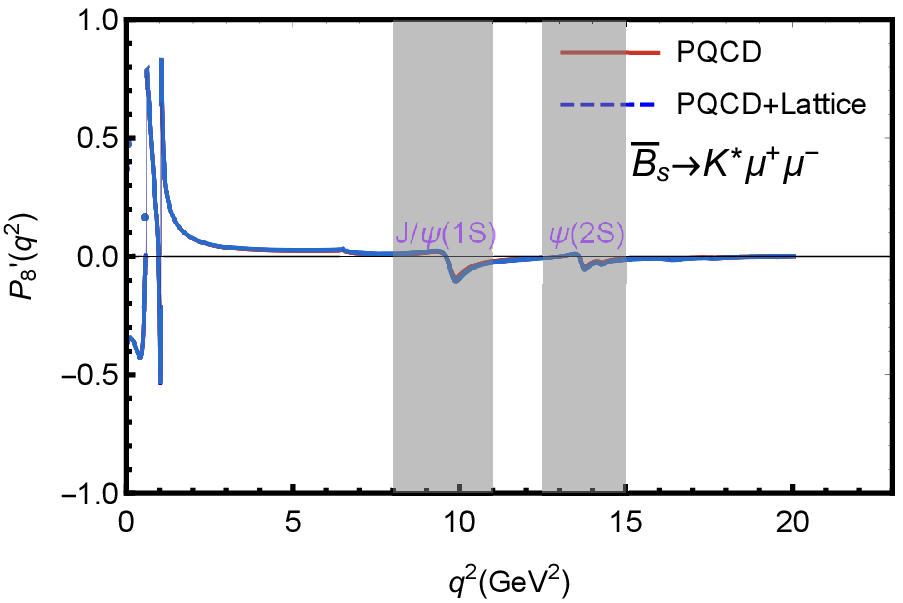}    \hspace{0.5cm}  \epsfxsize=7cm\epsffile{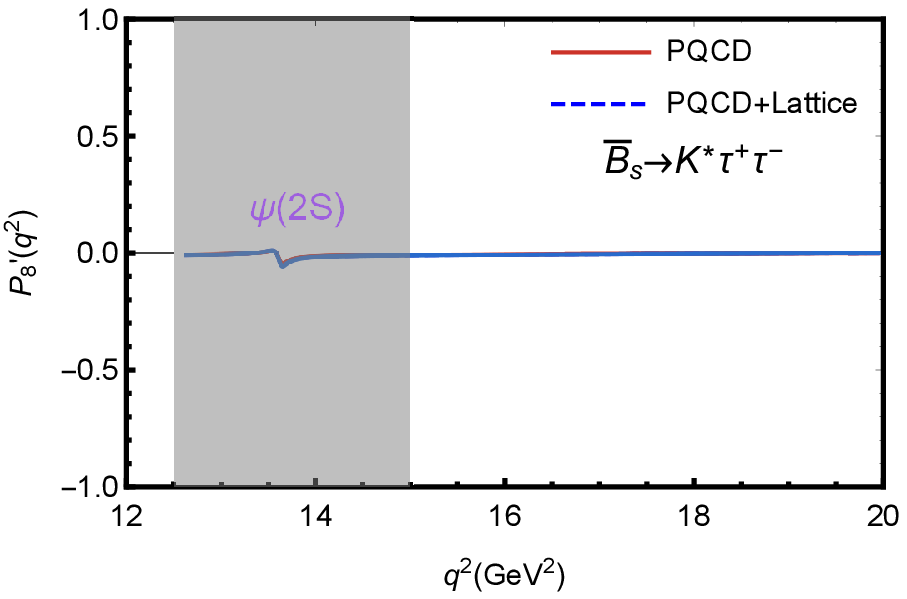} }
\end{center}
\caption{The theoretical predictions for the $q^2$-dependence of the angular observables $P'_j ~(j=4,5,6,8)$
for the decays  $\bar{B}_s \to K^\ast l^+ l^- $ ($\ell=\mu,\tau$)   in the PQCD and ``PQCD+Lattice" approaches.
For details see the text.}
\label{fig:fig8}
\end{figure}

In Fig.~\ref{fig:fig7} and   \ref{fig:fig8},  we show the $q^2$-dependence of the angular observables   $P_i ~(i=1,2,3)$ and $P'_j ~(j=4,5,6,8)$
for the considered semileptonic decays $B_s \to K^\ast l^+ l^- $ with $\ell= (\mu,\tau)$, respectively.
Since the relevant figures for the electron mode are very similar with those for the muon mode, we do not draw them in Figs.~\ref{fig:fig6}-\ref{fig:fig8}.
The symbols in these two figures have the same  meaning  with those in  Fig.~\ref{fig:fig5}.

From the numerical predictions as given in Tables     \ref{tab:table11} and  \ref{tab:table12} and in Figs.~\ref{fig:fig6}-\ref{fig:fig8},
we find the following points about the physical observables of   the $\bar{B}_s \to K^\ast l^+ l^- $ ($\ell= e,\mu,\tau$) decays:
\begin{enumerate}
\item[(1)]
For the considered decay modes,   the  PQCD and ``PQCD+Lattice"   predictions for  $\calb(\bar{B}_s \to K^\ast l^+ l^-)$  with  $\ell=( e,\mu,\tau)$
do agree well  with each other within the errors.  The  ``PQCD+ Lattice" predictions of  $\calb(\bar{B}_s \to K^\ast l^+ l^-)$
have   smaller errors than  those of  the PQCD predictions.
Both PQCD and ``PQCD+Lattice" predictions of  $\calb(\bar{B}_s \to K^\ast \mu^+ \mu^-)$  do agree well
with the LCSR prediction $(2.85\pm 0.72)\times 10^{-8}$ \cite{Kindra:2018ayz} and with the currently available  LHCb measured value
$ (2.9 \pm 1.1)\times 10^{-8}$ \cite{Aaij:2018jhg}.
For the electron and muon mode, on the other hand, we have to wait for the future experimental measurements.

\item[(2)]
For the ratio  $R_{K^\ast}^{e\mu }$,  the theoretical predictions from both PQCD and ``PQCD+Latatice" approach are almost the same one,  with
a  tiny $\sim 1\%$ error because of  the great cancellation  of the errors in the ratio of  the branching ratios.
For the ratio  $R_{K^\ast}^{\mu \tau}$,  however,   the remaining error of the theoretical predictions from both PQCD and ``PQCD+Latatice" approach are
still around  $10\%$.  These two ratios should be measured in the future experiments.

\item[(3)]
For the direct  CP asymmetries $\cala_{CP}$ of the considered decays, they are always very small as expected: less than $2\%$ in magnitude.
For  physical observables $\cala_{FB}$ and $F_L^{K^*}$,  the differences between the central values of the PQCD and  ``PQCD+Lattice"
are about $20\%$ in magnitude, while the errors of the theoretical predictions are less than $10\%$.

\item[(4)]
For  the angular observables $P_i$ and $P'_j$,   the   PQCD and ``PQCD+Lattice "  predictions  for each mode  are consistent within errors.
The values of  $P_3$ and $P'_{6,8}$ are close to zero: $\sim 10^{-2}$ for $P'_8$  and   $\sim 10^{-3}$ for $P_3$ and $P'_6$.
For the remaining $P_{1,2}$ and $P'_{4,5}$,  their magnitudes are small:   $-1 < (P_{1,2},P'_5 ) < -0.2$  while   $P'_4\sim 1$.

\item[(5)]
One can see from the curves in Figs.~\ref{fig:fig7}  and \ref{fig:fig8}  that  most angular observables   $P_i $ and $P'_j$ have weak  $q^2$-dependence
in the major region of  $q^2$ due to the large cancellation of  $q^2$-dependence in the ratios.

\end{enumerate}

For the semileptonic decays $\bar{B}_s \to K^{\ast}\ell^+\ell^-$ ($\ell=e,\mu$),   some regions of $q^2$ do correspond to some resonance states,
such as the charmonium $\jpsi, \psi(2S), etc,$ ,  and  should be removed for the sake of date analysis.
Following Ref.~\cite{Kindra:2018ayz}, we here also present the binned value of the observables as a function of lepton-pair momentum $q^2$
covering two $q^2$ regions: $[0.1-0.98]$ GeV$^2$ and $[1.1-6]$GeV$^2$ and consider the mass effect in the final state.
We employ the PQCD and ``PQCD+Lattice" approach to evaluate the form factors and compare the resultant results.
Analogous to Ref.~\cite{Kindra:2018ayz},  we also define the $q^2$-binned observables  in following form:
\beq
 \langle P_1 \rangle_{bin} &=& \frac{ \int_{bin}dq^2 (I_{3})}{2 \int_{bin}dq^2 (I^s_{2})}, \quad
\langle P_2 \rangle_{bin} =\frac{\int_{bin}dq^2 (\beta_\ell I^s_{6})}{8 \int_{bin}dq^2 (I^s_{2})}, \quad
 \langle P_3 \rangle_{bin} =-\frac{\int_{bin}dq^2 (I_{9})}{4 \int_{bin}dq^2( I^s_{2})},
\eeq
\beq
\langle P'_4 \rangle_{bin} &=&\frac{\int_{bin}dq^2 (I_4)}{\sqrt{-\int_{bin}dq^2 (I^c_2 I^s_2)}} , \quad
 \langle P'_5 \rangle_{bin} =\frac{\int_{bin}dq^2 (\beta_\ell I_5)}{2\sqrt{-(\int_{bin}dq^2 (I^c_2 I^s_2))}} ,
 \eeq
 \beq
\langle P'_6 \rangle_{bin} &=&-\frac{\int_{bin}dq^2 (\beta_\ell I_7)}{2\sqrt{-(\int_{bin}dq^2 (I^c_2 I^s_2))}}, \quad
\langle P'_8 \rangle_{bin} =-\frac{\int_{bin}dq^2 (I_8)}{\sqrt{-(\int_{bin}dq^2 (I^c_2 I^s_2))}} ,
\eeq
\beq
\langle \calb \rangle_{bin}  &=&  \int_{bin}dq^2 \frac{d\calb(B_s\to K^{*} \ell^+\ell^-)}{dq^2} , \\
\langle R_{K^\ast} \rangle_{bin} &=& \frac{\langle BR \rangle_{bin}(\ell=\mu)}{\langle BR \rangle_{bin}(\ell=e)}  ,
\eeq
\beq
\langle \mathcal{A}_{\rm FB}(\ell) \rangle_{bin} &=& { \frac{ \int_{bin}dq^2 b_\thl(q^2)}{ \int_{bin}dq^2 d\Gamma/dq^2} }
=  {\frac{3\int_{bin}dq^2 I^s_{6}}{4\Gamma_{bin}}} , \\
\langle F_L^{K^\ast} \rangle_{bin} &=& \frac{3\int_{bin}dq^2 (I^c_{1}-I^c_{2})}{\int_{bin}dq^2 [3\left( I^c_{1} + 2I^s_{1}\right) -\left(  I^c_{2}+ 2I^s_{2}\right)]}.
\eeq

\begin{table}[tbp]
\caption{ The binned values of observables for the process $\bar{B_s}\to K^{\ast}\mu^+\mu^-$ and $B_s\to \bar{K^{\ast}}\mu^+\mu^-$ at $\mu=m_b$ scale using the
PQCD and ``PQCD+Lattice" factorization  approaches.  The uncertainties shown are due to errors in determination of form factors and CKM parameters.
The LCSR predictions for  $B_s\to K^{\ast}\mu^+\mu^-$  decay as given in Ref.~\cite{Kindra:2018ayz} were added as a comparison.
For details see the text. }
\label{tab:table13}
\centering
\setlength{\tabcolsep}{6pt} 
\renewcommand{\arraystretch}{1} 
\begin{tabular}{l|l|cc|cc}  \hline \hline
\multicolumn{2}{c|}{Decay mode}& \multicolumn{2}{c|}{$\bar{B_s}\to K^{\ast}\mu^+\mu^-$} &  \multicolumn{2}{c}{$B_s\to \bar{K}^{\ast}\mu^+\mu^-$}\\
\cline{3-4}	\cline{5-6}	
\multicolumn{2}{c|}{Obs./Bin}& [0.1 -- 0.98] GeV$^2$ & [1.1 -- 6] GeV$^2$&  [0.1 -- 0.98] GeV$^2$ & [1.1 -- 6] GeV$^2$ \\ 		\hline
$\langle P_1 \rangle$ & PQCD &${0.004(2)}$ & ${-0.157(16)}$ & ${0.023(2)}$ & ${-0.163(16)}$\\
&PQCD+Lat.&                  ${0.003(2)}$ & ${-0.279(25)}$ & ${0.021(3)}$ & ${-0.292(28)}$\\
& LCSR~\cite{Kindra:2018ayz}&$0.012(129)$ & ${-0.081(111)}$ & $0.011(135)$ & ${-0.075(108)}$\\ \hline
$\langle P_2 \rangle$ & PQCD& ${0.127(2)}$ & ${-0.127(6)}$& ${0.139(2)}$ & ${-0.223(5)}$\\
&PQCD+Lat.&                   ${0.128(2)}$ & ${-0.101(18)}$& ${0.141(2)}$ & ${-0.155(16)}$\\
& LCSR~\cite{Kindra:2018ayz}& $0.118(13)$ & $0.112(80)$& ${0.112(13)}$ & ${0.142(79)}$\\ \hline
$\langle P_3 \rangle \times 10^2$ & PQCD& ${0.1\pm 0.1}$ & ${0.4\pm 0.2}$& ${0.2\pm 0.2}$ & ${0.5\pm 0.2}$ \\
&PQCD+Lat.&                   ${0.1\pm 0.1}$ & ${0.8\pm 0.2}$& ${0.2\pm 0.2}$ & ${0.9\pm 0.2}$\\
& LCSR~\cite{Kindra:2018ayz}& $0.1\pm 0.2$ & $0.4\pm 1.0$& ${0.1\pm 0.7}$ & ${0.3\pm 1.0}$ \\ \hline
$\langle P_4^{\prime} \rangle$ & PQCD& ${-0.131(2)}$ & ${0.817(14)}$ & ${-0.603(2)}$ & ${0.849(13)}$\\
&PQCD+Lat.&                           ${-0.131(3)}$ & ${0.854(7)}$ & ${-0.604(3)}$ & ${0.890(9)}$\\
& LCSR~\cite{Kindra:2018ayz}    & $-0.593(58)$ & $0.464(165)$ & ${-0.650(60)}$ & ${0.379(172)}$\\ \hline
$\langle P_5^{\prime} \rangle$ & PQCD& ${0.711(3)}$& ${-0.608(3)}$ & ${0.394(2)}$& ${-0.650(4)}$\\
&PQCD+Lat.&                          ${0.715(2)}$& ${-0.486(40)}$ & ${0.392(2)}$& ${-0.524(39)}$\\
& LCSR~\cite{Kindra:2018ayz}       & $0.547(53)$& $-0.286(133)$ & ${0.543(55)}$& ${-0.273(140)}$\\ \hline
$\langle P_6^{\prime} \rangle\times 10^2$ & PQCD& ${-0.4\pm 0.2}$& ${-0.2\pm 0.2}$ & ${-0.3\pm 0.2}$& ${-0.2\pm 0.2}$\\
&PQCD+Lat.&                            ${-0.3\pm 0.5}$& ${-0.2\pm 0.2}$ & $-0.3\pm 0.2$& ${-0.2\pm 0.2}$\\
& LCSR~\cite{Kindra:2018ayz}       & $-10.4\pm 1.7$& $-9.5\pm 1.1$ & ${-6.9\pm 0.5}$& ${-7.8\pm 0.4}$\\ \hline
$\langle P_8^{\prime} \rangle $ & PQCD& ${0.042(2)}$ & ${0.050(2)}$ & ${0.044(2)}$ & ${0.057(2)}$\\
&PQCD+Lat.&                            ${0.041(2)}$ & ${0.053(2)}$ & ${0.045(2)}$ & ${0.062(2)}$\\
& LCSR~\cite{Kindra:2018ayz}         & ${0.015(16)}$ & ${0.040(17)}$ & $0.044(16)$ & $0.034(19)$\\ \hline
$\langle \calb \rangle \times 10^{9}$ & PQCD & $1.44\pm 0.47$ & ${4.65}^{+1.52}_{-1.39}$ & ${1.72}^{+0.69}_{-0.49}$ & ${4.99}^{+2.09}_{-1.65}$\\
&PQCD+Lat. & ${1.43}^{+0.48}_{-0.42}$ & ${4.64}^{+1.54}_{-1.41}$ & ${1.71}^{+0.66}_{-0.69}\pm0.48$ & ${4.98}^{+2.13}_{-1.71}$\\
& LCSR~\cite{Kindra:2018ayz}       & $3.81\pm0.46$ & $7.80\pm 1.79$ & $4.41\pm0.57 $ & $8.39\pm1.89$\\ \hline
$\langle R_{K^\ast} \rangle$ & PQCD &${0.983(1)}$& ${0.995(1)}$ & ${0.984(1)}$& ${0.996(1)}$\\	
&PQCD+Lat.&                          ${0.982(1)}$& ${0.996(1)}$ & ${0.984(1)}$& ${0.997(1)}$\\
& LCSR~\cite{Kindra:2018ayz}        &${0.940(9)}$& ${0.998(4)}$ & $0.942(8)$& $0.998(4)$\\	 \hline
$\langle A_{FB} (\ell) \rangle$ & PQCD & ${0.110(2)}$ & ${-0.067(5)}$& ${0.076(2)}$ & ${-0.087(5)}$\\
&PQCD+Lat.&                             ${0.110(2)}$ & ${-0.034(4)}$ & ${0.077(2)}$ & ${-0.053(3)}$\\
& LCSR~\cite{Kindra:2018ayz}         & ${-0.060(8)}$ & ${-0.029(22)}$ & $-0.056(8)$ & $-0.036(22)$\\ \hline
$\langle F_L^{K^\ast} \rangle$ & PQCD & ${0.297(5)}$ & ${0.741(11)}$ & ${0.543(8)}$ & ${0.738(10)}$ \\
&PQCD+Lat.& ${0.297(10)}$ & ${0.769(16)}$ & ${0.543(13)} $& ${0.769(16)} $\\
& LCSR~\cite{Kindra:2018ayz} & ${0.453(68)}$ & ${0.853(39)}$ & $0.464(65)$ & $0.851(39)$ \\ \hline  \hline
  \end{tabular}
\end{table}

In  Table \ref{tab:table13},  we listed the PQCD and ``PQCD+Lattice"  predictions for  the binned values of all eleven physical observables
considered in this paper for the $\bar{B}_s\to K^* \mu^+\mu^-$ and  the $B_s\to \bar{K}^* \mu^+\mu^-$ decays.
The theoretical errors from the input parameters are combined in quadrature in the tabulated error estimates.
As  a  comparison,  we also insert an extra row of the results from the LCSR approach  \cite{Kindra:2018ayz} into the table, for each physical observable.
It is necessary to note that there exist  three differences between  our predictions and the LCSR results as given in  Ref.~\cite{Kindra:2018ayz}:
\begin{enumerate}
\item[(1)]
The sign definition of the forward-backward asymmetry $A_{FB}$ in Ref.~\cite{Kindra:2018ayz} is opposite to ours as given in Eq.~(\ref{eq:AFB}).

\item[(2)]
Our choices of the $q^2$ bin are  $ [0.1-0.98]$GeV$^2$ and $[1.1-6]$GeV$^2$, instead of  the $[0.1-1]$GeV$^2$ and $[1-6]$GeV$^2$ in Ref.~\cite{Kindra:2018ayz},
because we try to remove  the possible contribution from the light resonance $\phi(1020)$.

\item[(3)]
The authors in Ref.~\cite{Kindra:2018ayz} considered the nonfactorizable   corrections like weak annihilation and spectator scattering in the bin $[1-6]$ GeV$^2$
while these effects in our analysis are very small and have been neglected.

\end{enumerate}

On the theoretical side, from the numerical results  as listed in Table \ref{tab:table13},  one can find the following points:
\begin{enumerate}
\item[(1)]
For  binned values of observables $ \langle P_{1,2} \rangle$ and $\langle A_{FB} (\ell) \rangle$,  the  differences between the PQCD and
 ``PQCD+Lattice" predictions are around $(20-40)\%$ of the central values.  For other eight physical observales, however,
 the PQCD and ``PQCD+Lattice predictions  agree very well within errors. The source of the difference come from
a little different $q^2$-dependence of the form factors of these two factorization approaches.

\item[(2)]
The differences between our results and LCSR predictions  \cite{Kindra:2018ayz} are generally not large in magnitude and could be understood if  one takes
the three differences between our approaches and the LCSR as specified in previous paragraph.   Current difference will be tested in the future when
the experimental measurements become available.

\item[(3)]
For observables $ \langle P_{3} \rangle$ and $\langle P_{6,8}^\prime \rangle$, their  SM  values are tiny,  about $10^{-3}$ to $10^{-2}$ in magnitude,  because
they are  basically driven  by the NLO contributions.
It is noted that the observable $P'_6$ stems from the absorptive part of $b\to d\gamma$, a small imaginary number.
Since these observables are not protected from hadronic uncertainties in general, their  values are more sensitive
to the choice of the method of calculating the form factors or  to the variations of the input parameters being used in calculations.

\item[(4)]
In this paper,  the possible  long-distance charm loop effects has been taken into account. The modification induced to $C_9$ is encoded in a shift where the
factorizable charm loop and nonfactorizable soft gluon are taken into account.
We also use a phenomenological model to account for light resonances  like $\rho(770)$ and $\omega(782)$ in the low-$q^2$ region.
It is interesting to note that such particular effect is difficult to estimate and can be large in size,  casting some doubts on the possibility
to exploit the bins between $J/\psi$ and $\psi(2S)$ for comparison with experiments.

\end{enumerate}

\section{Summary and Conclusions} \label{sec:6}

In the framework  of the SM,  we here studied the rare semileptonic decays $\bar{B}_s \to K^{(*)} \ell^+ \ell^-$ with $l^-=(e^-,\mu^-,\tau^-)$ by using the PQCD
and ``PQCD+Lattice" factorization approaches and provided the theoretical predictions for the thirteen kinds of physical observables:
the branching ratios $\calb( \bar{B}_s \to K^{(*)} \ell^+ \ell^-)$,  $\calb( B_s \to \bar{K}^{(*)} \ell^+ \ell^-)$ and their CP-averages,
the ratios of the branching ratios  $R^{e\mu }_{K, K^*}$  and  $R^{\mu\tau }_{K, K^*}$,
the lepton FB asymmetry $\cala_{FB}(l)$,  the longitudinal polarization asymmetry  of the leptons $P_L$ and the quantity $F_L^{K^*}$,
the  angular observables $P_i $ with $(i=1,2,3)$ and $P^\prime_j$ with $(j=4,5,6,8)$.
In the PQCD factorization approach,  specifically, we first evaluated the relevant form factors $F_{0,+,T}(q^2)$, $V(q^2)$,  $A_{0,1,2}(q^2)$
and $T_{1,2,3}(q^2)$ in the low $q^2$ region and then extrapolate them to the whole $q^2$ region using the  BCL  parametrization method.
In the ``PQCD+Lattice"  approach,  we  also take those currently available Lattice QCD results for the relevant
form factors at the end point $q^2_{max}$ as additional input to improve  the extrapolation  of  the form factors  from the low $q^2$ region
to  the whole range of $q^2$.

Based on  our  numerical calculations and the phenomenological analysis, we find the following main points:
\begin{enumerate}
\item[(1)]
For all physical observables considered in this paper,   the PQCD and ``PQCD+Lattice"  predictions do agree well within one standard deviation.
The theoretical errors of  the ``PQCD+Lattice"  predictions for the branching ratios  become much smaller than those of the PQCD
predictions.

\item[(2)]
For $\bar{B}_s \to (K, K^\ast) \mu^+ \mu^-$  decays, for example,  the PQCD and ``PQCD+Lattice"  predictions for the CP-averaged branching ratios
are the following:
\beq
\calb(\bar{B}_s \to K \mu^+ \mu^-)|_{\rm CP-av.} =\left \{  \begin{array}{ll}
(1.28^{+0.52}_{-0.48})\times 10^{-8}, &  {\rm PQCD},  \\
(1.06^{+0.22}_{-0.29})\times 10^{-8}, &  {\rm PQCD+Lattice}, \\  \end{array} \right. \label{eq:brf01}
\eeq
\beq
\calb(\bar{B}_s \to K^\ast \mu^+ \mu^-)|_{\rm CP-av.}=\left \{  \begin{array}{ll}
(3.17^{+0.95}_{-0.78})\times 10^{-8}, &  {\rm PQCD},  \\
(2.48^{+0.56}_{-0.50})\times 10^{-8}, &  {\rm PQCD+Lattice}, \\ \end{array} \right. \label{eq:brf02}
\eeq
Our theoretical predictions for the $\calb(\bar{B}_s \to K^\ast \mu^+ \mu^-)$
do agree well with the measured one $(2.9\pm 1.1 )\times 10^{-8}$ as reported by LHCb collaboration \cite{Aaij:2018jhg}.

\item[(3)]
For the ratios  $R_{K^\ast}^{e\mu }$ and $R_{K^\ast}^{\mu \tau}$,    the PQCD and ``PQCD+Lattice" predictions agree very well and have
a small error less than $10\%$ due to the cancellation of the theoretical uncertainties in the ratios of the branching ratios.
For the direct CP asymmetries $\cala_{CP}$, they  are always very small: less than $5\%$ in magnitude.
For  physical observables $\cala_{FB}$ and $F_L^{K^*}$,  the differences between the central values of the PQCD and  ``PQCD+Lattice" are about $(10 \sim 30) \%$
in magnitude, while the errors of the theoretical predictions are less than $10\%$.

\item[(4)]
For  the angular observables $P_{1,2,3}$ and $P^\prime_{4,5,6,8}$,     the PQCD and ``PQCD+Lattice "  predictions  for each lepton $l^-$
are consistent within errors.
The theoretical predictions of  $P_3$ and $P'_{6,8}$ are  tiny, say less than $10^{-2}$ in absolute value, and thus hardly  to be measured.
For the remaining $P_{1,2}$ and $P'_{4,5}$,  on the other hand,  their magnitudes are larger than $0.2$ and therefore could be
measured by future LHCb and Belle-II experiments.

\item[(5)]
For the sake of data analysis, we also defined eleven $q^2$-binned observables and presented our theoretical  predictions
of  the binned values of all considered observables  with the choice of two $q^2$-bins $[0.1-0.98]$GeV$^2$ and $[1.1-6]$GeV$^2$.
The PQCD and ``PQCD+Lattice"  predictions generally agree with each other  and  are also consistent with most LCSR results within errors.

\end{enumerate}

In general,  we believe that most physical observables considered in this paper could be measured in the future LHCb or Belle-II experiments.
Any clear deviations from above SM  predictions might be a signal of new physics beyond the SM.

\begin{acknowledgments}

This work was supported by the National Natural Science Foundation of China under Grant  No.~11775117 and 11235005.

\end{acknowledgments}


\appendix

\section{Relevant functions}\label{sec:app1}

The threshold resummation factor $S_t(x)$ is adopted from \cite{Kurimoto:2001zj}:
\beq
\label{eq-def-stx} S_t=\frac{2^{1+2c}\Gamma(3/2+c)}{\sqrt{\pi}\Gamma(1+c)}[x(1-x)]^c,
\eeq
and we here set the  parameter $c=0.3$. The hard functions
$h_1$ and $h_2$ come form the Fourier transform and can be written as
\beq
\begin{aligned}
h_1(x_1,x_2,b_1,b_2)&=K_0(\beta_1 b_1)
[\theta(b_1-b_2)I_0(\alpha_1 b_2)K_0(\alpha_1 b_1)\\
&+\theta(b_2-b_1)I_0(\alpha_1 b_1)K_0(\alpha_1 b_2)]S_t(x_2),
\end{aligned}
\eeq
\beq
\begin{aligned}
h_2(x_1,x_2,b_1,b_2)&=K_0(\beta_2 b_1) [\theta(b_1-b_2)I_0(\alpha_2 b_2)K_0(\alpha_2 b_1)\\
&+\theta(b_2-b_1)I_0(\alpha_2 b_1)K_0(\alpha_2 b_2)]S_t(x_2),
\end{aligned}
\eeq
where $K_0$ and $I_0$ are modified Bessel functions, and
\beq
\alpha_1 = m_{B_s}\sqrt{x_2r \eta^+},\quad  \alpha_2=m_{B_s}\sqrt{x_1 r \eta^+ - r^2 +r_d^2},\quad
\beta_1 = \beta_2=m_{B_s}\sqrt{x_1x_2 r \eta^+},\quad
\eeq
where $r=m_{K^{(*)}}/m_{B_s}, r_d=m_d/m_{B_s}$.

The factor $\exp[-S_{ab}(t)]$ contains the Sudakov logarithmic
corrections and the renormalization group evolution effects of both
the wave functions and the hard scattering amplitude with
$S_{ab}(t)=S_B(t)+S_M(t)$ \cite{Kurimoto:2001zj},
\beq
S_B(t)&=&s\left(x_1\frac{m_{B_s}}{\sqrt{2}},b_1\right)
+\frac{5}{3}\int_{1/b_1}^{t}\frac{d\bar{\mu}}{\bar{\mu}}\gamma_q(\alpha_s(\bar{\mu})),\\
S_M(t)&=&s\left (x_2\frac{m_{B_s}}{\sqrt{2}} r\eta^+,b_2 \right )
+s\left ((1-x_2)\frac{m_{B_s}}{\sqrt{2}} r \eta^+,b_2 \right)
+2\int_{1/b_2}^{t}\frac{d\bar{\mu}}{\bar{\mu}}
\gamma_q(\alpha_s(\bar{\mu})),
\eeq
with the quark anomalous dimension $\gamma_q=-\alpha_s/\pi$.
The explicit expressions of the functions $s(Q,b)$ can be found for example in Appendix A of Ref.~\cite{Lu:2000em}. The hard scales $t_i$ in
above equation are chosen as the largest scale of the virtuality of the internal particles in the hard $b$-quark decay diagrams,
\beq
t_1=\max\{\alpha_1, 1/b_1, 1/b_2\},\quad  t_2=\max\{\alpha_2,1/b_1, 1/b_2\}.
\eeq


\end{document}